\documentclass[prx,nofootinbib,twocolumn,showkeys,groupaddress,preprintnumbers,floatfix]{revtex4-1}

\usepackage{etex}
\usepackage{ifpdf}
\usepackage[hidelinks]{hyperref}
\usepackage{dcolumn}
\usepackage{url}
\usepackage{amsmath}
\usepackage{amscd}
\usepackage{amsfonts}
\usepackage{amssymb}
\usepackage{bm}   
\usepackage{bbm}
\usepackage{verbatim}
\usepackage{stmaryrd}
\usepackage{amsthm}
\usepackage{xcolor}
\usepackage{setspace}
\usepackage{braket}
\usepackage{empheq}
\usepackage{perpage} 
\usepackage{xspace}
\MakePerPage{footnote}

\usepackage{tikz}
\usetikzlibrary{arrows}
\usetikzlibrary{automata}
\usetikzlibrary{decorations.pathreplacing}
\usetikzlibrary{positioning}
\usetikzlibrary{plotmarks}
\usetikzlibrary{calc}
\usetikzlibrary{patterns}
\usetikzlibrary{external}
\usepackage{pgfplots}
\pgfplotsset{compat=newest}

\usepackage{graphicx}

\theoremstyle{plain}    
\theoremstyle{plain}    
\theoremstyle{plain}    
\theoremstyle{plain}    
\theoremstyle{plain}    
\theoremstyle{plain}    
\theoremstyle{plain}    
\theoremstyle{plain}    
\theoremstyle{plain}    
\theoremstyle{plain}    
\theoremstyle{plain}    
\theoremstyle{plain}    

\usepackage{wrapfig}
\newcommand{\eM}{$\epsilon$-machine}

\newcommand{\Alphabet}{\mathbb{A}}

\newcommand{\ShiftSpace} {\mathcal{X}}

\newcommand{\Koop}{\mathcal{K}}
\newcommand{\PF}{\mathcal{P}}
\newcommand{\Reals}{\mathbb{R}}

\newcommand{\past}{\overleftarrow{x}}

\newcommand{\future}{\overrightarrow{x}}
\newcommand{\Future}{\overrightarrow{X}}

\newcommand{\Benard}{B\'{e}nard\xspace}
\newcommand{\SecondLaw}{$2^{nd}$ Law\xspace}
\newcommand{\FirstLaw}{$1^{st}$ Law\xspace}
\newcommand{\dbar}{d\hspace*{-0.08em}\bar{}\hspace*{0.1em}}

\begin{document}

\title{On Principles of Emergent Organization}

\author{Adam Rupe}
\email{adam.rupe@pnnl.gov}
\affiliation{Pacific Northwest National Laboratory, 
Richland, WA. USA}

\author{James P. Crutchfield}
\email{chaos@ucdavis.edu}
\affiliation{Complexity Sciences Center and Physics Department,
University of California at Davis, One Shields Avenue, Davis, CA 95616}

\date{\today}
\bibliographystyle{unsrt}

\begin{abstract}
After more than a century of concerted effort, physics still lacks basic
principles of spontaneous self-organization. To appreciate why, we first state
the problem, outline historical approaches, and survey the present state of
the physics of self-organization. This frames the particular challenges arising from mathematical intractability and the resulting need for computational
approaches, as well as those arising from a chronic failure to define
structure. Then, an overview of two modern mathematical formulations of
organization---intrinsic computation and evolution operators---lays out a way
to overcome these challenges. Together, the vantage point they afford shows how
to account for the emergence of structured states via a statistical mechanics
of systems arbitrarily far from equilibrium. The result is a constructive path
forward to principles of organization that builds on mathematical identification of
structure. 
\end{abstract}


\preprint{arxiv.org:2310.XXXXX [cond-mat.stat-mech]}

\maketitle
\begin{spacing}{1.0}
\tableofcontents
\end{spacing}



\setstretch{1.1}

\section{Genesis}
\label{sec:Genesis}


Consider a fluid contained within a box. If isolated from its environment, the
fluid will be in thermodynamic equilibrium, a state of maximal symmetry with
uniform temperature and velocity fields unchanging in time.

Now, heat the box from below. When the temperature gradient is low, thermal
energy is transported up through the fluid via conduction, and the fluid
velocity remains everywhere zero. Symmetry is broken vertically, in the
direction of heat flux, but maintained in the orthogonal directions. As the
temperature gradient is further increased it will eventually reach a critical
value and a moment of magic occurs. Symmetry now breaks in the orthogonal
directions, in a spectacular manner. All of a sudden, and all at once, the
fluid conspires to form convective columns in, for example, a hexagonal
lattice, as convective heat transport becomes favored over conduction.


Symmetry-breaking in the orthogonal directions signifies the onset of
\emph{pattern formation}, a special case of the more general phenomenon of
\emph{spontaneous self-organization} that is our topic of discussion.


The fluid-in-a-box scenario just described is known as the B\'{e}nard
instability (sometimes Rayleigh-\Benard convection), first studied by Henri
\Benard in 1900, initiating the formal study of pattern
formation~\cite{Bena01a}. Over the following century it has become one of the
most studied phenomena in physics, both theoretically and
experimentally~\cite{Rayl16a,Chan68a,Buss78a,Fens79a,Stei85}. Despite the
intense study, deep mysteries persist with this system and with
self-organization more generally. 




Self-organization is ubiquitous in natural systems, at essentially all
scales---from patterning in quantum wave functions at sub-Plank-length scales
\cite{Zure01a} to biological morphogenesis \cite{Turi52} to the mass
distribution at the largest scales of the universe
\cite{Weyg09a,Schi14a,Newe19a}. 
For additional examples and phenomenology, see e.g. Refs.~\cite{Cros09a,Hoyl06a}.

The myriad of self-organizing systems found throughout nature emphasizes a basic question---What actually \emph{is} organization? 
Certainly a highly organized system will also be highly
correlated, but organization is a property beyond correlation---something
not quantifiable with a simple scalar value. We can recognize that the Great
Red Spot of Jupiter has an intricate level of organization, a blend of
randomness and regularity, but how do we specify this? How do we ``write down''
the Great Red Spot?

The fundamental question of formalizing organization, as a system characteristic, has largely been side-stepped in the study of the physical process of self-organization. For analytical tractability, historically much of the focus has been on exact symmetries and small perturbations from them. Only a small subset of this might we consider as directly addressing forms of organization. To understand how organization spontaneously arises, surely we first need a deeper understanding of what organization actually is.

\section{Narrative and Roadmap}
We start our journey in Section~\ref{sec:NonlinearDynamics} with an overview of
pattern formation theory, which currently provides the best framework for the
physics of self-organization. The present work addresses two shortcomings of
pattern formations theory: (i) it is not a ``thermodynamic'' theory of
self-organization and (ii) mathematical intractability prevents its application
to complex forms of organization that emerge far from equilibrium. As we will
show, only the latter is an actual shortcoming, the former only apparently so. 

We examine thermodynamic considerations of self-organization in Sections
\ref{sec:Equilibrium} and \ref{sec:nonequilibrium}. A continued distinction in
contemporary literature between pattern formation theory and the theory of
dissipative structures by Prigogine et al suggests that pattern formation is
generally not seen as a ``thermodynamic'' theory of self-organization. This
despite the two theories sharing the same core goal: Compute the
exchange of stability of nonequilibrium steady states as a function of an
external driving parameter. Since the theory of dissipative structures
approaches this problem by constructing an entropy-based Lyapunov function, one
must conclude that pattern formation theory is not considered a thermodynamic
theory as it does not directly invoke entropy or any of its nonequilibrium
generalizations. 

Given the persistent confusion and conceptual difficulty surrounding entropy
and its possible nonequilibrium generalizations, it is necessary in
Section~\ref{sec:Equilibrium} to set the stage with an in-depth review of
organization in thermodynamic equilibrium. We emphasize that entropy is not a
measure of ``disorder'' and detail its information-theoretic interpretation.
Section~\ref{sec:nonequilibrium} then begins with a review of irreversible
thermodynamics for local equilibrium field theories. While entropy and the
\SecondLaw provide a variational selection principle in equilibrium, as
described in Section~\ref{sec:Equilibrium}, Section~\ref{sec:nonequilibrium}
reviews the arguments and counterexamples against entropy production
variational principles for nonequilibrium steady state selection. Such a
selection principle, if one exists, would provide a base level understanding of
pattern formation---a patterned state arises because it is the unique state
that satisfies the selection principle.  

While historically notable, dissipative structures must be relegated to history
as a failed theory of self-organization. Section~\ref{sec:nonequilibrium} D
recalls the arguments invalidating dissipative structures by showing the
proposed ``excess entropy production'' is not a Lyapunov function and thus has
no bearing on the stability of nonequilibrium steady states. Given more than
four decades ago, these arguments disproving dissipative structures seem to be
largely forgotten. Section~\ref{sec:nonequilibrium} E uses the subadditivity
property of entropy to show that the differential entropy balance equation, key
in the development of dissipative structures, cannot hold for organized
systems. This provides theoretical insights into why dissipative structures
does not hold in general.

In addition, Section \ref{sec:nonequilibrium} A shows that the nonequilibrium
generalizations of entropy and the \SecondLaw are not necessary to derive the
nonequilibrium transport equations used by pattern formation theory and
dissipative structures. It is these equations that determine nonequilibrium steady states and their stability. Therefore, entropic quantities do not
explicitly appear in pattern formation theory as they are simply not needed.
One of our primary goals is to convey that pattern formation exactly supplants
the role of a thermodynamic theory of self-organization as attempted by
dissipative structures.

Having addressed the first shortcoming of pattern formation theory---by showing it is not actually a shortcoming---we then turn to address the limits of its applicability. Section~\ref{sec:Intractability} details how pattern formation theory fits into the reductionist paradigm of constructionism for explaining physical phenomena. In this, it is necessary to formulate mathematical equations to model the phenomenon of interest and then deduce physical consequences from the model. There are two main challenges hindering the extension of pattern formation theory beyond simplified systems near equilibrium. 

First is the mathematical tractability of computing the stability of steady
states (or time-dependent states) from more complicated sets of coupled partial
differential equations with difficult boundary conditions. We review
uncomputability and \textbf{P}-Completeness from the theory of computation, in
the context of constructionism, to highlight formal limits on our ability to
deduce physical consequences from complicated mathematical models. This places
a fundamental restriction on the scope of phenomena that can be explained using
the tools of pattern formation theory. 

Second, analyzing a particular form of organization via pattern formation
theory---think of the Great Red Spot---requires a way to ``write it down'' such
that the organization can be substituted into the governing equations of motion
and evolved. However, such complex organization eludes traditional
representational bases like Fourier modes or wavelets. Any theory that aims to
explain self-organization far from equilibrium, whether a direct extension of
pattern formation theory or otherwise, must have some way to mathematically
represent the complex organization that arises. This is the forefront of the
physics of self-organization. 

Section~\ref{sec:FormalizingOrganization} addresses this frontier with a review
of data-driven methods to formalize complex organization: evolution operators
and intrinsic computation. Necessarily, these are outside the constructionist
paradigm. Rather than attempt to write down organization in terms of a
pre-specified basis templates, intrinsic representations are learned directly
from system behaviors. The section begins with general computation-theoretic
motivation for learning representations of organization through compression---of
representational resources, not of a gas by a piston, for example.

Evolution operators provide Hilbert space dynamics for classical field
theories. Koopman operators evolve system observables, analogous to the
Heisenberg picture of quantum mechanics; while Perron-Frobenius operators
evolve densities, analogous to the Schr\"{o}dinger picture. Eigenfunctions of
these operators provide a natural representational basis of system behaviors
that are intrinsic to the system's dynamical evolution. Compression is achieved
by retaining a finite set of eigenfunctions---functions that capture any
organization present in the system. The leading Koopman modes---projections
onto Koopman eigenfunctions---provide a statistical notion of organization for
time-independent systems. Almost-invariant sets---leading eigenfunctions of
Perron-Frobenius operators---and related coherent sets identify coherent
structure and organization related to transport in time-dependent systems.
Data-driven algorithms implementing the operators then provide rigorous
approximations that can be deployed in practice on real-world systems. 

In a complementary way, intrinsic computation achieves compression using the
idea of predictive equivalence: if two past histories of system behaviors lead
to the same possible set of future behaviors, those past histories are
considered equivalent. And, they are grouped together as they give the same
predictive information about the system's future.

For discrete statistical field theories, predictive equivalence provides the
unique minimal model for optimal prediction of system behaviors. This model has
a semigroup algebraic structure that generalizes exact symmetries and their
group algebra. Intrinsic computation rigorously identifies structure and
organization as generalized symmetries. It considers organization to be a
predictive regularity. And, in so doing, it covers the full spectrum from the
perfect regularity of exact symmetries to the absence of regularity for
independent random events. Interesting organization, like Jupiter's Great Red
Spot, is found somewhere in between, with a mixture of order and randomness.

For classical field theories, predictive equivalence of lightcones identifies
coherent structures and organization through generalized spacetime symmetries
and local deviations from them. We show there is a close relation, both
conceptually and in data-driven approximation, between coherent structures
identified through predictive equivalence and through evolution operator
coherent sets. 

Although we now have representations in hand for complex organization, these
cannot be ``plugged in'' to governing equations, as would be done with pattern
formation theory---evolution operators and intrinsic computation fall outside
the constructionist paradigm. Data-driven scientific understanding of emergent
behavior beyond constructionism is a large and ongoing endeavor. As such, it is
not fully clear what ``principles of organization'' will ultimately look like
in this new paradigm.

That said, Section~\ref{sec:StatMechOrganization} closes with an intriguing
path forward, bringing together evolution operators and predictive equivalence
to provide a statistical mechanics of emergence. While higher-level emergent
behaviors cannot (generally) be deduced from their lower-level governing
equations, our statistical mechanics provides the physical foundations for a
complete and self-contained dynamics governing the emergent
behaviors---dynamics that is consistent with the lower-level physics. Thus,
evolution operators and intrinsic computation represent complex organization as
emergent higher-level degrees of freedom while simultaneously providing the
complete and consistent dynamical laws for their evolution. This again is all
behavior-driven and so outside the constructionist paradigm: higher-level
degrees of freedom cannot be written out analytically and their dynamics are
not (generally) expressed by any closed-form equations of motion. 


\section{Nonlinear Dynamics}
\label{sec:NonlinearDynamics}

Dynamical systems theory provides a geometric view of structures in a system's
state space---structures that guide and constrain nonlinear
behaviors, amplifying fluctuations and eventually attenuating them into
macroscopic behaviors and patterns. A canonical example is turbulence---a
dynamical explanation for which occupied much of the 70s and 80s.

\subsection{Structure in Turbulence}
\label{sec:Turb}

Fluid turbulence was long recognized as the pre-eminent problem in
thermodynamic self-organization; important too in nonlinear physics according
to Heisenberg \cite{Heis67a}---whose dissertation was on turbulence. Landau had
introduced the idea that turbulent behavior consisted of a collection of
incommensurate oscillators---a sufficient number of which could produce the
experimentally-observed power spectra. This was superseded by the mathematical
discovery in the 1950s of chaotic attractors and the 1970s conjecture that they
were the state of turbulent flow \cite{Ruel71a}. This was experimentally
verified only in the 1980s \cite{Bran83}, overthrowing the Landau theory.

The implicated mechanism was one that amplified microscopic fluctuations
exponentially-fast to macroscopic scale. As they reached observable scales they
were guided by complex state-space structures, eventually condensing into
complex spatial patterns as nonlinearities attenuated their growth. This
understanding was eventually articulated in the subfield of pattern formation
\cite{Turi52,Cros93a,Hoyl06a,Cros09a}.

While describing and predicting the state-space structures that guide and
constrain the behavior of complex systems is clearly an essential insight from
dynamical systems, this too falls short of leading to a principle of emergent
organization. Patterns emerge, but what exactly are they, and what complex
behavior do they exhibit?

\subsection{Pattern Formation Theory}
\label{sec:PaternFormation}

Patterns are born out of conflict and compromise. Whenever patterns form (out of
equilibrium) there are two competing forces at play: the inexorable pull towards
thermodynamic equilibrium---dissipation due to the \SecondLaw---and an external push away from
equilibrium---drive given by gradients in intensive quantities and fluxes in extensive quantities.
To monitor the competition, we define a dimensionless \emph{bifurcation parameter}
$R$ that is the ratio of the competing forces:
\begin{align*}
R \propto \frac{\mathrm{Drive}}{\mathrm{Dissipation}}
  ~.
\end{align*}
$R$ can be thought of as a proxy for the distance from thermodynamic equilibrium.

To monitor the B\'{e}nard instability in fluid flow, one uses the Rayleigh
number $R$ which is proportional to $\Delta T / \nu \kappa$, where $\Delta T$
is the temperature difference across a system, $\nu$ is the kinematic
viscosity, and $\kappa$ the thermal diffusivity. (Additional constants are
included to nondimensionalize $R$.) From $R$ we see that driving and
dissipation are crucial ingredients for nonequilibrium organization---a
constant supply of energy, and potentially matter, must be continuously pumped
into the system and simultaneously dissipated away to maintain nonequilibrium
organization. We explore these thermodynamic considerations shortly.


We assume the system is governed at the macroscopic level by an effective field
theory, given as a set of nonlinear partial differential equations
\cite{Cros09a}:
\begin{align}
\partial_t X(\mathbf{r},t)
  = F \big(X(\mathbf{r},t), \partial_r X(\mathbf{r},t), ... ; R \big)
  ~.
\label{eq:gen_dyn}
\end{align}
The system state is given by $X$, which varies smoothly and continuously in
space $\mathbf{r}$ and time $t$. Time evolution, as given by
Eq.~(\ref{eq:gen_dyn}), is governed by local interactions (finitely many
spatial derivatives of $X$) that are applied uniformly in space and time; i.e.,
they do not have explicit space or time dependence. Additionally, the dynamics
of the full system depends on external conditions as encapsulated by the
bifurcation parameter $R$.

For fluid flows, these are the Navier-Stokes equations (or some approximation
thereof), given as 
\begin{align*}
    \frac{\partial \mathbf{u}}{\partial t} + \left(\mathbf{u} + \nabla \right)\mathbf{u} = - \nabla p + \frac{1}{Re} \nabla^2 \mathbf{u}
    ~,
\end{align*}
for incompressible flows, where $\mathbf{u}(\mathbf{r}, t)$ is the fluid velocity field, $p(\mathbf{r}, t)$ is the pressure field, and $Re$ is the Reynolds number.
We consider Eq.~(\ref{eq:gen_dyn}) as a \emph{macroscopic} description
of the system due to the presence of dissipation in the form of $R$.
Dissipation means that macroscopic state space volumes decrease over time. By
contrast, a \emph{microscopic} description---think of the collective state of
the fluid's constituent molecules---obeys Liouville's theorem that says
microstate state space volumes are preserved over time.
Reconciliation of dissipative macroscopic dynamics arising from conservative
microscopic dynamics---the thermodynamic arrow of time---has a long and storied
history, starting with Boltzmann~\cite{boltz1896a,brush16a}; see e.g. Refs.~\cite{Mack92a} and \cite{lebo93a} for modern reviews. 


With fixed, gradient-free boundary conditions $R=0$ and the system reaches
thermodynamic equilibrium with $X$ uniform in space and time. If the boundary
conditions are fixed with nonzero gradients, $R$ is nonzero and the system
reaches a \emph{nonequilibrium steady-state}. Close to equilibrium, with $R$
small but nonzero, the nonequilibrium steady-state is known as the \emph{base
state}, which we denote as $\bar{X}$. The base state shares the symmetries of
the boundary conditions. And so, if the boundary conditions are
time-independent, then the base state will also be time independent:
\begin{align*}
\partial_t \bar{X}(\mathbf{r},t) = 0
  ~.
\end{align*}

Below a critical $R$ value $R_c$, the base state is stable. All infinitesimal
perturbations---e.g., thermal fluctuations---exponentially decay as:
\begin{align*}
\delta X(\mathbf{r},t) = Ae^{\sigma t}e^{ik\cdot \mathbf{r}}
  ~.
\end{align*}
For $R < R_c$, the growth rate $\sigma$ of perturbations is negative. At $R_c$,
though, the growth rate of a perturbation at wavenumber $k_c$ becomes zero for
the first time. Linear stability analysis shows that this critical mode begins
to grow as the system moves through $R_c$. Perturbation theory (e.g., amplitude
equations~\cite{Newe74a,Cros09a}) shows how this growing mode saturates just
above $R_c$ to create a patterned state. Which mode grows and how it saturates
are dictated, in this close-to-equilibrium regime, by the geometry of the
boundary conditions.


This simple exposition of \Benard convection illustrates the mathematics of
\emph{bifurcation theory}: The base ``conduction'' state becomes unstable at
the critical Rayleigh number when convection takes over from heat conduction
and saturates to form a patterned state. There is an exchange of stability, from the previously-stable base state to the now-stable patterned state, as $R$ moves through $R_c$. Similar bifurcation-theoretic analyses
have successfully predicted many similar pattern-forming phenomena that closely
agree with experiment \cite{Cros93a, Cros09a}.\footnote{An analogous
explanatory framework for some pattern formation phenomena falls under the
rubric of ``symmetry-breaking''. As symmetry plays such a pivotal role in
physics, the symmetry-breaking perspective is a fundamental one.
Mathematically, though, it is a special case of dynamical system's bifurcation
theory. And, in any case, this is a bit of a distraction, since organization is
more than a collection of broken symmetries, as we will show.}

So, are pattern formation and self-organization, more generally, ``solved''?
Clearly this note would not be necessary if this were the case. What mysteries
remain then?

Bifurcation theory has been most successful in describing and predicting the
near-equilibrium ``primary bifurcation'', as just described above with small
$R$ such that the base state going unstable during the bifurcation is the
nonequilibrium steady-state closest to equilibrium. Continuing on, though, the
patterned state that is the stable steady-state after the primary bifurcation
may itself become unstable as $R$ is increased and as the system is driven
further from equilibrium. Many important natural occurrences of
self-organization exist in these far-from-equilibrium regimes. For example,
hurricanes cannot be explained as finite-amplitude instabilities as bifurcation
theory would attempt \cite{Eman91a}.

In short, today no general theory predicts what patterns and organization may
emerge far from equilibrium. In contrast to equilibrium phase transitions and
near-equilibrium bifurcations, there does not appear to be any notion of
universality in the forms of organization that appear. As stated by Harry
Swinney~\cite{Swin00a}:
\begin{quote}
Far beyond the primary instability, each system behaves differently.
Details matter... There is no universality ....
\end{quote}
Or, from Philip Ball~\cite{Ball99a}:
\begin{quote}
... the patterns of a river network and of a retinal nerve are both the
same and utterly different. It is not enough to call them both fractal, or even
to calculate a fractal dimension. To explain a river network fully, we must
take into account the complicated realities of sediment transport, of changing
meteorological conditions, of the specific vagaries of the underlying bedrock
geology--things that have nothing to do with nerve cells.
\end{quote}

Compounding the challenge, even in the near-equilibrium regime, where
bifurcation theory has been so successful, there are still mysteries. When the
base state goes unstable, why should the newly-stable state exhibit any degree
of organization, let alone organization not present in the base state? Is there
a general \emph{selection principle} that dictates the necessity of intricate
organization in the system?

This question takes us to our next field of study, equilibrium statistical
physics, where there \emph{is} a general state selection principle in
thermodynamics' \SecondLaw. It also provides a framework that has been used to
try to understand another mystery. The nonlinear dynamics analysis above
addresses the macroscopic level, in terms of continuum field theories.
However, we know this is only an approximation to the collective motion of an
enormous number of discrete particles. How do the water particles conspire to
organize together into convective columns in \Benard convection, for example?
The macroscopic patterns that emerge live on a scale that is about one million
times larger than the characteristic length scales of individual water
molecules. 

\section{Equilibrium Statistical Physics}
\label{sec:Equilibrium}


The phenomenology of pattern-forming systems like \Benard convection certainly
has thermal aspects, as does the bifurcation analysis just described. The
instability is thermally driven by a temperature gradient, and the bifurcation
parameter was described in thermodynamic terms as a distance from equilibrium.
Yet, most do not consider the nonlinear dynamics theory of pattern formation as
a \emph{thermodynamic theory}. The most likely reason it is not considered so
is that it does not invoke the most central quantities of
thermodynamics---energy and entropy---nor the cornerstone Laws governing
thermal systems---the \FirstLaw and \SecondLaw. While the dynamical framing of
some forms of self-organization may involve variables and parameters that refer
to energy and entropy or equations of motion that refer to thermodynamic laws,
these are neither always necessary nor foundational to the dynamical analyses.
In short, dynamical systems theory stands on its own without mention of entropy
or the \SecondLaw. Its purview is behavior arising in any state space governed
by general dynamics---with energetics or not, inside or outside physics.

\Benard convection has been the guiding example thus far and, arguably, the
\FirstLaw \emph{is} central in this case. The equations of motion on which
bifurcation analysis is carried out are derived based on energy and mass
conservation. Further, Chandrasekhar analyzed the primary \Benard instability in
terms of energy balance between viscosity and buoyancy; ``Instability occurs at the minimum temperature gradient at which a balance can be steadily maintained between the kinetic energy dissipated by viscosity and the internal energy released by the buoyancy force''~\cite[pg. 34]{Chan68a}.

To stress the point here, though, there are pattern-forming real-world
phenomena whose mathematical models are constructed with regard to neither
energy nor entropy. One might imagine recent attempts to describe emergent
social order \cite{vics95a,tone5a,taji01a,aren08a}, where energy is irrelevant
and not defined, in any case. Consider an even simpler setting---the
well-studied arena of cellular automata (CAs)
\cite{Gras83a,Grass86b,Crut93a,Hans95a}. Much of their investigation rather
directly arose due to the wide range of intricate patterns that they readily
generate. Pattern formation in CAs, though, does not occur through the kind of
bifurcation considered here. Generally, there is no system invariant quantity
that is the analog of energy to be conserved.

To make the distinctions clearer, the setup so far may be described as ``physical
pattern formation'', particularly since the bifurcation parameter $R$ has indeed
been formulated in energetic terms. The nonphysical self-organizing examples serve
to highlight that pattern formation arises more broadly in dynamical systems.

While energy is thus not necessary for pattern formation, the conservation of energy provides a simplification in analyzing physical dynamical systems, as well as a unified theoretical framework. All physical systems, from steam engines to motor proteins, obey the \FirstLaw. Similarly, all physical systems obey the venerated \SecondLaw. This universality makes thermodynamics one of the most powerful and widely-employed theories in science. It is inevitable then, to wonder if there are thermodynamic principles of self-organization in physical systems given in terms of energy and entropy. 

As we will now detail, much of the confusion surrounding self-organization, and
our understanding of it, stems from the subtleties of entropy and the
\SecondLaw for equilibrium systems and from attempts to generalize these to
out-of-equilibrium systems.  The following first contends with entropy and the
\SecondLaw before moving to survey historical attempts at general
nonequilibrium Laws and related attempts at thermodynamic theories of
self-organization.

Particular emphasis is given to the theory of \emph{dissipative structures},
often treated as synonymous with thermodynamics of self-organization. We recall
the counterexamples and arguments invalidating the theory, which appear to have
been largely forgotten. We then show that the generalizations of the
$1^{\text{st}}$ and $2^{\text{nd}}$ Laws given by the classical theory of
irreversible thermodynamics, the starting point of dissipative structures,
simply yield the nonequilibrium transport equations of Eq.
(\ref{eq:gen_dyn})---the starting point of dynamical systems pattern formation
theory just reviewed.

\subsection{Equilibrium Formalism}

Due to the subtleties of thermodynamics, both in equilibrium and particularly in
attempts to generalize out of equilibrium, it is worth stepping back to firmly
ground ourselves in its formalism and terminology. Equilibrium thermodynamics is a
phenomenological theory that formalizes empirical observations of allowable
manipulations for macroscopic systems. A thermodynamic \emph{system} is
characterized by a collection of relevant thermodynamic \emph{properties}, some of
which are \emph{extensive} and scale linearly with system size---such as, volume and
particle number---while others are \emph{intensive} and are invariant to system
size---such as temperature and pressure.

A \emph{transformation} of a thermodynamic system is a change in its properties.
Central to equilibrium thermodynamics is the notion of a \emph{state variable}: a
property whose changes do not depend on the particular transformations taken. State
variables may be defined more precisely as properties with exact differentials
\cite{cart01a}, but the following does not require this level of detail. More
loosely, the \emph{state} of a thermodynamic system is uniquely specified by a set
of properties, and an \emph{equilibrium state} is one in which the system's
intensive properties are uniform in space and all properties do not change over
time. State variables are the necessary and sufficient properties for completely
describing an equilibrium thermodynamic state. 

A \emph{path} is a continuous sequence of states through which the system passes,
and a \emph{process} specifies a path that maintains a particular relation among
state variables---called an \emph{equation of state}. For example, an isothermal
process is a path along which the temperature does not change. A \emph{quasistatic
process} is one for which the system is instantaneously always in an equilibrium
state, so that if external conditions are suddenly fixed the system is
instantaneously in equilibrium at the current state. A \emph{reversible process} is
one for which the direction along the path can change due to an infinitesimal change
of external conditions. For example, heat flow between two systems with temperatures
infinitesimally close is reversible since an infinitesimal change in temperature
will change which system is hotter than the other. 
Note that reversible processes are
quasistatic, but not all quasistatic processes are reversible.
Finally, a \emph{cycle} is a path
that starts and ends in the same state. Since changes in
state variables are path independent, they remain unchanged after a complete cycle. 

To ground the formalism, consider two thermodynamic systems at different
temperatures, $T_1$ and $T_2$, that only exchange thermal energy $U$ between each
other in the form of heat $Q$. It is observed that heat flows between the two
systems, from high temperature to low temperature, until the two temperatures are
equalized: $T_1 = T_2$. The equilibrium state for the composite system---which as a
whole is an isolated system---is achieved when the intensive property of temperature
is uniform; i.e., the two subsystems reach the same temperature. The net flux of the
extensive property $U$ vanishes.

Transformations between equilibrium states are governed by the First and Second
Laws of Thermodynamics. The \FirstLaw formalizes the observation that energy is
conserved in isolated systems---those that have no external interactions. In
particular, the \FirstLaw is typically stated to show that thermal energy, in
the form of heat $Q$, and mechanical energy, in the form of work $W$, may both
contribute to changes in a system's internal energy $U$:
\begin{align}
    dU = \dbar Q + \dbar W
    ~.
    \label{eq:firstlaw}
\end{align}
The notation $\dbar$ indicates inexact differentials whose values are path-dependent. Since the internal energy $U$
is a state variable, it has an exact differential and the difference $\Delta U = U_2
- U_1$ between two thermodynamic states does not depend on the particular path taken
between them. Whereas, heat and work are not state variables and typically have
inexact differentials. Therefore, their values $\dbar Q$ and $\dbar W$ generally do
depend on the particular path taken between the two thermodynamic states. 

Conservation of energy, the \FirstLaw of Thermodynamics, is one of the most powerful
theories in physics and expresses a fundamental symmetry of nature. The \FirstLaw
originated as a mathematical formulation of empirical observations of energy
conservation, particularly Joule's (at the time surprising) results on ``the
mechanical equivalent of heat'' \cite{joule1850a}. The mathematical formulation in
Eq.~(\ref{eq:firstlaw}) necessitated the introduction of the \emph{internal energy}
property, a state variable, to express the observed conservation law. 

The \FirstLaw expresses a symmetry through time and, thus, cannot indicate a
preferred direction in a system's temporal evolution. If a process conserves
energy in forward time, it must necessarily conserve energy in reverse time. In
our simple composite-system heat exchange example, energy is conserved if the
heat flowing out of one subsystem equals the heat flowing into the other. As
far as the \FirstLaw is concerned, it does not matter whether heat flows from
hot to cold or cold to hot. For macroscopic systems, it is empirically observed
that in such scenarios, there \emph{is} a preferred direction of heat flow.
Heat flows ``spontaneously'' from hot to cold and never the other way around.
Spontaneous here means there are no other sources of energy change, such as
work, that may drive a heat flow from cold to hot; e.g., no refrigeration. 

The \SecondLaw was introduced as a means to formalize the empirical observation
of thermodynamic processes that occur spontaneously in one direction but not
the reverse, even though both directions conserve energy and thus satisfy the
\FirstLaw. As the state variable of internal energy $U$ was introduced to
mathematically encode the \FirstLaw, a new state variable \emph{entropy} $S$
was introduced to formalize the \SecondLaw. Building on the earlier work of
Carnot \cite{carn1824a}, Clausius recognized the significance of the ratio
$\dbar Q / T$ in governing the direction of thermodynamic processes
\cite{crop86a}.  Using this ratio, he mathematically formulated his statement
of the \SecondLaw:
\begin{align}
    \oint \frac{\dbar Q}{T} \leq 0
    ~.
    \label{eq:clausius_inequality_int}
\end{align}
That is:
\begin{quote}
\textit{it is impossible to construct a device that operates in a cycle and
whose sole effect is to transfer heat from a cooler body to a hotter body}.
\end{quote}
Several other equivalent ``impossibility statements'' of the \SecondLaw follow
from the Clausius inequality, such as the impossibility of creating a
perpetual-motion machine. 

Equality in Eq.~(\ref{eq:clausius_inequality_int}) is achieved if and only if
the process is reversible, allowing for the introduction of the entropy state
variable whose change is given by:
\begin{align}
    dS = \frac{\dbar Q_r}{T}
    ~, 
    \label{eq:clausius}
\end{align}
where $\dbar Q_r$ is a reversible heat flow. The Clausius inequality
Eq.~(\ref{eq:clausius_inequality_int}) then takes its more familiar form:
\begin{align}
    dS \geq \frac{\dbar Q}{T}
    ~,
    \label{eq:clausius_inequality}
\end{align}
for arbitrary heat flow $\dbar Q$. The most common statement of the \SecondLaw 
follows: 
\begin{quote}
\textit{The entropy of an isolated system ($\dbar Q = 0$) increases over time
for irreversible processes or remains constant for reversible processes.}
\end{quote}

The Clausius entropy in Eq.~(\ref{eq:clausius}) is sometimes referred to as
\emph{physical entropy}, since reversible heat flow and temperature are quantities
that can be measured in the lab. Strangely, while entropy was introduced to formalize
the observed irreversibility of some thermodynamic processes, precise values of
(changes in) entropy can only be calculated for reversible processes. Equilibrium
thermodynamics is thus sometimes seen as a theory of reversible processes.

There is no paradox however, owing to the path-independence of state variables. Due
to path-independence, the entropy change between two equilibrium states is the same
for an irreversible path between those states as for a reversible path between them.

To avoid confusion, we emphasize that the Clausius inequality does \emph{not} state
that the entropy change is different for reversible and irreversible processes. It
is the relationship between changes in entropy (path-independent) and heat flow
(path-dependent) that differs between reversible and irreversible processes. This
point will be clarified shortly.

The power of path-independence is also seen when combining the \FirstLaw and
\SecondLaw. Note that the statement of the \FirstLaw in Eq.~(\ref{eq:firstlaw}) has
a path-independent change in a state variable on the left hand side given in terms
of path-dependent quantities of heat and work on the right hand side. For reversible
processes, we can substitute the physical entropy along with $\dbar W_r = -P dV$,
where $P$ is the pressure and $V$ the volume, to get the fully path-independent
relation:
\begin{align}
    dU = TdS - PdV
    ~.
    \label{eq:state_relation}
\end{align}
Due to path-independence, this important relation is fully general for reversible
and irreversible processes alike. It is only for reversible processes that the right
hand terms are identified as heat and work. 

There is an additional path-independent contribution to changes in internal energy
that occurs due to exchange of particles, sometimes known as \emph{chemical work} in
analogy with the mechanical work $\dbar W_r = -P dV$. Including this term gives one
of the most important relations in classical thermodynamics, the \emph{thermodynamic
identity}:
\begin{align}
    dU = T dS - P dV + \sum_k\mu_k \; dN_k
    ~,
    \label{eq:thermo_identity}
\end{align}
where $\mu_k$ is the intensive chemical potential and $N_k$ the extensive number of
particles of species $k$. Rearranging the identity, we see that the entropy of a
thermodynamic system changes due to exchanges of internal energy (thermal
interactions), volume (mechanical interactions), or particles (diffusive or chemical
interactions). In this way, the thermodynamic state of a large class of systems
(e.g., those without magnetic interactions) is fully specified by the three state
variables $(U,V,N)$. 

\subsection{Equilibrium Organization}

Entropy is one of the more subtle and apparently mysterious concepts in the physical
sciences. While it may be convenient to side-step this subtlety by referring to
entropy as a ``measure of disorder'', this does a great disservice, muddying key
issues. After all, if entropy is disorder, then the \SecondLaw states that disorder
always increases. And, if self-organization is the spontaneous formation of order,
is not the phenomenon at odds with the venerated \SecondLaw?

Sadly, when Boris Belousov discovered an oscillating chemical reaction in the
early 1950s he was unable to publish precisely due to this misconceived
conflict; this despite his providing the recipe for others to reproduce his
experiment~\cite{Winf84a}. It was perfectly acceptable for, say, clear chemical
reagents to mix and turn to opaque homogeneity. It was thought impossible at
the time, due to this misunderstanding of entropy and the \SecondLaw, for these
reagents to apparently unmix and the solution turn clear again, as happens with
the ``chemical clock'' discovered by Belousov.

Today, the Belousov-Zhabotinsky (BZ) chemical reaction-diffusion system
\cite{Zhab91a} has become a prime example of pattern formation, much like
\Benard convection in hydrodynamics. There are two key reasons why spontaneous
self-organization, like that observed by Belousov, does not exist in opposition
to the \SecondLaw.  First, as hopefully made clear above in
Sec.~\ref{sec:NonlinearDynamics}, self-organization in the BZ reaction and in
\Benard convection occurs outside of equilibrium, where the \SecondLaw does not
apply. In the case of the BZ reaction, a chemical concentration gradient is
the thermodynamic force maintaining the system out of equilibrium. Second,
entropy is \emph{not} strictly a measure of disorder. In point of fact,
organization may emerge in equilibrium systems, such as strongly correlated
electron materials~\cite{kive98a,zhen17a,Qian17a} and nematic phases of
superconductors~\cite{fern14a}. More on this momentarily. 

Self-organization in equilibrium systems occurs \emph{due} to the
\SecondLaw, not in opposition to it. Consider, for example, the role of free energy
in the Landau theory of phase transitions \cite{Land69a}. Analogous to the conflict
and compromise of driving and dissipation out of equilibrium, there is
conflict and compromise in balancing free energy between internal energy and
entropy. The system evolves to minimize energy and reach its ground state, but is
prevented doing so by thermal fluctuations. Self-organization in equilibrium systems
is dictated by the \SecondLaw, meaning that the organization which emerges
represents the highest entropy state of the system commensurate with the given
constraints. 

As for the \SecondLaw itself, it must be emphasized that it only applies to isolated
systems in equilibrium. A subtlety arises when considering composite systems---the
simplest case being a \emph{system of interest} and an \emph{environment} with which
it interacts. The system is thus not isolated from the environment, but the
environment is suitably defined such that the composite \emph{system + environment}
is isolated---it does not interact with anything else. The \SecondLaw then states
the total entropy of \emph{system + environment} is nondecreasing. Yet, it cannot
say anything specifically about the system itself. Moreover, it has nothing to say
about the ``order'' that may or may not arise in the system of interest.

We can further clarify this point using the expression for entropy changes from
classical irreversible thermodynamics~\cite{Prig68a,DeGr62a}. For a system interacting
with an environment in equilibrium, the change of system entropy $S$ is:
\begin{align}
dS =d_e S + d_i S
  ~.
\label{eqn:dS}
\end{align}
Here, for simplicity, $d_e S = dQ / T$ denotes a change due to energy exchange
(under fixed volume and no particle exchange) and $d_i S \geq 0$ are changes due
to ``irreversible processes''. 
Equation~(\ref{eqn:dS})'s decomposition nicely encapsulates many well-known
properties of entropy changes under thermodynamic transformations:
\begin{itemize}
      \setlength{\topsep}{0pt}
      \setlength{\itemsep}{0pt}
      \setlength{\parsep}{0pt}
\item \emph{Clausius inequality}: \\ $dS \geq dQ / T$;
\item \emph{Reversible transformation}:\\
	$d_i S=0 \Rightarrow dS = dQ / T$;
\item \emph{Reversible cycle}: \\$dS = 0$ and $d_i S = 0 \Rightarrow dQ / T = 0$;
\item During an \emph{irreversible cycle} the system dumps waste heat into
	environment: \\$dS=0 \Rightarrow d_e S = - d_i S < 0 \Rightarrow dQ<0$; and
\item \SecondLaw: \\For an isolated system $d_e S= 0 \Rightarrow dS \geq~0$.
\end{itemize}

Note that for general transformations, the entropy of a system interacting with
an environment may increase or decrease: $dS$ may be positive or negative
depending on the balance of driving $d_e S$ and dissipation $d_i S$. Thus, even
if entropy \emph{were} a measure of disorder, spontaneous self-organization
(in equilibrium) would not be in conflict with the \SecondLaw. Spontaneous
order in a system is allowed, as long as it is paid for with a
compensating increase in the environment's entropy. 

\subsection{What \emph{is} Entropy?}
If entropy is not a measure of disorder, what is it and why is it so often conflated
with disorder? In short, as Jaynes emphasized early on \cite{Jayn57a}, entropy is a
measure of uncertainty in a system's microscopic state, given the constraints
specified by the macroscopic state. This uncertainty, applied too broadly to the
\emph{organization} of system configurations, is the origin of entropy's
disorderedness interpretation. A system we describe as ``more ordered'' on the
macroscopic scale will often, but not always, provide more constraint on the
microscopic state.

The prime example of this mismatch is a phase transition with the higher-entropy
state coinciding with a more ordered configuration \cite[pg. 56-57]{Gran08a}. This
occurs, for example, in the nematic transition of liquid crystals studied by Onsager
\cite{Onsa49}. Consider a collection of many cylindrical rods contained in a
two-dimensional box. Now, vary the density of the rods by changing the box's size.
At higher density the rods get jammed into random ``disorderly'' orientations.  As
the box's size increases and the rods are free to move around, they prefer to orient
themselves into more ``orderly'' raft-like shapes. The decrease in orientational
degrees of freedom is more than compensated for by the increase in translational
degrees of freedom in the lower-density state. Therefore, the seemingly ordered
state actually has higher entropy than the randomly-arranged disordered state.
Similarly, gravitational systems may also evolve to ordered configurations of higher
entropy \cite{Moor03a}.

While Jaynes' information-theoretic interpretation of entropy in statistical
mechanics has been elaborated upon at length
elsewhere~\cite{Jayn57a,Naim08a,Gran08a}, it is worth going into some detail
here. Recall from above that classical equilibrium thermodynamics is a
phenomenological theory introduced to explain, most broadly, empirical
observations of irreversible processes. The theory applies to what we now call
\emph{macroscopic systems}. It was only later that atomic theory was introduced
and experimentally verified. In this, macroscopic systems are seen as being
composed of a large number of universal microscopic ``building blocks''---what
we will simply call \emph{particles}---that obey the laws of quantum mechanics.
That noted, the classical laws of Newtonian mechanics will suffice for our
purposes here.

The thermodynamic state is thus called the \emph{macrostate} in the context of
statistical mechanics, since state variables are macroscopic properties. Roughly,
macroscopic properties are those we experience on the human scale, whereas
microscopic properties---i.e., positions and momenta of constituent particles---we
do not experience. For example, we feel heat flow when touching a surface at a
different temperature from our skin, while we do not feel the collisions of
individual particles from that surface. 

Clausius, Maxwell, and Boltzmann were key contributors linking macroscopic thermodynamics to the microscopic kinetic behavior of particles \cite{brush76a}, and it was Boltzmann who first made the connection between macroscopic entropy and microscopic kinetics \cite{boltz1909,brush76a}. 
His key observation was
that typically there will be many different microstates that correspond to a single
macrostate. That is, there may be many different collections of particle positions
and momenta that give rise to the same total internal energy. For a given
thermodynamic macrostate $(U,V,N,\ldots)$, the number of corresponding microstates
is a state variable called the \emph{multiplicity}, denoted as
$\Omega(U,V,N,\ldots)$. Boltzmann's formulation of entropy as:
\begin{align}
    S = k \ln \Omega
\label{eq:boltzmann}
\end{align}
follows from three key assumptions:\\[-20pt]
\begin{enumerate}
      \setlength{\topsep}{-2pt}
      \setlength{\itemsep}{-2pt}
      \setlength{\parsep}{-2pt}
	\item All (accessible) microstates are equally probable,
	\item Entropy is a function of $\Omega$, and
	\item Entropy is \emph{additive}.
\end{enumerate}
Additivity states that when combining two systems $1$ and $2$ together, the entropy
of the full composite system is equal to the sum of entropies of the two constituent
systems: $S_{12} = S_1 + S_2$. Additivity in particular is a strong assumption that
we examine in more detail shortly, as it has immediate implications for a
thermodynamic theory of organization. 

Our statement of the \SecondLaw above is the most general form of the
``impossibility'' versions of the \SecondLaw. It forbids thermodynamic processes
that decrease entropy of an isolated system, such as heat flowing from a cold
subsystem to a hot subsystem. However, there is an alternative, more constructive,
perspective one can take of the \SecondLaw that is most salient to thermodynamic
theories of organization. As first formulated by Gibbs \cite{gibbs1875a}, the
\SecondLaw is seen as a variational selection principle:
\begin{quote}
    \textit{For given fixed external conditions, the equilibrium state of a
	thermodynamic system is that which maximizes its entropy consistent with the
	external conditions.}
\end{quote}
We elaborate on the Gibbs perspective and its connection to information theory
momentarily, but let us first examine the Boltzmann entropy expression
Eq. (\ref{eq:boltzmann}) in light of the \SecondLaw as a variational selection
principle. 

As is standard in introductory statistical mechanics texts (e.g., Refs.
\cite{path11a,schr21a}), the variational statement of the \SecondLaw follows as a
statistical argument from the Boltzmann entropy and the assumption of
equally-probable microstates. Simply, the argument is that if all microstates are
equally likely, then a macrostate with a multiplicity significantly larger than
other macrostates will be observed with probability close to $1.0$ due to the vast
majority of microstates contributing to its multiplicity.

In statistical kinetic theory, macroscopic thermodynamic behavior is seen to arise
in the \emph{thermodynamic limit} of infinitely-large systems. In this limit, it is
assumed that the multiplicity of a single macrostate does indeed dominate. The
canonical toy example is flipping many fair coins, with each particular sequence of
heads or tails being the microstate and the total number of heads being the
macrostate. As the number of coin flips increases, the overwhelmingly most likely
macrostate is that consisting of microstates each with equal number heads and tails.
Shortly, we will see the information-theoretic justification for this assumption is
the \emph{asymptotic equipartition theorem}. 

Gibbs' ensemble approach is somewhat inverted from Boltzmann's, although they are
ultimately closely related. Boltzmann starts with the assumption that all
microstates are equally likely and that each microstate has an associated
macrostate, with many microstates being associated to each macrostate. Then, in the
thermodynamic limit, the unique equilibrium thermodynamic state is that with the
(overwhelmingly) largest multiplicity. From the Boltzmann expression of entropy in
Eq. (\ref{eq:boltzmann}), the \SecondLaw naturally follows.

Gibbs takes the variational statement of the \SecondLaw as the starting point. It is
\emph{not} assumed that all microstates are equally probable, and we emphasize
microstates are associated with a thermodynamic system, perhaps in contact with
other systems that have their own distinct microstates. Rather, particular
parameterized families of microstate probability distributions, called
\emph{ensembles}, are determined as a function of the equilibrium thermodynamic
macrostate. Thus, the \SecondLaw determines the equilibrium macrostate, which then
determines the microstate ensemble distribution. 

In the Gibbs formulation, the entropy of a given microstate distribution is:
\begin{align}
    S = -k \sum_i p_i \ln p_i
    ~,
    \label{eq:entropy}
\end{align}
where $p_i$ is the probability of microstate $i$. We assumed a countable number of
accessible microstates to avoid measure-theoretic complications and make more direct
connections to information theory. The equilibrium macrostate is that which
maximizes the entropy in Eq. (\ref{eq:entropy}), subject to fixed external
constraints. In modern use, the Gibbs approach is applied to find the microstate
distribution for a given equilibrium macrostate. That is, the macrostate is taken as
a given and used as the constraint. The notion of ``external constraint'' is not
clearly specified and, as is typical with classical thermodynamics, careful bookkeeping
is required. 

\subsection{Equilibrium Selection Principles}
To further refine our appreciation of entropy, it is first helpful to clarify
three distinct types of thermodynamic equilibrium. As described above in
connection to the thermodynamic identity Eq. (\ref{eq:thermo_identity}), the
entropy can change due to thermal, mechanical, and diffusive interactions. Each
form of thermodynamic interaction has an associated thermodynamic equilibrium.
For two systems that exchange only internal energy as heat, they reach thermal
equilibrium when they have the same temperature. Similarly, two mechanically
interacting systems reach mechanical equilibrium when they have the same
pressure, and diffusive equilibrium is reached when two systems have the same
chemical potential. See TABLE~\ref{tab:thermo}, adapted from Ref.
\cite{schr21a}. The extensive variables exchanged in these thermodynamic
interactions fully specify the thermodynamic state in most cases. Thus, the
thermodynamic state is typically given as $(U,V,N)$. 

\begin{table}[t]
\centering
\begin{tabular}{c|c|c|c}
Interaction  & Exchange & Equilibration &  Formula\\
       & {\small (extensive)} & {\small (intensive)} & \\
       \hline
        thermal & $U$ & $T$ & $\frac{1}{T} = \big( \frac{\partial S}{\partial U}\big)_{V,N}$\\
        mechanical & $V$ & $P$ & $\frac{P}{T} = \big(\frac{\partial S}{\partial V}\big)_{U,N}$\\
        diffusive & $N$ & $\mu$ & $\frac{\mu}{T} = - \big(\frac{\partial S}{\partial N}\big)_{U,V}$
    \end{tabular}
\caption{Thermodynamic Interactions and Equilibria}
\label{tab:thermo}
\end{table}

Despite being directly related by Legendre transforms, one reason the extensive
exchange variables are typically chosen over the intensive equilibration variables
is that the extensive exchange variables remain well-defined on the microscopic
level, whereas the intensive variables are not necessarily defined. In the Gibbs
ensemble approach, these key macroscopic state variables are defined as ensemble
averages over their microscopic counterparts:
\begin{align}
    U &= \langle H \rangle = \sum_i p_i H(i) \label{eq:U-avg}\\ 
    V &= \langle V \rangle = \sum_i p_i V(i) \label{eq:V-avg}\\ 
    N &= \langle N \rangle = \sum_i p_i N(i) \label{eq:N-avg}
    ~,
\end{align}
where $H(i)$ is the Hamiltonian, $V(i)$ is the volume, and $N(i)$ is the number of particles, of the $i^{\text{th}}$ microstate. 

The standard maximum entropy derivation of the microstate ensemble distributions in
the Gibbs approach proceeds as follows \cite{Jayn57a,wehr78a,Gran08a}. Each type of
thermodynamic equilibrium, or combinations thereof, has an associated ensemble---a
parameterized family of microstate distributions. The most common is the
\emph{canonical ensemble}, associated with only thermal interactions and equilibria.
For the given ensemble, the corresponding extensive exchange variable is assumed
known and its expectation value is then given as a constraint. In addition, the
normalization condition:
\begin{align}
    \sum_i p_i = 1
    \label{eq:normalize}
\end{align}
is always also given as a constraint. The microstate distribution $\{p_i\}$ is then
determined by extremizing the entropy in Eq. (\ref{eq:entropy}), while maintaining
the given constraints, using Lagrange multipliers. For example, the canonical
ensemble is found by extremizing entropy subject to Eq. (\ref{eq:U-avg})'s and Eq.
(\ref{eq:normalize})'s constraints, resulting in:
\begin{align}
    p_i = \frac{1}{Z} e^{-\beta H(i)} \;\;\;\; \text{(canonical)}
    ~,
\label{eq:canonical}
\end{align}
with Lagrange multiplier $\beta = (kT)^{-1}$ and normalization given by the \emph{partition function} $Z(\beta) = \sum_i e^{-\beta H(i)}$. 

There are several points to expand upon. First, note that if a system does
\emph{not} have a particular thermodynamic interaction with its environment
then the corresponding exchange variable is a fixed constant for the system,
and the associated expectation value in Eqs.~(\ref{eq:U-avg}) -
(\ref{eq:N-avg}) is simply given by that constant value.

For example, if the system has an immovable physical boundary, it does not
exchange volume with its environment and so its volume remains fixed at some
value $V_0$. Thus, every microstate has the volume $V(i) = V_0$, for all $i$.
The expectation value constraint in Eq. (\ref{eq:V-avg}) then simply reduces to
the normalization constraint in Eq. (\ref{eq:normalize}). In particular, if the
system is isolated from its environment, it has \emph{no} thermodynamic
interactions and so exchanges none of the extensive quantities. The only
constraint then is normalization, and extremizing the entropy subject to this
constraint yields the \emph{microcanonical ensemble} of Boltzmann with all
microstates being equally probable:
\begin{align}
    p_i = \frac{1}{\Omega} \;\;\;\; \text{(microcanonical)}
    ~,
    \label{eq:microcanonical}
\end{align}
with $\Omega$ the total number of microstates---i.e., the multiplicity. 

Second, notice that in the example of the canonical ensemble the Lagrange
multiplier $\beta$, which enforces the expectation value constraint of the
exchanged extensive variable $\langle H \rangle = U$, is proportional to the
associated intensive equilibration variable formula in TABLE~\ref{tab:thermo},
up to a Boltzmann factor $\beta = (kT)^{-1}$. This is of course no accident.
The appropriate ensemble distributions are recovered only when the physically
appropriate mathematical extremization constraints are used to express the
thermodynamic ``external constraints'' describing the type of thermodynamic
equilibrium. This point is critical to seeing the Gibbs version of the
\SecondLaw as a selection principle for the equilibrium thermodynamic state
$(U,V,N)$. In deriving the canonical ensemble, it appears as though we must
specify the state variable $U$ as a constraint. Note, though, that the
resulting expression in Eq. (\ref{eq:canonical}) does not require the specific
value of $\langle H \rangle = U$. Either $U$ or the Lagrange multiplier $\beta$
can act as the independent variable that determines the other. 

The Gibbs \SecondLaw is thus seen as a selection principle for the
thermodynamic state in the following way. Consider a thermodynamic system
interacting with its environment. First, the type(s) of thermodynamic
interaction(s) must be specified, as determined by which of the extensive state
variables are exchanged between the system and environment. Exchanged variables
have an associated expectation value constraint Eqs. (\ref{eq:U-avg}) -
(\ref{eq:N-avg}), and the entropy Eq.  (\ref{eq:entropy}) is extremized subject
to these constraints together with the normalization constraint Eq.
(\ref{eq:normalize}). The result is a parameterized microstate distribution
with either the expectation value(s) or Lagrange multiplier(s) as independent
variables. Recall that the system and environment exchange extensive quantities
until they reach the same value of the corresponding intensive quantity. Thus,
the equilibrium state $(U,V,N)$ is selected by specifying the Lagrange
multiplier as the independent variable, whose value is given by this
equilibrium condition. In the simplest case, the environment is a
\emph{reservoir} such that its intensive quantities remain unchanged through
the interactions and so immediately specifies the equilibrium value for the
system.

\subsection{Principle of Maximum Entropy}
The Gibbs entropy in Eq.~(\ref{eq:entropy}) is identical in form to the Shannon
entropy of information theory \cite{Shan48a,Cove06a}. The Shannon entropy of a
probability distribution provides a scalar measure of uncertainty represented by
that distribution. The information-theoretic interpretation of the Gibbs entropy
follows from viewing the extensive state variables as the independent parameter
during constrained entropy maximization, as just described. The Gibbs entropy is
then a measure of the uncertainty in the precise microstate, given the corresponding
macrostate. 

The connection between Gibbs and Shannon entropy was first elaborated upon by Jaynes
\cite{Jayn57a}, who additionally advocated for a deeper logical Principle of Maximum
Entropy (PME) that states the distribution most logically consistent with known
constraints is that which maximizes the Shannon entropy subject to those constraints. The
argument, in brief, is that the PME procedure results in a distribution that is as
uniform as possible while still satisfying the known constraints. This is
interpreted as avoiding any bias in the distribution not justified by the
constraints.

Note that the PME argument applied to thermodynamic equilibrium constraints is
precisely the same as used above for the Gibbs \SecondLaw \cite{Jayn57a}. Many
contemporary practitioners are cautious of how foundational the PME may be for
statistical mechanics. The standard objection is that including other
constraints---such as, e.g., the second moment $\text{Var}(H)$ of $U$---results in a
distribution different from the desired ensemble. The correct ensemble results
from the correct constraint, which appears to be applied in hindsight. This
particular argument---that the correct ensemble results only from the correct
constraints---applies to the Gibbs ensemble approach independent of the Jaynes PME.
The same constraints Eq. (\ref{eq:U-avg}) and Eq. (\ref{eq:normalize}) are used in
the standard statistical mechanics ensemble derivation of the canonical ensemble Eq.
(\ref{eq:canonical}) \cite[3.2]{path11a}. Similarly, the derivation of the canonical
ensemble starting from a system + reservoir in a microcanonical ensemble---joint
system+reservoir microstates are equally likely---the correct distribution Eq.
(\ref{eq:canonical}) results only if just the linear term is kept in the Taylor
expansion of the (logarithm of the) reservoir multiplicity \cite[3.1]{path11a}.
Including the second-order term results in the same ``incorrect'' ensemble the PME
arrives at with a nontrivial $\text{Var}(H)$ constraint added. 

We will return to the PME briefly below, to discuss its potential as a foundational approach to nonequilibrium statistical mechanics. 

\subsection{Asymptotic Equipartition}
Regardless of the PME's foundational status for statistical mechanics, it is
universally recognized that the Gibbs entropy Eq. (\ref{eq:entropy}) of statistical
mechanics is a Shannon information entropy. Mathematical results from information
theory that apply to entropy of the form Eq. (\ref{eq:entropy}) provide useful
results for statistical mechanics, in addition to conceptual insights provided by
information theory. 

The first such result we discuss here highlights the universality of the
microcanonical ensemble (all microstates equally probable) and the Boltzmann
entropy Eq. (\ref{eq:boltzmann}). Stated in statistical mechanics terms, the
\emph{asymptotic equipartition theorem} defines a \emph{typical set}, denoted
$\tilde{\Omega}$, such that for a large number of microstates $x$ the
probability of finding a microstate in the typical set is close to unity:
$\Pr(x \in \tilde{\Omega}) > 1 - \epsilon$. Moreover, for a large number of
microstates, the microstates in the typical set are nearly equiprobable
\cite[Theorem 3.1.2]{Cove06a}. Both results follow from the weak law of large
numbers.

Thus, for large systems---of order at least $10^{23}$ particles for the
macroscopic systems considered in classical thermodynamics---the assumptions
used for the microcanonical ensemble and Boltzmann entropy follow naturally
when we consider the ``accessible number of microstates'' to be the size of the
typical set for a given macrostate. (See also Ref. \cite[4.1]{Gran08a} for a
more in-depth discussion. The typical set is referred to there as the ``high
probability manifold''.) In light of the asymptotic equipartition theorem, it
is not surprising that the canonical ensemble limits to the microcanonical
ensemble in the thermodynamic limit, a well-known result. The asymptotic
equipartition theorem applies, however, to arbitrary microstate distributions,
which also converge to the microcanonical ensemble in the limit of
infinitely-many microstates. 

\subsection{Subadditivity}
The most important mathematical property of the Gibbs entropy Eq.
(\ref{eq:entropy}) for a thermodynamic theory of organization is
\emph{subadditivity} \cite{wehr78a}. Classically, thermodynamic entropy is
given as an extensive function of the extensive variables:
\begin{align*}
    S(\lambda U, \lambda V, \lambda N) = \lambda S(U, V, N)
    ~.
\end{align*}
(See, e.g., Ref. \cite{Call85a}.)
Practitioners commonly exchange extensivity for \emph{additivity}. 

Consider two
interacting thermodynamic systems $A$ and $B$.
Entropy is additive if the entropy of the joint $AB$ system is given as the sum of
the entropies of the subsystems $A$ and $B$: $S_{AB} = S_A + S_B$. While general
expositions state the caveat that additivity applies for independent or ``weakly
interacting'' systems, this conditional is often neglected in practice, as we will
discuss below in Section~\ref{sec:nonequilibrium}. In general, entropy is
\emph{subadditive} \cite{wehr78a} such that:
\begin{align}
    S_{AB} \leq S_A + S_B
    ~.
    \label{eq:subadditivity}
\end{align}
The equality---i.e., strictly additive case---is given in the limit of
weakly-interacting or independent subsystems. 
 
From the Boltzmann microcanonical perspective, entropy is additive when multiplicity
is multiplicative:
\begin{align}
    S_{AB} = S_A + S_B \iff \Omega_{AB} = \Omega_A\Omega_B
    ~.
    \label{eq:boltzmann_add}
\end{align}
This is a statement of subsystem independence: microstate configurations of the
subsystems do not affect each other. More generally from the Gibbs perspective,
entropy is additive when microstate probabilities are multiplicative:
\begin{align}
    S_{AB} = S_A + S_B \iff \Pr(AB) = \Pr(A)\Pr(B)
    ~. 
\end{align}
That is, additivity follows when the \emph{joint} probability distribution $\Pr(AB)$
over microstates is given by the product of the \emph{marginal} distributions of the
subsystems, $\Pr(A)$ and $\Pr(B)$. Note that, again, the microcanonical condition in
Eq. (\ref{eq:boltzmann_add}) follows as a special case when the joint and marginal
microstate probabilities are uniformly distributed. 

Subadditivity follows from the mathematical form of Gibbs entropy in Eq.
(\ref{eq:entropy}). The information theory connection here, however, is
conceptually very useful. For the simple two-subsystem case considered so far,
the joint entropy decomposes as:
\begin{align}
    S_{AB} = S_A + S_B - I_{AB}
    ~. 
\end{align}
The \emph{mutual information} $I_{AB} \geq 0$ between subsystems $A$ and $B$
tracks the discrepancy between the true joint system entropy and the sum of
subsystem marginal entropies. Thus, the mutual information is zero when $A$ and
$B$ are independent, and so the mutual information quantifies the amount of
information each subsystem contains about the other \cite{Cove06a}. ($I_{AB}$
is symmetric with respect to $A$ and $B$.) In the context of statistical
mechanics, $I_{AB}$ quantifies the reduction in microstate uncertainty of one
subsystem if the microstate of the other is known. Clearly, this quantity is
zero if the subsystems are independent. Mutual information can be considered as
an information-theoretic generalization of a correlation measure. 

We cannot overemphasize the importance of entropy subadditivity in the context
of the thermodynamics of organization. Entropy is only additive for independent
subsystems. This is the antithesis of an organized collection of subsystems. As
we discuss in more detail below in Section \ref{sec:nonequilibrium},
irreversible thermodynamics considers equilibrium field theories such that the
joint system is composed of infinitely-many subsystems. It additionally treats
densities of extensive thermodynamic quantities as additive so that their joint
full-system values are given by volume integrals. In particular, the full
system entropy is $S = \int_V s(\mathbf{r}) dV$. \emph{Irreversible
thermodynamics, at its very starting point, considers the systems' volume
elements to be independent and therefore not organized}.

For an organized system like \Benard convection, the volume elements are clearly not
independent. The fluid volumes move together in coherent motion to form the
macroscopic convective patterns. Thus, it must be that $S < \int_V s(\mathbf{r})
dV$.  A useful multivariate generalization of mutual information is known as the
\emph{total correlation} $C$ and is given by the difference between the full system
joint entropy and the sum (integral) over the marginal entropies \cite{Jame11a}. In our case, this is:
\begin{align}
    C = \int_V s(\mathbf{r}) dV - S
    ~,
\label{eq:total_correlation}
\end{align}
which can be seen as an \emph{entropy of organization}. It quantifies the
amount the true entropy of an organized system is overestimated if the system
volume elements are considered independent and, hence, not organized. 

Before moving on to nonequilibrium thermodynamics, there is a final point
regarding entropy subadditivity in equilibrium statistical mechanics. Reference
\cite{jayn65a} shows that the Clausius form of physical entropy differentials
Eq. (\ref{eq:clausius}) is recovered from the entropy of statistical mechanics
Eq. (\ref{eq:entropy}) only when the full joint distribution over microstates
is considered. The Boltzmann approach that considers single-particle marginal
distributions is insufficient for capturing thermodynamic properties that arise
from interparticle forces---pressure, in the case shown in Ref. \cite{jayn65a}.
The Boltzmann approach neglects interparticle forces implicitly through the
assumption of independence by using single-particle marginal probabilities---his
famous \emph{Stosszahlansatz} assumption of \emph{molecular chaos}
\cite{ehren90a}. Thus, in very general physical circumstances, a proper
accounting of the entropy contributions to organization is crucial to connect
statistical mechanics entropy with physical entropy.

As a side note, Ref. \cite{gao19a} shows a related result connecting the Gibbs
and Clausius entropies. In essence, this shows the expectation value
constraints when maximizing the Gibbs entropy in statistical mechanics (i.e.,
using the PME) are the correct constraints to recover classical thermodynamics
and Clausius entropy. The resulting microstate distributions are called
generalized Boltzmann distributions there.

We emphasize, though, that the arguments in Ref. \cite{jayn65a} should not be
interpreted as saying that the ``correct'' physical entropy of Clausius is
recovered from the more fundamental statistical mechanics entropy of Gibbs.
Rather, it should be seen as a consistency between the two perspectives. Since
physical entropy changes can be measured in a laboratory setting, by virtue of
absolute temperature and heat exchanges being macroscopically measurable, it is
seen as the more physically-grounded formulation of entropy.

It is important to remember that the relation between thermodynamic entropy, heat, and temperature were devised by Clausius with certain assumptions and limitations \cite{crop86a}. In a sense, Clausius' arguments have been lost in time, as they are not standard in modern thermodynamic developments. It is simply taken that Eq. (\ref{eq:clausius}) is \emph{the} definition of physical entropy, without any qualification. This is not to say the Clausius definition Eq. (\ref{eq:clausius}) is wrong, although it is likely incomplete. The most obvious way it can be incomplete is that Clausius derived the expression without the consideration of particle exchange. 
Including diffusive interactions into Clausius' arguments, outlined in Ref.
\cite{crop86a}, results in a different physical entropy differential. This new
form of physical entropy is just as physically grounded, but it is appropriate
to physical circumstances not originally considered by Clausius. 

The lengthy discussion on the foundations and assumptions of entropy is
necessary as we move away from equilibrium systems. Thermodynamic theories of
organization in nonequilibrium systems necessarily must be built on a
foundation of nonequilibrium thermodynamics. Such a foundation is not at all
well-established. It is not even clear where to start when formulating theories
of nonequilibrium thermodynamics. In assessing proposed and potential theories,
it is therefore essential to understand the starting points and assumptions
going into the theory. For example, the presumed additivity of entropy in local
equilibrium theories is immediately suspect as a foundation for a thermodynamic
theory of organization. 

\section{Nonequilibrium Statistical Physics}
\label{sec:nonequilibrium}



The spontaneous patterns in Belousov's and \Benard's experiments emerge in
manifestly nonequilibrium systems. As such, they cannot be described in terms
of equilibrium transformations. Why? When the driving $d_e S$ is removed during a
quasistatic equilibrium transformation, the system response immediately halts. By
assumption the system is always instantaneously at equilibrium, including at the
moment when driving stops.

In contrast, the Belousov and \Benard patterns are the result of
\emph{nonequilibrium steady-state processes}. When driving is turned off, the
processes do not immediately stop, but rather begin to relax to the equilibrium
state determined by the now-uniform and fixed boundary constraints. For example,
convection in the \Benard experiment is driven by a temperature gradient. Thus,
no-driving corresponds to zero gradient and there is a single boundary
temperature that determines the subsequent equilibrium state. There is active
heat transport in this nonequilibrium steady-state process, but no net
transport in an equilibrium state.


The previous section recounted the central role entropy plays in equilibrium theory.
When it comes to explaining pattern formation and self-organization, we said that a
``thermodynamic theory'' should be an entropic theory. In equilibrium, this holds.
Although spontaneous self-organization in thermodynamic equilibrium at first blush
appears at odds with the \SecondLaw, we saw this is not the case. Organization in
equilibrium is due to the \SecondLaw, not in spite of it. From the perspective of
the \SecondLaw, entropy plays a crucial role as a selection principle: equilibrium
states, including organized states, are those that maximize entropy consistent with
environmental constraints.

Classical irreversible thermodynamics, and in particular the theory of dissipative
structures built upon it, aims (in part) to provide a generalization of the
\SecondLaw as a selection principle for nonequilibrium steady-states. The
\emph{entropy production}, introduced shortly, is called upon to play a role out of
equilibrium analogous to that of entropy in equilibrium. Historically, entropy
production in classical irreversible thermodynamics is most notable for the various
proposed extremum principles to determine nonequilibrium steady-states. As will be
described in more detail, though, to date no entropy production extremum principle
is generally accepted as a valid selection criterion for determining nonequilibrium
steady-states.

To appreciate this, we first introduce the basics of entropy production (density) in
the context of classical irreversible thermodynamics and argue that this quantity
does not generally have a firm physical basis. Despite their claimed primacy, we
show that entropy production and the entropy balance equation---that generalizes
the \SecondLaw---are, in fact, largely superfluous in classical irreversible
thermodynamics. The main question raised in this Section is: What role, if any, does
entropy (or an appropriate generalization) play in a ``thermodynamic'' theory of
self-organization out of equilibrium? 

\subsection{Irreversible thermodynamics of local equilibrium}
Equilibrium theory describes irreversible changes between equilibrium states, with
calculations performed along idealized reversible paths. This assumes that the system
is always \emph{instantaneously} in equilibrium. In contrast, to describe dynamical
transport processes, classical irreversible thermodynamics, pioneered by
Onsager~\cite{Onsa31a,Onsa31b} and Prigogine~\cite{Prig68a}, instead assumes
\emph{local equilibrium}.

In this, one considers a spatially-extended system in contact with an environment,
such as the box of fluid in contact with two heat baths for \Benard convection. The
system consists of many mesoscopic ``elementary volume units'' $dV$ such that they
are differential volume elements from the macroscopic perspective; i.e., the total
volume is given as $V = \int dV$.  However, from the microscopic perspective they
contain sufficiently-many particles that their momenta follow a Maxwell
distribution. Thus, local equilibrium assumes intensive quantities---e.g.,
temperature---become fields and extensive quantities are replaced by their
densities, with all varying over time and space. For these quantities to be
well-defined on the mesoscale, the thermodynamic identity Eq.
(\ref{eq:thermo_identity}) must hold locally in each volume element $dV$.

In the local equilibrium field theories of classical irreversible
thermodynamics, the \FirstLaw and \SecondLaw are given in the form of
\emph{balance equations} \cite{DeGr62a}. Following Ref. \cite[Box
15.1]{Kond14a}, a general balance equation for an extensive quantity $Y$ is
derived as follows.

Let $y$ be the density of $Y$ and $\mathbf{J}_Y$ the current density of $Y$,
and consider an arbitrary volume $V$. The change of $Y$ in the volume due to
the current flow is given by the surface flux $\int_{\partial V} \mathbf{J}_Y
\cdot d\mathbf{n}$. Let $P[y]$ be the amount of $Y$ produced or destroyed per
unit volume per unit time. Hence, the change of $Y$ in $V$ due to internal
changes is $\int_V P[y] dV$. (Consider, for example, molecular concentrations
that may change internally due to chemical reactions.) The time rate of change
of $Y$ in the volume is given by the integral balance equation:
\begin{align*}
    \int_V \biggl( \frac{\partial y}{\partial t}\biggr) dV = \int_V P[y] dV - \int_{\partial V} \mathbf{J}_Y \cdot d\mathbf{n}
    ~.
\end{align*}
Using Gauss's divergence theorem, the surface flux integral can be changed to a
volume integral, giving:
\begin{align*}
    \int_V \biggl( \frac{\partial y}{\partial t}\biggr) dV = \int_V P[y] dV - \int_V (\nabla \cdot \mathbf{J}_Y) dV
    ~.
\end{align*}
Since the volume $V$ is arbitrary, the integrands on both sides must be equal,
giving the differential balance equation:
\begin{align}
    \biggl(\frac{\partial y}{\partial t}\biggr) + (\nabla \cdot \mathbf{J}_Y) = P[y]
    ~.
\end{align}

Using the differential balance equation and the form of the \SecondLaw in Eq.
(\ref{eqn:dS}) we now review the derivation of the entropy balance equation \cite{DeGr62a,Kond14a}. We
will then examine what role entropy balance and entropy production may, or may not,
play in nonequilibrium theory.

First, we express the time derivatives of the terms in Eq. (\ref{eqn:dS}) by volume
and flux integrals as:
\begin{align}
\frac{d S}{dt}   & = \int_V \frac{\partial s}{\partial t} \; dV \label{eq:dSdt}\\
\frac{d_e S}{dt} & = - \int_{\partial V} \mathbf{J}_S \cdot d\mathbf{n} \label{eq:Js}\\
\frac{d_i S}{dt} & = \int_V \sigma \; dV
    ~.
\label{eq:sigma}
\end{align}
Since entropy is an extensive quantity, it is assumed to have a well-defined
density $s$ that obeys the thermodynamic identity. Second, we recall that $d_e
S$ referred to the change of entropy due to energy and material transport, and
so we express its time rate of change in terms of an \emph{entropy current
density} $\mathbf{J}_S$. Finally, the internal generation $P[s]$ of entropy is
defined as the \emph{entropy production} $\sigma$. Since we postulated that
$d_i S \geq 0$ and recalling the volume $V$ is arbitrary, it must be that
$\sigma \geq 0$. That is, entropy may be internally generated, but not
destroyed, during irreversible processes. The righthand terms in Eqs.
(\ref{eq:dSdt}) - (\ref{eq:sigma}) are those of an integral balance equation.
The differential entropy balance equation is:
\begin{align}
\frac{\partial s}{\partial t} + \nabla \cdot \mathbf{J}_S = \sigma
    ~.
\label{eq:entropy_balance}
\end{align}

Specific forms of entropy current density and entropy production are found using the local thermodynamic identity:
\begin{align}
    T ds = du + Pdv - \sum_k \mu_k dn_k
    ~,
\end{align}
along with balance equations for energy density $u$ and chemical concentration
$n_k$. It is found that entropy production has a general bilinear form~\cite{DeGr62a,Kond14a}:
\begin{align}
    \sigma = \sum_i F_i \mathbf{J}_i
    ~,
    \label{eq:bilinear}
\end{align}
with products of thermodynamic fluxes $\mathbf{J}$ and thermodynamic forces $F$.

This expression is in line with an intuition that thermodynamic force gradients
in intensive quantities drive fluxes in extensive quantities during
irreversible processes. For example, the force $\nabla(1/T)$ drives a heat flux
$\mathbf{J}_u$ in \Benard convection.

That said, the practical utility and physical significance of Eq.
(\ref{eq:bilinear}) is questionable. In particular, while the extremum of
entropy provides a selection criterion for equilibrium states,
Section~\ref{sec:nonequilibrium} C below will establish that nonequilibrium
steady-states do not correspond to extrema of $\sigma$. More alarming, the
usefulness of the entropy balance equation Eq. (\ref{eq:entropy_balance})
itself is unclear! We show shortly that entropy balance is superfluous for
deriving nonequilibrium transport equations---the central endeavor of classical
irreversible thermodynamics. This is surprising, given that the entropy balance
equation is the nonequilibrium generalization of the \SecondLaw, the
cornerstone of equilibrium theory. 

As Ref. \cite{DeGr62a} points out, the balance equations for entropy, energy, and
species concentration are insufficient to determine the time evolution of all
relevant variables, even if the bilinear form of entropy production Eq.
(\ref{eq:bilinear}) is used. The missing relations between variables comes in the
form of \emph{phenomenological laws}, sometimes also called \emph{constitutive
relations}. These laws assume that thermodynamic fluxes are linear functions of
thermodynamic forces, with a general form given as:
\begin{align}
    J_i = \sum_k L_{ik} F_k
    ~,
    \label{eq:phenom}
\end{align}
where the $L_{ik}$ are known as the \emph{phenomenological coefficients}. 

Familiar transport phenomena like Fourier's heat law, Fick's law of diffusion,
and Ohm's law of electrical conduction employ linear phenomenological laws.
Additionally, Eq. (\ref{eq:phenom}) encompasses cross-effects in transport,
such as the Seedbeck and Peltier thermoelectric effects \cite{thom74b,Kond14a}.
The celebrated Onsager reciprocal relations \cite{Onsa31a,Onsa31b} place
symmetric or anti-symmetric constraints on cross coefficients $L_{ik} = \pm
L_{ki}$. Inserting Eq. (\ref{eq:phenom}) into the bilinear form of entropy
production Eq. (\ref{eq:bilinear}) gives a nonnegative-definite constraint to
the matrix of phenomenological coefficients. Relying on linear laws, classical
irreversible thermodynamics is often called \emph{linear nonequilibrium
thermodynamics}. 

To demonstrate the superfluousness of entropy balance in deriving transport
equations, let's now examine the canonical example of heat conduction and
Fourier's law. Start with the phenomenological law:
\begin{align}
    \mathbf{J}_u = L_{uu} \cdot \nabla \biggl(\frac{1}{T}\biggr)
    ~,
    \label{eq:cond_phenom}
\end{align}
that takes the more familiar form:
\begin{align}
    \mathbf{J}_u = -\kappa \nabla T
    ~,
    \label{eq:heat_cond}
\end{align}
where $\kappa = L_{uu}/T^2$ is the conductivity coefficient or tensor for
anisotropic systems. Relation Eq. (\ref{eq:heat_cond}) expresses the familiar
observation that heat flows down a temperature gradient, from hot to cold.
Next, consider the balance equation for internal energy. In integral form, this
is:
\begin{align*}
    \frac{d}{dt} \int_V u \; dV = - \int_{\partial V} \mathbf{J}_u \cdot \mathbf{n} \; dV
    ~,
\end{align*}
and in differential form:
\begin{align}
    \frac{\partial u}{\partial t} + \nabla \cdot \mathbf{J}_u = 0
    ~. 
    \label{eq:u_balance}
\end{align}
Appealing to local equilibrium thermodynamics, the internal energy density is
related to the temperature field through:
\begin{align}
    \partial u = c \rho \; \partial T
    ~,
    \label{eq:heat_capacity}
\end{align}
where $c$ is the specific heat and $\rho$ the mass density. Combining relation Eq.
(\ref{eq:heat_capacity}) with internal energy balance Eq. (\ref{eq:u_balance}) and
the phenomenological law Eq. (\ref{eq:heat_cond}), we arrive at Fourier's heat law:
\begin{align}
    \frac{\partial T}{\partial t} = k \nabla^2 T
    ~,
\end{align}
with $k = \kappa/c\rho$.

We emphasize that entropy production and entropy balance were not necessary for
deriving the self-contained equation of motion for the temperature field---or,
equivalently, internal energy density. Notably, the two most widely-used
textbooks on classical irreversible thermodynamics \cite{DeGr62a,Kond14a} both start
their sections on heat conduction with the bilinear form of entropy production:
\begin{align*}
    \sigma = - \frac{1}{T^2} \mathbf{J}_u \cdot \nabla T
    ~.
\end{align*}
They then proceed, as we have above, without further reference to it. For a
more complicated example of deriving nonequilibrium transport
equations-of-motion Eq. (\ref{eq:gen_dyn}) without entropy production or
entropy balance, see Ref.~\cite[Chapter IV, Section 4]{DeGr62a}. It derives the
equations of motion for a single chemical component isotropic fluid, which it
calls ``thermo-hydrodynamics''. 

\subsection{Reprise}
Let us pause to reflect on the status of entropy production and entropy
balance. Recall that equilibrium thermodynamics introduces the entropy state
function and the \SecondLaw to formalize the observation that some processes
occur spontaneously in only one direction, despite both directions being
consistent with the \FirstLaw. The derivation of Fourier's heat law and other
nonequilibrium transport laws demonstrates that entropic quantities are not
necessary to produce a full set of equations-of-motion for nonequilibrium
transport processes. Furthermore, entropy balance---the generalization of the
\SecondLaw---is not necessary to capture irreversibility, as this is already
inherently contained within the resultant equations-of-motion. For these
reasons, entropy production and entropy balance (as defined and used above) are
superfluous in the general theory of classical nonequilibrium thermodynamics. 

In the context of self-organization, this is exactly how the nonlinear dynamics
approach to pattern formation proceeds. The governing local equilibrium transport equations
Eq. (\ref{eq:gen_dyn}) are derived (and, typically, simplified) and it is from these
equations that the bifurcation analysis is performed. We emphasized that a
``thermodynamic'' theory generally implies an entropic theory. We now see why
entropic quantities are not present in the nonlinear dynamics analysis---they are
simply not needed.

No surprise then that entropy balance, in particular, is largely absent from
contemporary work in nonequilibrium thermodynamics. A notable exception is the
recent review of the \SecondLaw for the climate system \cite{Sing22a}. However, the
authors note that the entropy balance equation ``contains no additional information
about the flow that is not already contained in the energy budget''. Entropy balance
is only used to provide nonnegative-definite constraints on entropy production.

Most contemporary research in nonequilibrium thermodynamics adopts the statistical
approach and uses stochastic processes, such as Langevin dynamics
\cite{Keiz87a,Seif12a}. Since the stochastic setting works with probability
distributions, Shannon entropy is commonly employed. And, then, the time rate of
change of Shannon information is often referred to as ``entropy
production''. Paralleling equilibrium entropy, the terminology ``entropy
production'' is overloaded and often leads to confusion.

\subsection{Variational Principles for Nonequilibrium Steady States}
Beyond providing certain nonnegativity constraints, that may be useful in some
situations, what is the utility of entropy production in classical irreversible
thermodynamics? Historically, its most important appearances are in attempts to
generalize the \SecondLaw as a selection principle, where extrema of entropy
production may determine nonequilibrium steady-states. 

The first instance of a nonequilibrium variational principle is found in 1848,
with Kirchhoff's study of current flow  in electrical networks \cite{Kirc48a}.
He identified the final network nonequilibrium steady state as that which
minimized ``dissipation'': current distributes itself so as to dissipate the
least possible heat for the given applied voltages. This predates by several
decades Gibbs' variational statement of the \SecondLaw for heterogeneous
equilibrium \cite{Gibb06a} and Maxwell's principles of minimum heat
\cite[407-408]{Maxw54a}. Helmholtz stated a minimum dissipation theorem for steady-state flow of viscous fluids at low velocities \cite{helm68a}. Similar theories of minimum dissipation of energy were proposed by Rayleigh \cite{Rayl77a} and Lorentz \cite{lore96a}.

To unify the known cases, Onsager~\cite{Onsa31a,Onsa31b} and
Prigogine~\cite{Prig68a} both attempted to derive a general variational principle
that determined near-equilibrium steady states. On the one hand, Onsager developed
the \emph{principle of least dissipation of energy}. Depending on the boundary
constraints, he showed this principle can be cast as \emph{maximizing entropy
production}. (Note that Onsager clearly distinguished energy dissipation and entropy
production.) Prigogine, on the other hand, advocated for the \emph{principle of
minimum entropy production}. Interestingly, it is not necessarily at odds with
Onsager's maximum entropy production. Prigogine's principle can actually be derived
from Onsager's, with additional restrictions. References
\cite{Land75a,Jayn80a,barb99a,Palf01a, Ross05a} and \cite[Ch. 12]{Gran08a} give
in-depth accounts of nonequilibrium variational principles and their shortcomings.
Helpfully, they include worked counterexamples demonstrating that well-known
steady-states, such as heat conduction, cannot be extrema of entropy production. 

Having the necessary expressions in hand, it is instructive to reproduce the heat
conduction counterexample here, as found in Refs. \cite{barb99a,Palf01a,Ross05a} and
\cite[Ch. 12]{Gran08a}. For a one-dimensional rod of length $L$, the total entropy
production is the integral of $\sigma$ along the rod:
\begin{align}
    \frac{d S_i}{dt} &= \int_0^L \mathbf{J}_u \cdot \nabla \biggl(\frac{1}{T(x)}\biggr) dx \nonumber \\ &= \int_0^L \frac{\kappa}{T^2(x)} \biggl(\frac{\partial T(x)}{\partial x}\biggr)^2 dx
    ~,
\label{eq:cond_entprod}
\end{align}
using the linear phenomenological law Eq. (\ref{eq:cond_phenom}) above. Recall that
the derivation of Fourier's heat law does not require entropy production, and it is
a self-contained equation-of-motion for the temperature field with steady-state
solution:
\begin{align}
    T^*(x) = T_h - (T_h -T_c)\frac{x}{L}
    ~,
    \label{eq:heat_conduction}
\end{align}
which solves Laplace's equation $\nabla^2 T(x) = 0$. 

To recover the steady-state solution Eq. (\ref{eq:heat_conduction}) as an
extremum of entropy production in Eq. (\ref{eq:cond_entprod}), a
``near-equilibrium'' approximation is required \cite{DeGr62a,Kond14a} such that
$L_{uu} = \kappa T^2(x) \approx \kappa T^2_{avg}$. However, the extremum of Eq.
(\ref{eq:cond_entprod}) can be directly computed, without approximation, by the
Euler-Lagrange equation \cite{Palf01a}:
\begin{align*}
    T \frac{\partial^2 T}{\partial x^2} = \biggl(\frac{\partial T}{\partial x}\biggr)^2
    ~, 
\end{align*}
which has an exponential solution:
\begin{align*}
    T_{ext}(x) = T_h e^{-a(x/L)}, \;\;\;\; a=\ln(T_h/T_c)
    ~.
\end{align*}
This exponential steady-state temperature distribution has not been observed in the
lab. And so, the true steady-state cannot be an extremum of entropy production.
Reference~\cite{Ross05a} argues that the approximation used above to recover the
true steady-state distribution is only valid in equilibrium, and so only equilibrium
solutions are extrema of entropy production. Despite clear and numerous
counterexamples, variational principles of entropy production, particularly minimum
entropy production, are regularly invoked or referenced as valid.

As Ref.~\cite{Jayn80a} points out, in the cases for which the steady-state solution
is compatible with an extremum of entropy production---e.g., examples in Refs.
\cite{DeGr62a} and \cite{Kond14a}---the variational principle is redundant. This is
because the final Euler-Lagrange equations reduce to the conservation and
phenomenological laws; e.g., Laplace's equation for heat conduction, with the
``near-equilibrium'' approximation used. And, the latter was already valid before
invoking a variational principle. This is not surprising, given that entropy balance
is similarly redundant. 

\subsection{Dissipative Structures}
While the nonlinear dynamics approach to pattern formation may be rightly
considered a thermodynamic theory, the notion of a thermodynamic theory of
self-organization continues, regrettably, to be synonymous with Prigogine et.
al.'s theory of \emph{dissipative structures}. Even more regrettable, it has
become so synonymous that several contemporary authors continue to cite
dissipative structures while lacking awareness of the many critiques and
counterexamples that disprove the theory. Worse still, dissipative structures
are sometimes cited without a basic understanding of the theory.

The theory of dissipative structures \cite{Glan71a,Nico77a,Kond14a} attempted
to create a ``thermodynamic''---i.e., entropic---analysis paralleling the
nonlinear dynamics' bifurcation theory. Both approaches center on the stability
and exchange of stability between nonequilibrium steady-states. The nonlinear
dynamics approach employs linear stability analysis---Lyapunov's first method.
Dissipative structures follows Lyapunov's second method, attempting to
create entropic Lyapunov functions that determine the stability of
nonequilibrium states. 

The theory of dissipative structures proposes two nonequilibrium steady-state
stability criteria, one for ``near-equilibrium'' and one ``far-from-equilibrium''---both of which do not hold in general. Recall that Section
\ref{sec:NonlinearDynamics} above introduced the bifurcation parameter $R$ as a
distance from equilibrium. Starting in equilibrium, we have $R=0$. As $R$ is
slowly increased from $R=0$ the system remains, for small $R$, in stable
nonequilibrium steady-states until the first, or primary, bifurcation occurs,
denoted $R=R_c$. The dissipative structure approach refers to the
``near-equilibrium'' values of $0 \leq R < R_c$ as the \emph{thermodynamic
branch}; see, for example, Ref. \cite[Chapter 18]{Kond14a}. (Above we referred
to this as the base state.) Recall that a ``near-equilibrium'' assumption is
necessary for an extremum of entropy production to coincide with the
steady-state solution of the nonequilibrium transport equations. (As we have
seen, though, these only coincide in general exactly at equilibrium.) 

The dissipative structure approach uses the proposed principle of minimum
entropy production to state that the thermodynamic branch is always stable
``near-equilibrium''. Recall though that nonequilibrium steady-states---states
that are any nonzero distance from equilibrium---do \emph{not} generally
correspond to states of minimum entropy production. 

\subsection{Universal Evolution Criteria?}
According to dissipative structures, minimum entropy production guarantees that steady-states ``near equilibrium'' are always stable (again, this is not true in general), and therefore pattern-forming instabilities must occur ``far-from-equilibrium''. 
In the far-from-equilibrium regime, the
phenomenological laws may no longer be linear. To assess the stability of states
arbitrarily-far from equilibrium, Prigogine and Glansdorf introduced the
``universal evolution criterion'' \cite{Glan71a}, based on the \emph{excess
entropy production} or \emph{second variation of entropy} $\delta^2 S$. They
showed that, as long as the local equilibrium assumption is valid, then
$\delta^2 S < 0$. With this, to be a Lyapunov function, the time derivative of
$\delta^2 S$ must additionally be positive for stable states and negative for
unstable states. They found the time derivative takes a bilinear form:
\begin{align*}
    \frac{d}{dt} \frac{\delta^2 S}{2} = \int_V \sum_i \delta F_i \delta J_i \; dV
    ~.
\end{align*}
Similar to entropy production density in Eq.~(\ref{eq:sigma}), this expresses excess entropy production in terms of (variations of) physical quantities that can be computed or measured. 

Recall that essentially the entirety of classical irreversible thermodynamics
was developed in the near-equilibrium linear regime. In this light, the
universal evolution criterion was a hugely ambitious attempt to address
arbitrarily-far-from-equilibrium states. Unfortunately, the universal evolution
criterion does not hold in general and the excess entropy production is not a
Lyapunov function. Following the theme here, the nonequilibrium transport
equations-of-motion---that, again, do \emph{not} require entropy
production---are valid arbitrarily far from equilibrium. And so, steady-states
and their stability may be assessed independently of excess entropy production;
i.e., using nonlinear dynamics bifurcation analysis.

In a series of back-and-forth letters to the \emph{Proceedings of the National
Academy}, Keizer and Fox provided counterexamples for which the excess entropy
production makes stability predictions that are counter to the established
linear stability analysis of the transport equations-of-motion
\cite{Keiz74a,Glan74a,Fox79a,Nico79a,Fox80a}. While Prigogine et al. initially
attempted to respond, a final counterexample was given by Fox \cite{Fox80a} for
which there is no rebuttal. Anderson and Stein similarly concluded that the
universal evolution criterion is nothing of the sort; ``As far as we can see,
in the few cases in which this idea can be given concrete meaning, it is simply
incorrect'' \cite{Ande87a}.

The latest version of Kondepudi and Prigogine's textbook softens the universal
evolution criterion to provide ``a necessary but not a sufficient condition for
instability''. And, it remarks more generally that ``in the
far-from-equilibrium nonlinear regime, there is no such general principle for
determining the state of the system'' \cite[Chapter 18]{Kond14a}. When
providing examples of pattern-forming instabilities, this textbook and Ref.
\cite{Reic80a} both revert to the standard linear stability analysis, after
introducing dissipative structures. Likely, this causes confusion, leading
readers to associate the theory of dissipative structures with nonlinear
dynamics' well-vetted bifurcation theory. We emphasize again, however, that the
theory of dissipative structures is auxiliary to the nonlinear dynamics
approach and does not provide a similar stability analysis since the excess
entropy production is not a Lyapunov function. 

Interestingly, in their back-and-fourth letters to PNAS, Keizer and Fox
advocate for a competing entropic stability criterion \cite{Keiz76a}. We are
not aware of similar counterexamples to their criterion, but it also does not
appear that it has been widely adopted. That said, given that it takes a
stochastic approach, based on fluctuation-dissipation relations, it is more in
line with contemporary nonequilibrium thermodynamics and, thus, may be
worthwhile revisiting. However, \emph{any} successful entropic stability
criterion adds nothing fundamentally new to the nonlinear dynamics analysis, as
their goals are the same. Perhaps, optimistically, some stability calculations
may be easier within an entropic stability theory.

One claimed advantage of an entropic stability criterion is to give ``a physical
meaning in terms of thermodynamic quantities of direct experimental interest to the
otherwise abstract predictions of the qualitative theory of differential equations''
\cite{Glan74a}. However, given the ample success of matching nonlinear dynamics
predictions with experiment, this is a dubious claim. (See the many examples
described in Ref. \cite{Cros93a}.) Yet again, entropic considerations appear quite
superfluous in nonequilibrium theories. 

\subsection{Entropy Productions}
Given the central role of entropy and the \SecondLaw in equilibrium theory, it
is quite surprising to find the generalizations of entropy production and
entropy balance do not play a similar role in nonequilibrium theory. Perhaps
shockingly, they appear to play almost no role at all in classical local
equilibrium field theories! This is by no means a novel insight, but it is not
something we have seen emphasized much in discussions of classical
nonequilibrium thermodynamics.

By way of contrast, other terms called ``entropy production'' feature
prominently in contemporary stochastic thermodynamics \cite{Gasp04a,Seif12a}.
While various terms may share the title of ``entropy production'', the physical
relationship between them is often unclear; particularly so between uses in
stochastic thermodynamics and what we have introduced above in classical
irreversible thermodynamics. 

Section \ref{sec:Equilibrium} gave a lengthy discussion on the nature of
entropy in equilibrium theory. It concluded that the Shannon information
interpretation of entropy is the best foundation available for understanding
what entropy actually is. It also emphasized that from this perspective, an
assumption of entropy additivity implies independence between components. Thus,
taking volume integrals of entropy density in local equilibrium field theories
implicitly assumes there are no interactions or correlations between mesoscopic
volume elements. While in restricted settings this may be a reasonable
assumption---as with, say, heat conduction through an isotropic material---from
the outset it precludes describing organized systems.

Furthermore, recall that the derivation of the differential form of entropy balance
Eq. (\ref{eq:entropy_balance}) required the volume integrals in Eqs. (\ref{eq:dSdt})
- (\ref{eq:sigma}). This makes the entropy balance equation suspect in the general
case. For instance, the time rate of change of total entropy cannot equal the volume
integral of the time rate of change of entropy density in Eq. (\ref{eq:dSdt}) if
total entropy is not equal to the volume integral of entropy density. Using the residual total correlation introduced above
in Eq.  (\ref{eq:total_correlation}), we have:
\begin{align*}
    \frac{d S}{dt} = \int_V \frac{\partial s}{\partial t} \; dV - \frac{d C}{dt}
    ~.
\end{align*}
The total correlation (degree of organization) term depends on the volume over which it is evaluated,
and recall that independence from the arbitrary volume was necessary to go from
integral balance equations to differential balance equations. 
An immediate consequence is that the differential entropy balance equation, on which dissipative structures is largely based, is not valid for organized systems. 
That is, dissipative structures was an attempt at a thermodynamic theory of self-organization, but was built on assumptions that do not hold for organized systems.

In the general subadditive case, it is not clear how to conceptualize an
entropy production density $\sigma$. And, while it may seem intuitive that
currents of matter and energy may ``carry entropy'', from an
information-theoretic perspective such ``flows of information'' are difficult
to properly define and interpret \cite{Jame16a}. Overall, there is not a solid
physical basis for assuming that information, and thus entropy, is locally
conserved \cite{Gran08a}. 

One can only conclude that (differential) entropy balance and entropy
production density do not have a solid physical basis in the general
subadditive case when organization is present in the system. In this light, it
is rather fortunate that they are not needed to derive the full set of
nonequilibrium transport equations Eq. (\ref{eq:gen_dyn}).

This is not to say, however, that entropic quantities should be banished from
nonequilibrium theories. Quite the opposite, in fact, as recent developments
have recognized the significance of information as a thermodynamic resource
\cite{Parr15a}. Entropy is a subtle concept that must be invoked with care.
While the information-theoretic approach is the most promising,
high-dimensional information theory is in its infancy \cite{Will10a,Jame17a}.
Contemporary work in stochastic thermodynamics typically concerns
low-dimensional systems. And, with a few exceptions \cite{Fala18a}, there
remains a disconnect between theoretical developments in stochastic
thermodynamics and the physical phenomenon of self-organization. 

\subsection{Principle of Maximum Caliber}
To close the discussion on thermodynamics, we briefly return to Jaynes'
Principle of Maximum Entropy. Recall that this approach builds up equilibrium
statistical mechanics starting from the perspective of information-theoretic
entropy, along with the logic that unbiased distributions should maximize
entropy subject to appropriate constraints. This foundation allows for a clear
and rigorous path to nonequilibrium statistical mechanics. The \emph{Principle
of Maximum Caliber} (MaxCal), a generalization, creates time-dependent
microstate distributions by maximizing uncertainty over space-time paths,
consistent with the constraints of the macroscopic process
\cite{Jayn85a,Gran08a,Press13a,Ghos20a}.

MaxCal suffers from the same objections as the Principle of Maximum Entropy,
though---that the resulting microstate distributions depend on the specific
macroscopic constraints given. That noted, when using appropriate macroscopic
constraints, MaxCal derives known phenomenological laws---e.g., Fick's and
Fourier's laws---from first principles using linear perturbation theory
\cite{Gran08a}. Given the prominence of phenomenological laws in nonequilibrium
theory, any potential derivation from first principles is worth consideration.
Furthermore, numerical evidence suggest that bifurcations in fluid
instabilities \cite{Atta12a} and lasers \cite{Hake12a,Hake16a} are consistent
with MaxCal. 

Ideally, MaxCal would offer a path to deriving as-yet undiscovered nonlinear
constitutive relations for far-from-equilibrium systems. To date this has not
been achieved, due to the intractability of the non-perturbative space-time
path integrals. Conceptually, a nonequilibrium statistical mechanics, such as
MaxCal, links microscopic dynamics with the macroscopic field theories that
nonlinear dynamics shows exhibit pattern-forming and self-organizing behavior.
So, technically speaking, this provides one answer to ``how'' microscopic
constituents organize into macroscopic patterns. However, nonequilibrium
statistical mechanics so far falls short of providing a constructive and direct
answer to this question due to the many layers of logic and complicated
mathematical operations separating the macroscopic from the microscopic. 
 
This separation and general intractability of nonequilibrium statistical
mechanics and high-dimensional information theory lead us to an important
interlude in the discussion of far-from-equilibrium thermodynamics and
self-organization more generally. There is reason to believe these hurdles are
not exceptional but rather typical for self-organization, in much the same way
nonlinear dynamics is the norm and linear dynamics the exception in mechanics.
This intractability is one likely reason why self-organization has stubbornly
resisted our understanding for so long. However, there are others.

\section{Intractability and Limits of Constructionism}
\label{sec:Intractability}

From one perspective, at this point we reached our destination in the search
for principles of emergent organization. Macroscopic equations of motion Eq.
(\ref{eq:gen_dyn}) are derived from conservation and phenomenological laws,
without the need for entropic quantities or generalizations of the \SecondLaw.
Their range of validity away from equilibrium depends on the range of validity
of the phenomenological laws. In many cases, those laws hold very far from
equilibrium---e.g., the Navier-Stokes equations governing fluid flow. From
such equations, emergent organization can be derived in principle from the
bifurcation analysis of their nonlinear dynamics. In another sense, though, the
journey has only just begun towards principles of organization. Much of the
challenge ahead hides behind two simple words used above: \emph{in principle}. 

\subsection{Constructionism}

Quantitative science relied on and flourished due to the reductionist
hypothesis---that the constituents of all systems, no matter how complicated,
are governed by the same fundamental laws. This, however, came at the cost of
an erroneous assumption---what the late Philip Anderson called
\emph{constructionism}---the ability to start from the fundamental laws and
reconstruct the universe \cite{Ande72a}. That is, although a macroscopic system
of interest is, \emph{in principle}, fully described by the Hamiltonian
dynamics of its constituent particles, its macroscopic dynamics cannot always
be constructed directly from the microscopic dynamics. Even though emergent
organization may be derived from bifurcation theory in principle, there are
intractability limits in practice. Reductionism and constructionism are
effective tools for uncovering the fundamental laws of the universe, but ``in
principle'' answers do not go far enough for understanding emergent phenomena.


Statistical mechanics considers constructionism as a single bridge between two
levels---``macroscopic'' transport equations Eq.~(\ref{eq:gen_dyn}) are
constructed from ``microscopic'' Hamiltonian particle dynamics~\cite{lebo99a}. Pattern
formation theory of nonlinear dynamics is then seen as an additional bridge,
with the ``macroscopic'' dynamics of Eq.~(\ref{eq:gen_dyn}) taken as the
starting point, from which organization and patterns that emerge are
constructed. To avoid confusion, let's refer to the ``microscopic'' description
of system behavior, in terms of Hamiltonian dynamics of particles, as Level I.
``Macroscopic'' continuum dynamics will then be Level II, and emergent
organization will be Level III. Statistical mechanics bridges Level I to Level
II, and pattern formation theory attempts to bridge Level II to Level III. 

Note that for some purposes the Hamiltonian dynamics of particles is sufficient
as the ``lowest'' Level. However, other purposes may require considering
quantum fields or strings as lower Levels of description. Oppositely, higher
Levels---relevant to, say, chemistry and biology---may be required.

Here, the term \emph{Level}~\cite{Crut92c} delineates a \emph{consistent} and
\emph{self-contained} mathematical formulation of a system's behavior
\cite{carr17a}. There is nothing different about the physical system and its
behavior when moving between Levels; what changes is how those behaviors are
described. Consider properties of the air in the room you are sitting in as you
read this. We typically think of the air as an ``ideal gas'', in the
thermodynamic sense, with properties of temperature, pressure, and volume.
These are the properties we experience on the human scale. Empirically, it was
discovered that these properties are related to one another in a consistent and
self-contained manner, encapsulated in what we now call the Ideal Gas Law.
Another consistent and self-contained description of air in a room is given by
the kinetic theory of gases, in which the air mass is seen as a collection of a
huge number ($\sim 10^{23}$) of air molecules whizzing about and colliding with
each other and with objects in the room. An early triumph of statistical
mechanics was to derive, or construct, the Level II Ideal Gas Law starting from
the Level I kinetic description. 

From Ref.~\cite[Ch. 1]{brush76a}, ``The kinetic derivation of the ideal gas law
is still frequently cited as a paradigm of scientific explanation by teachers
and philosophers, despite (or perhaps because of) the fact that it is hard to
find such simple connections between microscopic models and macroscopic
properties in modern physical science.'' Indeed, the derivation of Fourier's
heat law reviewed above is entirely thermodynamic; i.e., it refers only on the
macroscopic Level. Direct construction of Fourier's law from microscopic
kinetics remains an open challenge~\cite{bone00a}. For the remainder of this
Section, we argue why such bridges between Levels are indeed quite hard to come
by in the constructionist paradigm. The failure of constructionism leaves a
void necessitating a wholly new paradigm for studying the mechanisms of
emergent organization.
 
To ground the discussion, let's return once again to \Benard convection and give an in-depth examination of a successful application of constructionism to emergent organization. 
This makes clear what is missing when constructionism cannot be relied upon. 

For \Benard convection, Level
I is the statistical description of particle motion given by the Boltzmann
equation, Level II is the continuum field theory of the Navier-Stokes
equations, and Level III describes the (in this case static) hexagonal \Benard
cells that emerge. The construction of Level II from Level I for fluid systems
has a long and storied history, including Hilbert's Sixth Problem---to extend
mathematics' axiomatic methods to physics and beyond \cite{grad63a,slem18a}.

In the following, though, the construction of Level III from Level II is what
concerns us. The derivation of the emergence of \Benard cells for the
near-equilibrium primary bifurcation highlights the power of constructionism in
explaining emergent organization---when it is successful. Different
constitutive relations in the Level II equations represent \emph{mechanistic
hypotheses} that underlie the organization that emerges. The correct mechanism
is found by identifying which Level II equations are able to \emph{construct} the
emergent organization that agrees with observation. 


The linear stability analysis performed by Rayleigh \cite{Rayl16a} was the
first theoretical attempt to explain the spontaneous self-organization
observed by \Benard. However, his calculated value of the critical Rayleigh
number that determines the exchange of stability from conduction to
convection---the primary instability---did not agree well with \Benard's
experimental value. Despite subsequent analyses and refined experiments, the
disagreement persisted. It turned out that Rayleigh made an incorrect
assumption about the underlying \emph{physical mechanism} driving the
instability. \Benard performed free-surface convection experiments, but
Rayleigh assumed surface-tension effects would be negligible compared to
buoyancy effects. 

M. J. Block \cite{Bloc56a} later posited that surface-tension effects would dominate
in free-surface convection, and this was subsequently confirmed by the
calculations of J. R. A. Pearson \cite{Pear58a}. He performed a linear
stability analysis similar to Rayleigh, but with surface-tension rather than
buoyancy effects, and this calculation agrees well with \Benard's and now
other's free-surface experiments \cite{Scha95a}.


The competing mechanistic hypotheses are instantiated in various
phenomenological laws that are then used as part of the Level II transport
equations on which the linear stability analysis is performed. Rayleigh assumed
that density varies linearly with temperature and that density variation is
only significant in the buoyancy force. He included no surface tension effects.
Pearson did the opposite, neglecting density variation while including surface
tension and its dependence on temperature. Once the hypotheses were formulated
into equations of motion, the same linear stability analysis is performed to
construct the resulting physical phenomenon of interest---deducing the critical
Rayleigh number at which convection sets in.

Discovering that surface-tension effects dominate the instability mechanism for
free-surface convection is a triumph of constructionism. Not only is the onset
of instability and pattern emergence predicted, but different physical
mechanisms encapsulated in the phenomenological laws predict different onsets
for free-surface and closed convection and these match experiment. We detailed
constructionism's success in explaining the primary \Benard instability to
raise a key question in the study of emergent organization: how do we explain
emergent behaviors if we cannot construct Level III from Level II? As Anderson
pointed out, and as we now detail, constructionism cannot always be relied
upon. There are cases (perhaps many) where Level III simply cannot be deduced
or derived or constructed directly from Level II. 

Bifurcation theory's success in quantitatively predicting the primary \Benard
instability was greatly facilitated by the problem's relative simplicity.  And,
this raises a critical challenge. How does the constructionist approach break
down for more complex systems? Complicated boundary conditions and base states
certainly make the analytical calculations more difficult, perhaps even
intractable. However, these challenges can be (approximately) overcome using
numerical solutions, as is commonly done now for many nonlinear systems. 

A more fundamental issue arises that cannot be easily side-stepped through
numerical solutions. In the convection example, it is possible to isolate a
single mechanistic hypothesis (as given by a particular phenomenological law)
and construct its emergent effect. For more complex systems, though, it may not
be possible to isolate the effect of a single physical mechanism on the
resulting behavior of interest within the tangle of nonlinear interactions and
feedbacks \cite{merl14a}.

In essence, successfully simulating a behavior does not mean we understand it.
For example, we do not have a complete picture of the physical mechanisms
underlying the formation and dynamics of hurricanes, despite being able to
simulate them quite well with sophisticated general circulation models
\cite{Eman87a,Wals97a,Wehn10a}. Similarly, turbulence remains one of the great
mysteries of physics, despite decades of successful and exquisitely detailed
numerical simulation \cite{Heis67a}.

For any given analysis that challenges us, is it really intractable or are we
simply insufficiently clever? For certain problems, we know there is no clever
solution to be found and that the problem is truly intractable in general. Such
problems are known as \emph{undecidable} or \emph{uncomputable}---there is no
algorithm or fixed procedure that yields an answer in finitely-many steps for
all instances of the problem \cite{Moor11a}. While some instances of an
undecidable problem may be solvable, there is no single procedure that always
gives an answer for all instances. 

Equilibrium statistical mechanics is the quintessential example of a bridge
between scales that computes Level II properties from the system's underlying,
``fundamental'', Level I description. Such a bridge does not always exist if
Level ($i+1$) properties are uncomputable starting from Level $i$. Reference
\cite{Gu09a} rigorously showed that for at least one case, the macroscopic
properties of an Ising-like model are uncomputable from the underlying
microscopic dynamics. There are cases then when knowing the lower-Level theory
really cannot reveal everything needed to know about emergent higher-Level
phenomena, even if the lower-Level theory is a consistent description, in
principle.

As an aside, if it seems strange that the theory of computation has something
to say about our fundamental ability to understand physical phenomena, recall
that uncomputability was first devised to answer a similarly fundamental
question in the foundations of mathematics~\cite{Turi36}.

\subsection{As Simple as Possible, But Not Simpler}
\label{sec:CAs}

It is illuminating to examine these issues in a markedly simpler setting than
hydrodynamics. The setting of \emph{cellular automata} (CA) models allows us to
make rigorous statements about formally-difficult problems in understanding
their behavior using the resource analyses provided by the theory of
discrete computation \cite{Lewi98a,Hopc06a,Sips14a}.

A CA deterministically evolves a fully-discrete classical field: A spatial
lattice whose sites take on values from a finite alphabet evolves in discrete
time steps according to a simple spatially-local deterministic update
rule---its \emph{lookup table}. Unlike continuum models, no approximation
scheme is required to numerically simulate CAs. Their numerical evolution is
exact; effectively mapping one large integer (the spatial configuration) to a
unique next integer. Therefore, there is nothing missing, nothing hiding, in CA
simulations. This makes them ideal for studying how governing equations, given
as the local update rule, produce particular temporal behaviors and spatial
configurations.

Since CA description Levels do not map onto the Levels of fluid flows used
above, we refer to the full CA system with dynamics governed by the lookup
table as Level A and emergent organization in the spatial configurations and
temporal behavior as Level B. The emergent Level B dynamics may be
deterministic \cite{Hans95a,Cook04a} or stochastic \cite{Gras83a,Hans90a}.
Reference \cite{Gu09a} showed that Level B dynamics that correspond to
thermodynamic properties for an Ising-like CA are uncomputable from the Level A
CA dynamics. We emphasize that while the Level B description cannot be directly
constructed in general, specific instances are always realized through
evolution---whether ``simulated'' by a computer or by hand---of the Level A
dynamics.  Uncomputability provides a hint of mystery for how Level B
arises from Level A, but due to the exactness of discrete CAs we know for
certain there is nothing extra-physical or magical in Level B that is not
consistent with Level A.

Beyond the physically-motivated CA studied by Ref. \cite{Gu09a}, many CAs, such
as Conway's Game of Life \cite{Conw70a}, are famously \emph{Turing complete}.
In a concrete sense, they are capable of producing arbitrarily complex
behaviors. In general, for a given look-up table and initial condition, one
cannot know whether a particular lattice configuration will ever be generated
by a Turing-complete CA (due to a reduction of the halting problem)
\cite{Kari94a}. How can we deduce the relation between mechanistic hypotheses
in the governing equations and resultant long-term behaviors if the latter are
not computable from the former? Relatedly, dynamical systems with uncomputable
long-time behavior represent a level of unpredictability above and beyond that
presented by chaotic systems \cite{Moor90a}.

Recall though that uncomputability means there is no general algorithm or
procedure that gives an answer (in finite time) for \emph{all} instances of
a problem. Also, it is the long-term CA behavior described above that is
uncomputable. For particular instances of finite-time behaviors, we can always
simulate a system's governing equations and observe what patterns and behaviors
emerge. In addition to undecidability, many CAs pose another formal challenge
that complicates mechanistic analysis for specific behaviors that arise in
finite time. For example, determining the configuration (system state) of
certain CAs from a given initial condition after $t$ time steps is
\textbf{P}-Complete \cite{Moor97c,Near06a}. Let's explain the consequences of
this.

Although not formally established, it is widely believed that there is a class
of ``inherently sequential'' problems that are in the computational complexity
class \textbf{P} but not in \textbf{NC}. Problems in \textbf{P} can be solved
in polynomial time on a serial computer, whereas problems in \textbf{NC} can be
solved in polylogarithmic time on a parallel computer. For example, there is a
class of ``quasi-linear'' CAs whose state prediction problem is in
\textbf{NC} \cite{Moor97b}. Algebraic properties of these CAs allow information
from an initial condition to be split onto multiple processors, and then each
processor performs a logarithmic-time computation on its own simplified
problem. If, as widely believed, \textbf{NC} $\neq$ \textbf{P}, the
\textbf{P}-Complete CA prediction problem is inherently sequential. As a
consequence, a prediction cannot be computed qualitatively faster than
explicitly evolving each state from its proceeding state, in sequence---which
is linear in time. 

Cellular automata directly and concretely demonstrate that very simple rules,
repetitively applied, can give rise to arbitrarily complicated behaviors.
Moreover, they can be formally hard to analyze, confounding traditional
constructionist methods. For many CAs deducing emergent Level B properties and
long-term behaviors from Level A dynamics is uncomputable. For inherently
sequential CAs, mechanistic hypotheses must be tracked through at each time
from an initial condition to a behavior at a later time. The number of relevant
interactions between initial and latter times (the lightcone) grows
exponentially with time. Of the $256$ ``elementary'' CAs---radius-$1$
interactions and binary alphabet---one, rule 110, is known to be Turing
complete \cite{Cook04a} and inherently sequential \cite{Near06a}. Both of these
properties arise through emergent Level B organization that can be mapped to a
model of universal computation. Why is it that the particular lookup table of
rule 110 is the only Level A dynamics of elementary CAs that produces such
complex Level B behaviors? Even though providing the bridge to explain Level B
behavior from Level A is formally difficult for rule 110, something we can do is to
formulate a complete and self-contained model for Level B behaviors, as has also been done for rule 54 \cite{Hans95a}.

To summarize, most approaches to pattern formation theory seek to provide a
bridge from the Level II theory provided by the nonequilibrium equations of
motion to the Level III organization that may spontaneously emerge. In this,
\emph{principles} of organization attempt to address ``how'' and ``why''
questions: e.g., how does fluid in a box spontaneously form hexagonal
convection cells, and why does it happen at different critical temperature
gradients for open- and closed-top boundary conditions? 

The previous section discussed how such questions are typically answered
from a constructionist approach. Mechanistic hypotheses of the underlying
physics are formulated as the Level II theory that then ``explains'' the
observed Level III behavior if said behavior can be constructed from the Level
II equations of motion. This constructionist paradigm breaks down if the
construction becomes intractable. Cellular automata models rigorously show that
constructing the higher Level (B) from the lower Level (A) can be uncomputable,
and thus formally difficult. Even though there is no mystery in the microscopic
constituent components and their interactions, there are cases where knowing
these cannot help us understand higher-Level organization that emerges from
the lower Level dynamics. 

These observations force turning attention away from constructionism to an
alternative approach to \emph{principles} of organization. Although we cannot
always bridge from one Level to the next, can we at least, as with rule 110,
formulate a complete and self-contained theory of the higher-Level organization
that emerges? The rigorous results of intractability in CAs suggests that
constructing Level III from Level II in physical systems, like fluid flows, may
be generally intractable. Can we, though, discover general principles common to
Level III descriptions of these systems, for example analogous to conservation
laws found in Level II descriptions?

While the computation-theoretic methods cannot be easily applied to continuum
models, cellular automata provide examples of self-organizing systems that are
provably difficult to study and for which simulation is very likely a
requirement for their study. Taken together, the arguments here point to
computation theory playing a central role in the study of self-organization
and, thus, in the formulation of general principles of organization. This is in
contrast to the typically analytically-oriented formulation of general
principles in physics. This is perhaps historically why principles of
organization have remained elusive there. The central contribution of
computation theory in our diagnosis of the failures of constructionism is that
intractability stems from the growth of the resources required to evolve
organized configurations. This suggests a duality between emergent organization
and the resources a system commandeers to evolve them. We now turn to modern
formulations of emergent organization using concepts and constructions from the
theory of computation.

\section{Organization Beyond Constructionism}
\label{sec:FormalizingOrganization}

If we seek general principles of organization, surely we must first lay out what organization \emph{actually is}. Bifurcation theory identifies organization with exact symmetries (and small deviations from them) using Fourier modes. Failing exact symmetries and their associated group algebra, how does one write down a mathematical expression for, say, a hurricane? 

As a concrete example, consider the dynamics of vortices in free-decay
two-dimensional turbulence. The Level II theory is the vorticity equation
(derived from Navier-Stokes) and the Level III theory describes the behavior of
the ``vortex gas'' that emerges from random initial conditions. In this,
like-signed vortices may undergo a pairwise merger if certain geometric
conditions are met \cite{carn91a}, resulting in a power-law decay of the number
of vortices over time. This behavior was discovered empirically, using a
specialized algorithm to identify vortices based on vorticity thresholding and
additional geometric considerations \cite{mcwil90a}. As a prerequisite to
general principles of organization and moving away from relying on specialized
approaches for specific systems, can we first identify general principles for mathematically identifying emergent organization?

In addition to solving nonlinear equations through numerical simulation,
computation---in the guise of computation theory---points a way to answering
this question, by providing a mathematical accounting of more general forms of
organization in---what it calls---\emph{formal languages}. This section
outlines two modern approaches for formalizing organization beyond exact
symmetries---approaches inspired by computation theory. The following section
then discusses their relationship and how they fit into a statistical mechanics
of self-organizing systems evolving arbitrarily far from equilibrium.

\subsection{Organization Through Compression}

One motivating intuition behind understanding organization is that of
\emph{compression}. Organized behaviors are higher-Level degrees of freedom
that emerge from \emph{collective motion} of lower-Level degrees of freedom.
\Benard cells and more general vortices, for instance, involve the coordinated
motion of many fluid parcels (differential volume elements). Since the fluid
parcels evolve together, they need not be tracked separately. They may be
considered equivalently and collectively as a single emergent degree of
freedom. The collective degrees of freedom then provide a natural
dimensionality reduction for a compressed representation of a system's behavior
and configurations.

Another perspective on the importance of compression is more easily seen
through the lens of ``pattern''. (For our immediate purposes ``pattern'' and
``organization'' are essentially synonymous.) One's typical mental image of a
``pattern'' is some exact symmetry; e.g., a repetitive tiling in two spatial
dimensions. The \Benard cells that have been our guiding example form a
symmetric hexagonal tiling, under the right circumstances. Such symmetries
provide a predictive regularity---one region of \Benard cells ``predicts'' other
parts. And, the regularity of exact symmetries can be predicted exactly.

As emphasized above, though, our goal is to articulate a more general notion of
organization beyond exact symmetries. One path to this is to account for
partial or noisy symmetries that describe approximate predictive regularities.
Like exact regularities, these noisy regularities can also be leveraged to
create a system's compressed representation. The latter will be constructed
using semigroup algebra which generalizes the group algebra of exact symmetries
to noisy symmetries. Before delving further into these two
approaches---predictive regularity and dimensionality reduction---it is
insightful to review several computation-theoretic ideas that motivated the use
of compression in the first place. 

Statistical mechanics' successful use of correlation functions, structure
factors, order parameters, and symmetry groups to describe regularities in
physical systems \cite{Binn92a} has for some time inspired the search for
mathematical descriptions of organization in the form of scalar quantities.
Historically, what came to be called measures of ``complexity'' \cite{Feld97a},
we will consider here as measures of ``degrees of organization''.

The most notable complexity measure motivated by compression is that introduced
by Kolmogorov and Chaitin. An object's Kolmogorov-Chaitin (KC)
complexity is the length of the shortest universal Turing machine program that
reproduces it \cite{Cove06a,Vita89a,Kolm65,Chai66}. KC complexity measures the
minimum computational resources---information storage and computational
operations---required to exactly describe an object. Intuitively, objects
consisting of more complex patterns require more description and so are less
compressible.

However, it is now well-appreciated that KC complexity is in fact a
quantitative measure of randomness, rather than organizational complexity
\cite{Brud83}. Related complexity measures---such as, \emph{sophistication} and
\emph{logical depth} \cite{Benn86,Kopp87a,Benn88,Kopp91a,Crut98d}---attempt to
separate out the ``meaningful'' complexity from randomness. And, in any
case, all approaches that rely on minimally compressed programs are
nonconstructive---the task of determining minimal programs is uncomputable. 

Although uncomputable, the true minimal program may be approximated in practice
using practical compression algorithms. A popular choice is the Lempel-Ziv
algorithm \cite{Lemp76a,Ziv78a} used for zip and Unix compress. Here too,
though, there are difficulties in separating organization from randomness. For
instance, the Lempel-Ziv algorithm doubles the length of random strings when,
being incompressible, they should be unchanged. 

Another compression-based method relies on determining an object's \emph{minimum
description length} (MDL) by encoding descriptions of both model and residual
error \cite{Riss78a}. Using this information-theoretic criterion, the MDL
principle selects the predictive model with the lowest generalization error
from a given set of candidates. So, while useful for training in a
parameterized model class---such as, autoencoder neural networks used for
dimensionality reduction \cite{hint93a}---it does not provide a constructive
basis for a first-principle approach to quantifying organization.

\subsection{Modes of Organization}

The structure of configurations generated by classical field theories is
commonly approached in physics and engineering through the notion of
compression via dimensionality reduction. Consider a (classical) spacetime
field $X(\mathbf{r}, t)$ and assume it can be decomposed through a linear
superposition as:
\begin{align}
    X(\mathbf{r}, t) = \sum_{i=1}^\infty a_i \phi_i(\mathbf{r},t)
    ~,
\end{align}
where $\mathbf{r}$ are spatial coordinates and $t$ is time. In principle these
coordinates are continuous, but in practice they are discretized for use with
numerical methods. The spacetime components $\phi_i(\mathbf{r}, t)$ of the
decomposition are referred to as \emph{modes}. 

If $X(\mathbf{r},t)$ is structured---e.g., there is collective motion among its
degrees of freedom through time---then this organization can be exploited to
produce a compressed representation $\tilde{X}_n(\mathbf{r}, t)$ using only a
finite number of $n$ modes:
\begin{align*}
    X(\mathbf{r}, t) &\approx \tilde{X}_n(\mathbf{r},t) \\
    & := \sum_{i=1}^n a_i \phi_i(\mathbf{r}, t) 
    ~.
\end{align*}
From the preceding discussion of organization through compression, the intuitive
idea is that an optimal truncation that minimizes the reconstruction error
$||X(\mathbf{r}, t) - \tilde{X}(\mathbf{r},t)||$ leverages any organization
present to create the modes $\phi_i(\mathbf{r}, t)$. The
leading modes of an optimal truncation individually contribute more to the full
system behavior if they contain collective, organized motion. (Though, not all
modes describe collective behavior.) This motivates seeking specific
constructions of $\{\phi_i(\mathbf{r},t)\}_1^n$ as \emph{modes of organization}. 

As a first step, it is typical to separate the space and time dependencies, so
that there are \emph{spatial modes} $\phi_i(\mathbf{r})$ with time-varying
coefficients $a_i(t)$:
\begin{align}
    X(\mathbf{r}, t) = \sum_i a_i(t) \phi_i(\mathbf{r})
    ~.
    \label{eq:discrete-decomp}
\end{align}
That is, the spatial modes $\phi_i(\mathbf{r})$ form a complete basis to
express a system's configurations---the spatial field $X(\mathbf{r},t^*)$, at
any (and every) time $t^*$ using the coefficients $a_i(t^*)$. That said, there may
not be a complete basis such that Eq.~(\ref{eq:discrete-decomp}) holds.  This
is the case, for example, with turbulent fluid flows. There is a natural
generalization to Eq.~(\ref{eq:discrete-decomp}) for these cases given below. 

The standard method to compute a finite set of spatial modes and their
time-varying coefficients is \emph{principal component analysis} (PCA), also
known as \emph{proper orthogonal decomposition} (POD), among other names
\cite{Tair17a}. In this, modes are computed using the singular value
decomposition or, equivalently, an eigendecomposition of the data covariance
matrix. From a statistical viewpoint, an organized system will have a high
degree of correlation among its degrees of freedom due to their collective
motion. POD computes modes that are linearly uncorrelated with each other, and
so capture the majority of correlations in the system.

In this way, POD finds a change of coordinates to an orthonormal basis such
that the first mode---the first dimension, in the new coordinates---captures
the majority of the system's variance, the second mode captures the majority of
the remaining variance, and so on. From a physics point of view, POD provides
an optimal finite decomposition since for any fixed number $n$ of modes the
reconstruction error $||X(\mathbf{r}, t) - \tilde{X}(\mathbf{r},t)||$ is
minimized by POD \cite{Alga69a}. If the spacetime field is, for example, the
velocity field of a fluid flow, POD is optimal in capturing the dominant
energetic contributions \cite{Rowl04a}. 

A finite decomposition as in Eq.~(\ref{eq:discrete-decomp}) can be used as a
reduced model of the system dynamics. In this, the partial differential
equation Eq. (\ref{eq:gen_dyn}) for the evolution of $X(\mathbf{r},t)$ is
replaced by a finite set of ordinary differential equations that govern the
evolution $\dot{a}_i(t)$ of the time-dependent coefficients. This can be
achieved through a Galerkin projection of the PDE evolution operator onto the
space spanned by the modes \cite{Holm12a} or through data-driven forecasting
of the coefficients \cite{maul20a,wang20a}. Note that POD mode orthogonality is
very useful in this.

\subsection{Evolution Operators} 

While POD has been an remarkably fruitful method for analyzing coherent
structures in fluid flows, it is an ad hoc empirical method. Its use in
investigating organization in fluid flows followed shortly after the notion of
coherent structures emerged in the study of turbulence. Let's recall this
history as a means to introduce a much richer and theoretically-grounded
framework for formalizing coherent organization in classical field theories
based on the Koopman and Perron-Frobenius \emph{evolution operators} \cite{Laso13a}. 

Statistical studies of turbulence in the latter half of the $20^{\text{th}}$ century found a separation of scales in fluctuations about the mean flow. There are smaller scale ``random'' fluctuations on top of coherent periodic fluctuations. The \emph{triple decomposition} \cite{reyn72a,Mezi13a} of the fluid velocity field formalizes this:
\begin{align*}
    X(\mathbf{r}, t) = \bar{X}(\mathbf{r}) + X_p(\mathbf{r}, t) + X_c(\mathbf{r}, t)
    ~,
\end{align*}
where $\bar{X}(\mathbf{r})$ is the mean flow, $X_c(\mathbf{r}, t)$ is the
chaotic component that generates the apparent small scale randomness, and
$X_p(\mathbf{r}, t)$ is the coherent component consisting of spatial modes that
evolve periodically in time as:
\begin{align*}
    X_p(\mathbf{r}, t) = \sum_i e^{\omega_i t} \phi_{i}(\mathbf{r})
    ~, 
\end{align*}
with $\omega_i \in \mathbb{C}$. Spatial structures that evolve collectively at
a single frequency, such as those in $X_p(\mathbf{r}, t)$, were first studied
in the context of linear stability where they were identified as \emph{global
modes} \cite{huer90a,Mezi13a}. 

\subsubsection{Koopman Operators}

Motivated in this way, we now review a general and rigorous framework for modal
decomposition of arbitrary dynamical systems using Koopman operators, as first
introduced in Ref.~\cite{Mezi05a}. Rather than analyze the system through the
geometric viewpoint of the nonlinear evolution of $X(t) \in \Omega$, it is
useful to analyze the linear dynamics of system observables given by Koopman
operators \cite{Koop31a}.

An \emph{observable} $g$ is simply a scalar-valued function of the system state
$g: \Omega \rightarrow \Reals$. Observables are elements of a function space
$\mathcal{F}$. The parameterized (semi)group of \emph{Koopman operators}
$\Koop^\tau: \mathcal{F} \rightarrow \mathcal{F}$ are defined through
composition with the dynamics $\Phi^\tau$:
\begin{align}
    \Koop^\tau g &:= g \circ \Phi^\tau \label{eq:Koopman}\\
     &= g \bigl(\Phi^\tau(X)\bigr) \nonumber
     ~,
\end{align}
where $\Phi^\tau$ is the flow map defined by Eq.~(\ref{eq:gen_dyn}) such that
$X(t_0+\tau) = \Phi^\tau\bigl(X(t_0)\bigr)$.

Koopman operators define a dynamic in the space of
observables generated by $\Phi^\tau$. That is, for an initial state $X(t_0)$
and observable $g_0$, the action of the Koopman operator $\Koop^\tau$ gives a
new time-shifted observable function $g_\tau$ defined by:
\begin{align}
    [\Koop^\tau g_0]\bigl(X(t_0)\bigr) &:= g_\tau\bigl(X(t_0)\bigr)\\
    &= g_0\bigl(X(t_0+\tau)\bigr)
    ~.
\end{align}
Considering discrete time steps for simplicity, Koopman operators give a flow
through the space of observables for a given observable $g_0$ and initial
system state $X_0 := X(t_0)$: 
\begin{align*}
    \{&g_0(X_0), \\
	&g_1(X_0) = [\Koop g_0](X_0),\\
	&g_2(X_0) = [\Koop^2 g_0](X_0), \\
	& \ldots, \\ &g_i(X_0) = [\Koop^i g_0](X_0), \\
	& \ldots\}
    ~.
\end{align*}

For dynamical consistency, the space of observables $\mathcal{F}$ must be
closed under the action of all Koopman operators: $\Koop^\tau g_i \in
\mathcal{F}$ for all $\tau$ and all $g_i \in \mathcal{F}$. For almost all
nontrivial cases, this means $\mathcal{F}$ is uncountable and the Koopman
operators $\Koop^\tau$ are infinite-dimensional \cite{brun16b}. If
$\mathcal{F}$ is a vector space, as is typical, then $\Koop^\tau$ are linear, even
if the original state-space dynamic $\Phi^\tau$ is highly nonlinear.
The Faustian bargain in this is to trade-away nonlinearity in low-dimensions for
linearity in infinite dimensions. 

It is common to take $\mathcal{F}$ to be a Hilbert subspace $\mathcal{F}
\subseteq L^2(\Omega)$ of square-integrable functions. Another common choice is
the space $L^\infty(\Omega)$ of essentially bounded functions. Note that since
Koopman operators are defined through composition with the system dynamic, if
the system state remains bounded over time---i.e., does not ``blow up''---then
the action of Koopman operators on bounded functions always returns another
bounded function. This gives the requisite dynamical consistency. 



With the operators in hand, let's consider the \emph{Koopman mode
decomposition} \cite{Mezi05a,brun22a} and its use to define organization in
time-independent dynamical systems. A Koopman eigenfunction $\varphi_i$ is a
scalar observable satisfying:
\begin{align*}
K^\tau \varphi_i(X) = e^{(\lambda_i \tau)} \varphi_i(X)
  ~,
\end{align*}
with
eigenvalue $\lambda_i$. If the full set of eigenfunctions forms a complete
basis of the function space $\mathcal{F}$, then an arbitrary scalar observable
$g \in \mathcal{F}$ can be expanded as:
\begin{align*}
    g(X) = \sum_i^\infty \nu_i \; \varphi_i(X)
    ~, 
\end{align*}
with expansion coefficients $\nu_i$. It is natural to consider vector-valued
observables, each component of which can be expanded as above:
\begin{align*}
\mathbf{g}(X) & =
	\begin{bmatrix} g_1(X) \\ g_2(X) \\ \vdots \\ g_m(X) \end{bmatrix} \\
	& = \sum_i^\infty \varphi_i(X)
	\begin{bmatrix} \nu_1 \\ \nu_2 \\ \vdots \\ \nu_m \end{bmatrix} \\
	& = \sum_i^\infty \varphi_i (X) \vec{\nu}_i
    ~.
\end{align*}
The vector-valued coefficients $\vec{\nu}_i$ are known as the \emph{Koopman modes} of the vector-valued observable $\mathbf{g}$. They are given as the projections of $\mathbf{g}$ onto the eigenfunctions:
\begin{align*}
    \vec{\nu}_i = \begin{bmatrix} \langle \varphi_i, g_1 \rangle \\ 
    \langle \varphi_i, g_2 \rangle \\ 
    \vdots \\
    \langle \varphi_i, g_m \rangle \\ \end{bmatrix}
    ~. 
\end{align*}

Emergent organization here is associated with the Koopman modes of one
vector-valued observable in particular---the \emph{state identity} observable
$\mathbf{g}_I(X) = X$. While Koopman operators (and Perron-Frobenius operators
introduced shortly) are defined for arbitrary dynamical systems, we now
consider the system state $X(\mathbf{r}, t)$ to have a spatial structure
$\mathbf{r} \in \mathcal{D}$ over spatial domain $\mathcal{D}$. In this case,
for fixed time, the system state $X(\mathbf{r})$ is a function from the spatial
domain $\mathcal{D}$ to the reals. It gives the value of the field
$X(\mathbf{r}^*) \in \Reals$ at each spatial location $\mathbf{r}^* \in
\mathcal{D}$. The components of the state identity observable are parameterized
by the spatial locations $\mathbf{r}$ in $\mathcal{D}$ and map from full
spatial fields $X(\mathbf{r}) \in \Reals^{\mathcal{D}}$ to the value
$X(\mathbf{r}^*) \in \Reals$ of the field at the specific location
$\mathbf{r^*}$: $g_{r^*}\bigl(X(\mathbf{r})\bigr) = X(\mathbf{r^*})$. 

The vector-valued Koopman modes have a component for each component of their
vector-valued observables. The vector-valued (or field-valued) state identity
variable has components indexed by the spatial locations $r \in \mathcal{D}$.
And so, the Koopman modes also have components indexed by spatial locations.
Said differently, the Koopman modes are themselves spatial fields. We can thus write the
expansion of $\mathbf{g}_I$ as:
\begin{align}
    \mathbf{g}_I\bigl(X(\mathbf{r})\bigr) = \sum_i^\infty \varphi_i\bigl(X(\mathbf{r})\bigr) \; \vec{\nu}_i(\mathbf{r})
    ~.
    \label{eq:koop_modes}
\end{align}
Using the definition of Koopman eigenfunctions, the evolution of $\mathbf{g}_I$
in the expansion takes the simple form:
\begin{align}
    \mathbf{g}_I\bigl(X(\mathbf{r}, t_0+\tau)\bigr) = \sum_i^\infty \varphi_i\bigl(X(\mathbf{r}, t_0)\bigr) \; e^{(\lambda_i \tau)} \vec{\nu}_i(\mathbf{r})
    ~.
    \label{eq:koop_mode_dyn}
\end{align}

In this view, Koopman eigenfunctions are a complete basis of $\mathcal{F}$ that
are intrinsic to the dynamics. And, importantly, Koopman modes are fields that
express the flow in this intrinsic basis evolving in time at a single
frequency---determined by $\lambda_i$. Therefore, they generalize the notion of
global modes, introduced for turbulent flows, to any classical field theory.

\begin{figure*}[t]
\centering
\includegraphics[width=1.0 \textwidth]{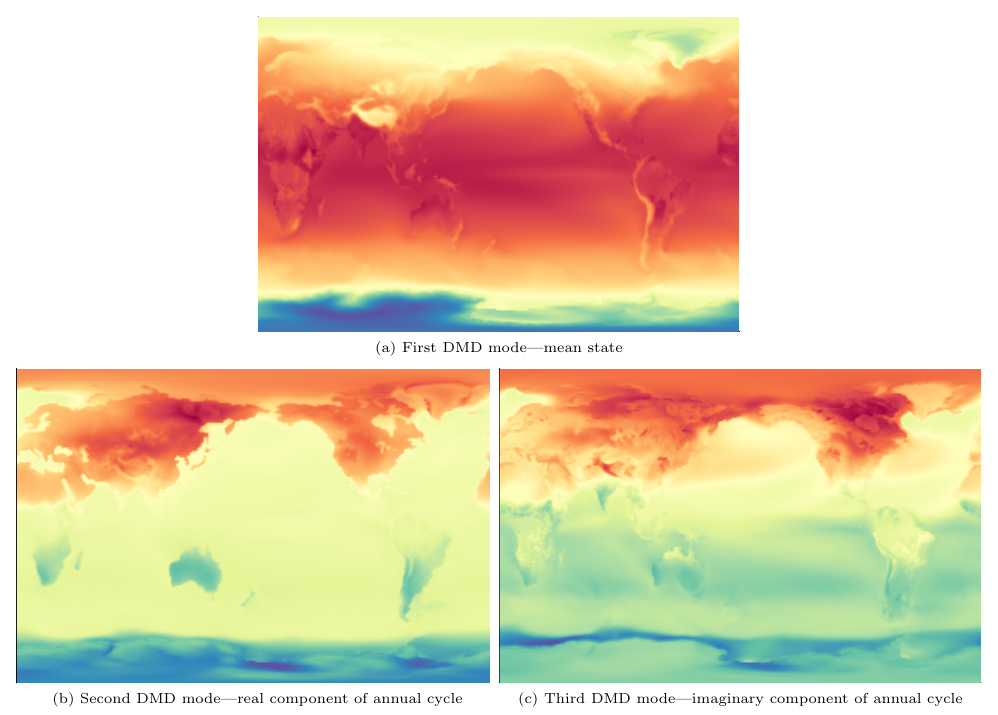}
\caption{The first three DMD modes of the temperature field from a Community
Earth System Model run. The first mode (a) has eigenvalue $\approx 1.0$ and
approximates the invariant field and corresponds to the climatological mean
state. The next two modes (b) and (c) are the real and imaginary components,
respectively, of the mode with an imaginary eigenvalue that corresponds to oscillations with a $12$ month period, and so correspond to the annual cycle. 
	}
\label{fig:dmd-modes}
\end{figure*}

Note that \emph{all} Koopman eigenfunctions, by definition, evolve at a fixed
frequency determined by their associated eigenvalue. When the eigenfunctions are a complete basis---e.g., Hamiltonian
(conservative) system dynamics---the dynamics can be fully decomposed into
Koopman modes. \emph{Koopman modes distill a particular notion of organization
in the system by decomposing the dynamics into components that evolve at single
frequencies in time}. The decomposition applies to all field theories, whose
behaviors can range between coherent collective motion and chaotic ``random''
motion. 

As with the generic and POD modal expansions described above, the sum in
Eq.~(\ref{eq:koop_mode_dyn}) may be well approximated with a finite sum; adding
an important level of constructiveness to the formalism. Consonant with this,
the leading Koopman modes of this truncation are considered as coherent spatial
structures in the flow, if any are present. In particular, system dynamics in
the Koopman basis, Eq.~(\ref{eq:koop_mode_dyn}), is given simply by the
exponential $e^{(\lambda_i \tau)}$, and so $\text{Re}(\lambda_i) \leq 0$ gives
the temporal decay rate of the contribution from the eigenfunction $\phi_i$ and
Koopman mode $\nu_i$. The leading Koopman eigenfunctions and modes are those
with largest $\text{Re}(\lambda_i)$---i.e., closest to $0$. These modes damp
out the slowest and so dominate the system dynamics at later times when other
modes have decayed away.

The popular \emph{dynamic mode decomposition} (DMD) algorithm provides a
practical and efficient method for approximating leading Koopman modes
\cite{Tu14a,Klus16a,brun22a}. In this, a best-fit linear dynamic is found
directly from a system's observed or measured behaviors. The eigenvectors of
the best-fit matrix are the dynamic modes that approximate Koopman modes. The
DMD modes are linear combinations of POD modes that evolve coherently with
single frequencies. Said simply, DMD is a POD spatial model combined with a
Fourier temporal model.

Example DMD modes for the temperature field of a $1$-degree resolution Community Earth System Model run are shown in Figure~\ref{fig:dmd-modes}. The first mode, shown in (a), has eigenvalue $\approx 1.0$ and so approximates the invariant climatological mean field. The next leading mode is imaginary with an imaginary eigenvalue corresponding to oscillations with a $12$ month period, and so is associated with the annual cycle. The real and imaginary components of this mode are shown in (b) and (c), respectively. The climatological mean and annual cycle are the primary statistical features that climate analysis is built upon. Dynamic Mode Decomposition is able to approximate Koopman modes that provide spatial fields associated with these statistical features. 


To complete the discussion of Koopman operators, we now examine the case when
the eigenfunctions do \emph{not} form a complete basis. Vector-valued
observables may still be fully decomposed, however the spectrum of Koopman
operators will have a continuous component in addition to the discrete point
spectrum of eigenvalues above. Using \emph{spectral measures} of Koopman
operators, we can expand the state identity observable as:
\begin{align}
     \mathbf{g}_I\bigl(X(\mathbf{r}, t_0+\tau)\bigr) &= \sum_j^\infty \varphi_j\bigl(X(\mathbf{r}, t_0)\bigr) \; e^{(\lambda_j \tau)} \vec{\nu}_j(\mathbf{r}) \nonumber \\
     & \quad + \int_{-\pi}^\pi e^{i \tau \omega} \tilde{\varphi}_{\omega} \bigl(X(\mathbf{r}, t_0)\bigr) d\omega
    ~,
\label{eq:spectral_measure}
\end{align}
where $\tilde{\varphi}_{\omega}$ can be thought of as
continuously-parameterized eigenfunctions. See Refs. \cite{colb21a} and
\cite{brun22a} for more information on spectral measures of Koopman
operators. 

If we separate out the mean value of the field from $\mathbf{g}_I$, then
Eq.~(\ref{eq:spectral_measure}) gives the generalization of the triple
decomposition to arbitrary classical field theories~\cite{Mezi13a}. The field dynamics are
given by (i) the mean, (ii) a term representing ``global modes'' that evolve at
single frequencies, and (iii) a chaotic component given by the continuous part
of the spectrum. Complex dynamics typically contain an amalgam of regularity
and randomness. And so, at its roots, extracting organization requires
disentangling the randomness from the regularity. In doing so, the Koopman
spectral decomposition shares a conceptual similarity with the intrinsic
computation approach described below. 

Finally, we note that the spectral measure with respect to an observable $g$ is
the Fourier transform of the dynamical autocorrelations $\langle \Koop^\tau g,
g \rangle$ \cite{colb21a,brun22a}. Intuitively, systems with high levels of
organization should be highly correlated. While simple scalar quantities like
two-point (auto)correlations do not directly capture the notion of organization
we seek, the above results show organization can be reconstructed from the
totality of dynamical multivariate autocorrelations using Koopman spectral
measures. 

\subsubsection{Organization Beyond Koopman}

The Koopman mode decomposition, as just described, is best seen as describing
organization statistically. Recall that the Koopman modes are full spatial
fields and, thus, are candidates for describing global organization, as opposed
to localized organization. Moreover, if the dynamics are time-independent, as
typically assumed when using Koopman mode decomposition, then the Koopman
operators are also time-independent. Hence, the Koopman eigenfunctions and
Koopman modes are time-independent. And so, organization from the Koopman mode
decomposition is presented in the form of Koopman mode fields that are
temporally-static and spatially-global. This is the case for any modal
decomposition that assumes a space and time separation of variables, given by
Eq.~(\ref{eq:discrete-decomp}).

As it generalizes global modes and the triple decomposition, such statistical organization is useful in the study of turbulence \cite{Mezi13a}. It has also proven a useful framework for organization in Earth's climate, typically treated as the statistical behavior of the Earth system \cite{luca23a}. 

Beyond the primary bifurcation, more complex and complicated organization may
emerge that cannot be adequately described in the statistical sense of the
Koopman mode decomposition. In particular, \emph{coherent structures}
\cite{Hall15a} may form that are localized in space, dynamic over time, and
ephemeral with finite lifetimes. Consider, for example, individual weather
systems such as hurricanes and atmospheric rivers. For fluid systems, the
Koopman mode decomposition is an Eulerian approach, whereas the Lagrangian
framework is better suited to capture coherent structures \cite{Hadj17a}.
Similarly, it is conceptually advantageous to utilize Perron-Frobenius
operators, which are dual to Koopman operators. Note that because the evolution
operators are dual to one another, they ultimately contain the same information
about the system. However, one may be more convenient than the other for
certain tasks~\cite{Klus16a}, and for coherent structures in time-dependent
flows Perron-Frobenius operators are more convenient than Koopman operators.

\subsubsection{Perron-Frobenius Operators}

In contrast to Koopman operators evolving system observables,
\emph{Perron-Frobenius operators} $\PF^\tau$ evolve probability distributions
over the system states. For a probability measure $\mu$, the probability
distribution given by $\mu$ evolves in time according to Perron-Frobenius operators
via:
\begin{align}
\mu_\tau & := \PF^\tau \mu \nonumber \\
	& := \mu \circ \Phi^{-\tau}
    ~,
\label{eq:PF}
\end{align}
where $\Phi^{-\tau}(A) := \{X \in \Omega \; : \; \Phi^\tau(X) \in A\}$ is the
preimage of a measurable set $A$ under the flow $\Phi^\tau$.

Comparing Eq.~(\ref{eq:PF}) to Eq.~(\ref{eq:Koopman}), Perron-Frobenius
operators are adjoint to Koopman operators in appropriately defined spaces:
\begin{align*}
\langle \PF^\tau f, g\rangle = \langle f, \Koop^\tau g \rangle
  ~.
\end{align*}
They provide classical dynamics on Hilbert spaces, analogous to quantum
mechanics. Perron-Frobenius operators are the classical equivalent of the
Schrodinger picture, and Koopman operators of the Heisenberg picture. 

Since Perron-Frobenius operators evolve probability distributions, they
historically have a closer association to statistical mechanics than Koopman
operators \cite{Mack92a,gasp05b}. The most immediate connection is to notice
that an eigenfunction of Perron-Frobenius operators with eigenvalue one is
an \emph{invariant distribution} $\mu^*$, such that:
\begin{align*}
    \PF^\tau \mu^* = \mu^*
    ~.
\end{align*}
Invariant distributions are abstractions of equilibrium and nonequilibrium
steady-state distributions from statistical mechanics. For ergodic dynamics,
there is a unique invariant distribution \cite[Theorem 4.5]{Mack92a}. For
Hamiltonian dynamics, Perron-Frobenius operators reduce to the familiar
Liouville operators of statistical mechanics \cite{Mack92a}. 

Let us now detail how Perron-Frobenius operators complement the Koopman mode decomposition for analyzing structure and organization in time-dependent systems. We consider a Lagrangian flow in which fluid parcels evolve according to the time-dependent flow map 
\begin{align}
    \mathbf{y}(t; t_0, \mathbf{y}_0) := F^t_{t_0}(\mathbf{y}_0)
    ~, 
    \label{eq:Lagrangeadvect}
\end{align}
where $\mathbf{y}$ is a fluid parcel or tracer, as opposed to a full spatial
field. The time-dependent dynamics $F^t_{t_0}$ define time-dependent evolution
operators, $\mathcal{K}_{t_0}^t$ and $\mathcal{P}_{t_0}^t$, that depend on the
initial time $t_0$ in addition to the evolution time $t$. Each initial time
$t_0$ thus has its own Koopman mode decomposition. (The latter may or may not
be unique; i.e., two different times may generate the same decomposition.) 

For Lagrangian flows, the (now time-dependent) Koopman modes are no longer
spatial fields, so their interpretation is not as clear. However,
Perron-Frobenius operators evolve distributions over the state space, which for
Lagrangian flow is the spatial domain. Recall, for analyzing statistical
organization through the Koopman mode decomposition above, we considered the
Koopman modes for a particular observable---the state identity observable. Now,
we do a similar development in which we consider the time-dependent evolution
of uniform distributions over smooth contiguous regions in the flow domain,
known as \emph{material surfaces}. The Lagrangian flow evolves all the points
on this surface, and so the action of the Perron-Frobenius operator $\mathcal{P}_{t_0}^t$
gives the dynamics and deformation of these material surfaces. 

Given that Koopman and Perron-Frobenius operators are adjoint, they share the
same spectrum. In both the time-independent and time-dependent cases, the
spectral decomposition of the evolution operators separates regular motion from
chaotic motion in the system. However, regular---i.e., not chaotic---motion does
not necessarily imply coherent collective motion. The interpretation of
Perron-Frobenius operators evolving material surfaces in Lagrangian flows
allows for a more precise definition of coherent structures, or more generally
\emph{coherent sets}, that emerge.

\begin{figure*}[t]
\centering
\includegraphics[width=1.0 \textwidth, trim={0 0 12cm 2cm},clip]{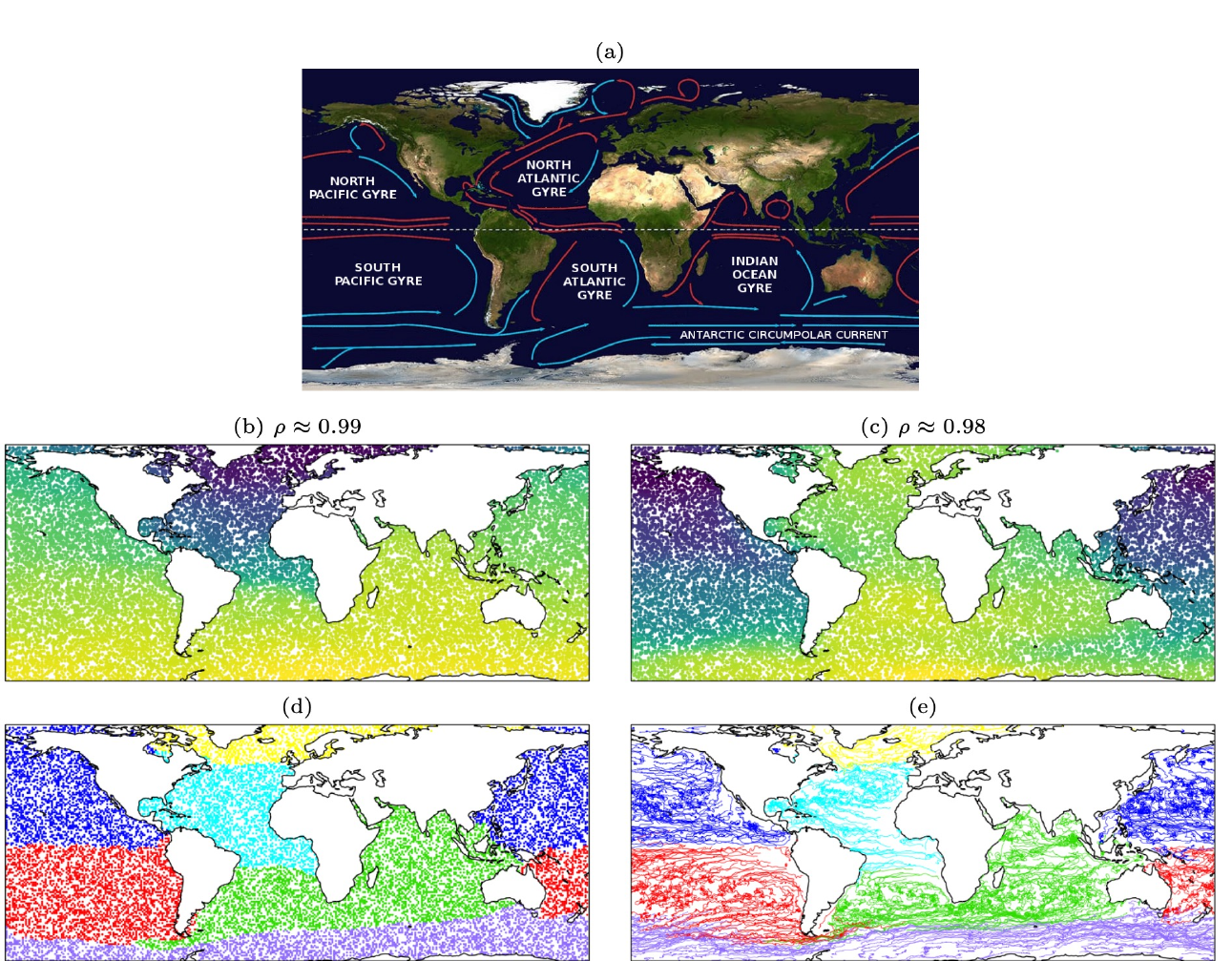}
\caption{(a) Illustration of the major ocean gyres (courtesy of NOAA). (b) First and (c) second eigenfunction. (d) $k$-means clustering of the first six eigenfunctions into six coherent sets. (e) Subset of the trajectories colored according to the coherent sets.
Reproduced, with permission, from Ref.~\cite{Klus19a}. 
	}
\label{fig:ocean-gyres}
\end{figure*}

The invariant distribution is one that does not change over time and is given as an eigenfunction of Perron-Frobenius operators with eigenvalue one. For Lagrangian flows, such an eigenfunction is a material surface that does not change with the flow. All of the fluid parcels originally in the surface, and only those, remain in the surface over time. This is the case of perfect collective motion, but is an idealization that is not realized in most physical scenarios of interest. Vortices in turbulent flows, for example, remain coherent, with a high degree of collective motion, but it cannot be expected that \emph{all} parcels remain in the vortices forever. 

Such behavior, of interest to real-world systems, is captured by eigenfunctions
of Perron-Frobenius operators with eigenvalue close, but not equal, to
unity---known as \emph{almost-invariant sets} \cite{Froy03a,Froy09a}.  These
correspond to \emph{metastable states} in statistical mechanics
\cite{Noe13a,Nuske16a}. Since the associated eigenvalues are not exactly one,
they are not asymptotically-invariant. However, since they are close to one, a
system tends to spend a long time in each before escaping. The simplest example
is found in overdamped Langevin dynamics in a double-well potential
\cite{Klus16a}. The asymptotic distribution is evenly split between the wells.
Before the asymptotic limit however, the system tends to spend a long time in
each well individually before leaving. Thus, distributions concentrated in each
individual well are metastable states. A more complicated example is given by
protein configurations \cite{Nuske16a}.

Returning to the Lagrangian flow case, recall that the Perron-Frobenius
operators evolve sets on the flow domain. The dynamic and ephemeral coherent
structures that interest us---as particular forms of emergent
organization---present several complications to the simple notion of
almost-invariant sets as eigenfunctions of Perron-Frobenius operators with
eigenvalues near one. First, they may move through space over time while
remaining coherent. Thus, the set will move over the flow domain and mostly not
return to itself. In this case, we can formulate \emph{coherent pairs} such
that the probability of Perron-Frobenius operators evolving the first set in
the pair to the second is close to one \cite{Froy10a,Froy10b}. Fluid parcels in
the initial configuration and location of a material surface are likely to end
up in the later configuration and location if that surface is a coherent set.
Finally, it has recently been shown how to extend the formulation of coherent
sets using time-dependent Perron-Frobenius operators to capture the ``birth''
and ``death'' of ephemeral structures with finite lifetimes \cite{froy21a}. 

An example of coherent sets is shown in Figure~\ref{fig:ocean-gyres}, reproduced from Ref.~\cite{Klus19a}. Dominant ocean gyres are identified from Lagrangian flow trajectories as coherent sets, reconstructed using a reproducing kernel Hilbert space approximation of evolution operators. The approximated coherent sets, corresponding to the ocean gyres, are shown in Figure~\ref{fig:ocean-gyres} (d). Lagrangian evolution of sample tracers are shown in (e), with colors corresponding to the coherent set they are assigned to at the initial time $t_0$. As expected from the definition of coherent sets, the trajectories in (e) largely stay contained within their original coherent set. However, there are some crossovers, for example between the red and green sets near the southern tip of South America. 

A note on terminology: Almost-invariant sets and coherent sets are closely-related mathematical constructions, both based on the Perron-Frobenius operator. Data-driven approximation of these provide tools that help identify coherent structures in physical systems. In the above example, the ocean gyres are coherent structures---physical organization in the real ocean that we are interested in---and the coherent sets shown in (d) are data-driven tools that approximately identify the gyres. 

\subsubsection{Recap}

Koopman and Perron-Frobenius evolution operators provide an alternative description of dynamical systems in terms of the evolution of system observables and probability distributions. Their spectral decompositions provide a means of mathematically identifying organization in classical field theories (and dynamical systems more generally). The continuous portion of their spectra is associated with chaotic, apparently random, motion. For time-independent systems, the eigenfunctions associated with the discrete spectrum (eigenvalues) evolve at single frequencies and their Koopman modes represent statistical organization of the system. For time-dependent field theories, coherent structures associated with localized collective motion are identified through eigenfunctions of time-dependent Perron-Frobenius operators defined from the Lagrangian flow map. In all cases, a finite set of eigenvalues, eigenfunctions, and their coefficients (modes) provide a compressed representation of the system and its dynamics, and hence its organization.

From their origins in the study of turbulence, today modal decompositions are
the standard approach to coherent structure analysis. They are commonly
employed in the fields of fluid mechanics \cite{Schmi10a,bagh13a,Mezi13a},
climate science \cite{hann07a,Tant15a,Klus19a,froy21b}, and computational chemistry
\cite{Noe13a,Nuske16a}, among others. This is due to the solid theoretical
foundations based on Koopman and Perron-Frobenius operators outlined here, as
well as the availability of effective algorithms to approximate them directly from
data \cite{Tu14a,Will15a,Klus16a,Arba17a}.

It is important to emphasize that, despite the rigor and their representational
power, the evolution operators do not lead to a direct definition of what
organization is. They are essential tools in this endeavor, but they skirt the
definitional challenge. Addressing the latter requires taking a different view
of the computation-theoretic concepts introduced above.

\subsection{Intrinsic Computation}

The computation- and information-theoretic foundations and practical
shortcomings of the approaches to organization through compression described
above in Section~\ref{sec:FormalizingOrganization} A led historically to the
development of our second framework for formalizing organization. Founded on
the idea that pattern and organization are predictive regularities,
\emph{computational mechanics} directly constructs optimally-predictive
compressed descriptions that represent a system's \emph{intrinsic
computation}---how a system stores and transforms information \cite{Crut12a}.
The following briefly reviews this framework and its arguments for formalizing
organization through intrinsic computation, emphasizing organization as
generalized symmetries.

Due to its origins in computation and information theory, we start in the
simplified setting of fully-discrete one-dimensional field theories. The
systems considered here are collections of indexed sequences of
symbols from a finite \emph{alphabet} $\Alphabet$; e.g., binary strings with
$\Alphabet = \{0,1\}$. Indices may correspond to locations in time or space.
Shifts in indices, given by the \emph{shift operator} $\sigma$, correspond to
translations in time or space. Shift-invariant collections of symbol sequences
are called \emph{shift spaces}, denoted $\mathcal{X}$. To connect to
statistical mechanics, they may be thought of as \emph{topological ensembles}.

For example, we may choose a binary alphabet of $\Alphabet = \{-1, 1\}$. The
field here represents an abstraction of lattices of two-state spins---up versus
down, say. In contrast to statistical mechanics, there is no probability
distribution over symbol sequences. For a given shift space (topological
ensemble), we are interested in mathematical representations of pattern and
organization present in that system. 

\subsubsection{Organization in Symbolic Dynamics}

The study of shift spaces is a rich subfield of nonlinear dynamics called
\emph{symbolic dynamics} \cite{Lind95a}. It has a long and venerable history
\cite{Mors38a} with origins that overlap with early studies in dynamical systems \cite{Smal67a}, computability
\cite{Turi37a}, decidability \cite{Chur36a}, and the logical foundations of
mathematics \cite{Post21a,Nage68a}.

An important distinction from statistical field theories is that the shift
operator provides a dynamical relation between the elements of a shift space.
The translations provided by the shift operator allow us to formalize
symmetries, and their generalizations, in shift spaces. Since shift spaces
are shift-invariant, equivalently they can be thought of as collections of
individual indexed-sequences related through the shift operator or as a single
bi-infinite sequence with $\sigma$ simply shifting the sequence indices. 

On one extreme, a sequence may posses an exact translational symmetry; e.g.,
$\{\ldots00100100100\ldots\}$. Such symmetries are described via a \emph{group
algebra}. If the symmetry's ``pattern'' is known---i.e., the group relations,
and the current phase, or group element is known---then the configuration at
any other location in the sequence can be exactly predicted. Exact symmetries
represent full predictive regularity. On the other extreme, if the sequence is
entirely random---e.g., generated by flips of a fair coin---then the symbols at
other locations cannot be predicted at all. There is no pattern or organization
present to leverage for any predictive regularity. It is commonly accepted that
these two extremes---fully predictable and fully random---represent the
``boundary conditions'' of vanishing complexity. That is, they represent null
organization \cite{ante96a,Feld97a}.

Most shift spaces lie somewhere between these extremes. They are neither
exactly predictable nor fully unpredictable. A generic sequence will posses
some limited or partial predictability, some nonzero degree of organization and
some nonzero amount of randomness. 

\begin{figure}[t]
\centering
\includegraphics[width=0.5\textwidth]{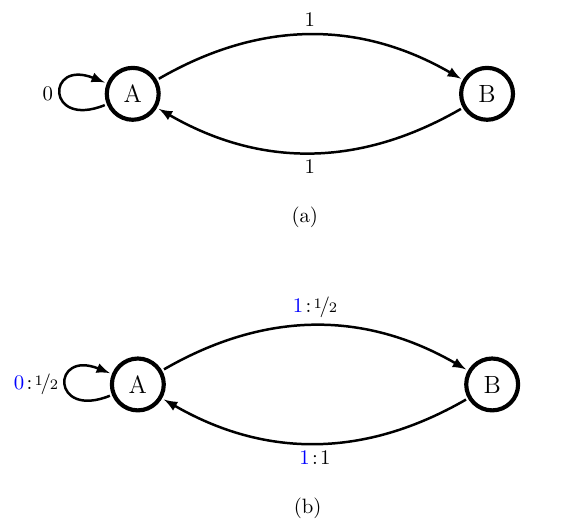}
\caption{(a) Semiautomaton presentation of the Even Shift with internal states
	$\{A, B\}$ and state-to-state transitions labeled with symbols from
	$\Alphabet = \{0, 1\}$. (b) Hidden Markov chain presentation of the
	uniform Even Process. Each transition leaving state $A$ has probability
	$1/2$ of occurring. That returning to $A$ emits a $0$ and that going to $B$
	emits a $1$. The transition leaving $B$ is taken with probability $1$ and
	emits a $1$.
	}
\label{fig:evenstuff}
\end{figure}

As we seek to formalize organization as a generalization of exact symmetries,
we restrict to the class of \emph{sofic shifts}, which are shift spaces
generated by finite semigroups \cite{Weis73}. Importantly, all sofic shifts can
be \emph{presented} by a certain class of finite-state machines \cite{Kitc86a}.
These finite-state machines, called \emph{semiautomata} \cite{ginz68a}, are
given as a finite collection of \emph{internal states} and \emph{symbol-labeled
transitions} between these states. Symbol sequences are generated from
semiautomata by following transition paths between the internal states and
recording the associated symbol for each transition. A semiautomaton is a
\emph{presentation} of a sofic shift if each sequence generated by the
automaton is in the shift space and if all sequences of the space can be
generated by the automaton.

Figure~\ref{fig:evenstuff} (a) shows a semiautomaton presentation of the
\emph{even shift}---the set of binary sequences such that there can only be
even-length blocks of $1$s enclosed by $0$s. Note the combination of randomness
and regularity; transitions from internal state $\mathsf{A}$ can emit $0$
or $1$, while transitions from state $\mathsf{B}$ must emit a $1$. The partial
predictive regularity is given by the constraint, already noted, that blocks of
$1$s bounded by $0$s must be an even number in length. 

For every sofic shift there is a \emph{unique} minimal presenting semiautomata,
which we will define shortly. Crucially, every symbol sequence with a finite
translation symmetry is a sofic shift \cite{rupe22a}. Moreover, the group
algebra of the translation symmetry is equivalent to the permutation symmetry
group of the minimal presenting semiautomaton of the sofic shift. Similarly,
fully ``random'' sequences without regularity are also sofic shifts. Since
sofic shifts encompass this full range of patterns, from no regularity to full
regularity, we argue that the \emph{minimal semiautomata presentations
are the mathematical representation of the organization contained in symbol
sequences.} Reference \cite{rupe22a} gives the details and proofs behind this
position.

We emphasize again the hypothesized duality between organization and
computation. Motivated by generalizing fully-symmetric sequences into a
spectrum of partially-predictable sequences, the mathematical representation we
found that does this is the class of models of discrete computation---the
finite-state machines of elementary computation theory
\cite{Lewi98a,Hopc06a,Sips14a}. 

\subsubsection{Organization Through Predictive Equivalence}

While viewing pattern and organization as generalized symmetries can be made
rigorous in the discrete one-dimensional setting of sofic shifts, the central
definition and construction of minimal presenting semiautomata---\emph{predictive equivalence}---is fully
generalizable to most cases of interest for self-organizing systems. These
include higher dimensions and continuous and statistical field theories. We now
give the simplest definition, for fully-discrete one-dimensional topological
and statistical ensembles. Spatially-extended and continuous versions are given later. Ultimately, we will find predictive equivalence to be the most central concept to date for principles of organization.

First, it is helpful to consider sequence indices as corresponding to time.  With this, a shift space represents a set of fully-discrete time series. The internal states of semiautomata (and finite-state machines more generally) are then defined as \emph{equivalent histories}. This idea, dating back to the early physics of computation \cite{Mins67} and logical machines \cite{Huff59a}, provides the definition for the unique minimal semiautomata presenting a sofic shift, as well as the means to a statistical mechanics generalization.

For every sequence $x \in \mathcal{X}$ in a given shift space, let index $t$
represent the \emph{present} time. The semi-infinite sequence $\past_t = \{x_t,
x_{t-1}, x_{t-2}, \ldots\}$ with indices less than or equal to $t$ is the
\emph{past} of the sequence at time $t$ and the \emph{future} $\future_t =
\{x_{t+1}, x_{t+2}, \ldots\}$ of the sequence at time $t$ is then the
complement---the semi-infinite sequence with indices greater than $t$.

For a given shift space $\ShiftSpace$, the \emph{follower set}
$F_{\ShiftSpace}(\past)$ of a past configuration $\past$ is the set of all future
configurations $\{\future\}$ such that the full sequence $x = \past
\future$---the concatenation of $\past$ and $\future$---is an element of the shift
space: $x \in \ShiftSpace$. Two pasts are considered \emph{equivalent}---i.e.,
they are equivalent histories---if they have the same follower set. The
central idea is that equivalence classes of unique follower sets
$F_{\ShiftSpace}$ are the internal states for the unique minimal semiautomaton
that presents the sofic shift $\ShiftSpace$.

Second, generalizing from topological to statistical ensembles of sequences is
straightforward. We consider two pasts to be \emph{predictively equivalent} if
they not only have the same follower set, but if they have the same
\emph{predictive distribution} $\Pr(\Future | \past)$ over the follower set.
(Note that the support of a predictive distribution is a follower set.)
Predictive equivalence defines the unique minimal \emph{hidden Markov chain}
representation for the sequence ensemble \cite{Shal98a}. Figure
\ref{fig:evenstuff} (b) displays the stochastic generalization of the
semiautomaton in Figure \ref{fig:evenstuff} (a). The hidden states of this minimal hidden Markov chain---the predictive equivalence classes---are known as \emph{causal states}
\cite{Crut12a}.

In the temporal setting, where indices correspond to time, these ensembles of
sequences are fully-discrete stochastic processes. And, the causal-state hidden
Markov chain is the unique model that optimally predicts the given process with
minimal computational resources \cite{Shal98a}. Predictive equivalence ensures
the dynamic over causal states is Markovian: the transition to the next causal
state depends only on the immediately preceding causal state.

Importantly, the internal-state dynamic is \emph{unifilar}: For each causal state $S_i$ there
is at most one causal state $S_j$ to which $S_i$ transitions for each  symbol
$a \in \Alphabet$. That is, there is at most one $S_j$ such that $S_{t+1} =
S_j$ given $S_t = S_i$ and $x_{t+1} = a$. Unifilarity ensures a one-to-one
correspondence between sequences of symbols (realizations of the stochastic
process) and the corresponding sequence of causal states. Summing over such
symbol-labeled transitions gives a semigroup of Markov transition operators
$M_\epsilon$ that governs the evolution of distributions over causal states:
\begin{align}
\Pr(S_{t}) = M_\epsilon^t \Pr(S_0)
~,
\label{eq:causalstate_dynamics}
\end{align}
with $M_\epsilon^{t_1} (M_\epsilon^{t_2} \Pr(S_0)) = M_\epsilon^{t_1 + t_2}
\Pr(S_0)$.

Computational mechanics merges the computation-theoretic tools of symbolic
dynamics with statistical mechanics. Its hidden causal-state Markov chain---the
\emph{\eM}---provides an explicit mathematical representation of the temporal
organization in fully-discrete processes. (However, shortly we will describe
how this generalizes to continuous field theories.) The hidden Markov chain
structure captures both organization as a generalized symmetry, as well as
organization quantitatively as the intrinsic computational resources used by
the system to convey the past to the future through the present \cite{Crut08a}.
It has been successfully applied to analyze the organization and information
processes in a wide range of complex systems \cite{Crut12a}.

\subsubsection{Local Causal States in Spatial Field Theories}

To adapt organization as intrinsic computation to the spatially-extended
systems of interest (classical field theories) \cite{Rupe23a}, we use
\emph{lightcones} as local notions of pasts and futures. Recall that
Eq.~(\ref{eq:gen_dyn}) imposed spatially-local interactions through
finitely-many spatial derivatives. Due to this, information propagates at a
finite speed $c$ through the system. For a spacetime point $(\mathbf{r}, t)$ in
a spacetime field $X(\mathbf{r}, t)$, the \emph{past lightcone} of
$(\mathbf{r}, t)$ is the set of all points at preceding times that could
possibly have influenced $(\mathbf{r}, t)$ through the local interactions:
\begin{align*}
\mathtt{L}^- = \{(\mathbf{r}', t') : t'\leq t , ||\mathbf{r}' - \mathbf{r}|| \leq c (t'-t)\}
    ~.
\end{align*}
Similarly, the \emph{future lightcone} of $(\mathbf{r}, t)$ is the set of all points at later times that $(\mathbf{r}, t)$ could itself possibly influence through the local interactions: 
\begin{align*}
\mathtt{L}^+ = \{(\mathbf{r}', t') : t' > t , ||\mathbf{r}' - \mathbf{r}|| \leq c (t'-t)\}
	~. 
\end{align*}

The \emph{local causal states} are then defined as the equivalence classes of
the \emph{local predictive equivalence relation}:
\begin{align}
    \ell_i^- \sim_\epsilon \ell_j^- \iff \Pr(\mathtt{L}^+ | \ell^-_j) = \Pr(\mathtt{L}^+ | \ell^-_i)
    ~,
    \label{eq:local_causal_equiv}
\end{align}
where $\ell^\pm$ are specific lightcone configurations or realizations of
lightcone random variables $\mathtt{L}^\pm$. Mirroring the one-dimensional
case, local causal states are sets of past lightcones with the same conditional
distributions over future lightcones. It is useful to state this equivalently
in terms of the \emph{$\epsilon$-function} that maps from past lightcones to
their associated local causal state:
\begin{align*}
    \ell^-_i \sim_\epsilon \ell_j^- \iff \epsilon(\ell^-_j) = \epsilon(\ell^-_i)
    ~.
\end{align*}

Since each point $X(\mathbf{r}, t)$ in spacetime has a unique past
lightcone and this past lightcone has a unique local predictive distribution
$\Pr\bigl(\mathtt{L}^+(\mathbf{r}, t) | \ell^-(\mathbf{r}, t)\bigr)$, each
point also has a unique local causal state, denoted $S(\mathbf{r}, t) =
\epsilon\bigl(X(\mathbf{r}, t)\bigr)$. For a full spacetime field $X$, we apply
the $\epsilon$-function to all its spacetime points $X(\mathbf{r}, t)$ to
create a corresponding \emph{local causal state field} with spacetime points
$S(\mathbf{r}, t) = \epsilon\bigl(X(\mathbf{r}, t)\bigr)$.

Since the local causal state field is created through the pointwise
$\epsilon$-function, it shares the same spacetime coordinate geometry as the
original field $X$. Therefore, characteristics of the field $X$ can be analyzed
through various properties of the associated local causal state field $S =
\epsilon(X)$.

In particular, organized \emph{coherent structures} are identified as
\emph{localized deviations from generalized spacetime symmetries} in $X$, which
are given as exact symmetries in $S$ \cite{Rupe18a,Rupe18b}. 

Clustering algorithms may be employed to approximate local causal
states in continuum field theories, such as those used to model fluid flows
\cite{Rupe19a,Rupe23a}. 
The local causal states can identify individual vortices in
two-dimensional free-decay turbulence and recover their power-law decay
behavior, as well as identify and track individual extreme weather events like
hurricanes and atmospheric rivers in high-resolution climate data
\cite{Rupe23a}.

\begin{figure}[t]
\centering
\includegraphics[width=1.0 \columnwidth, trim={1.25cm 0cm 3.8cm 0.8cm}, clip]{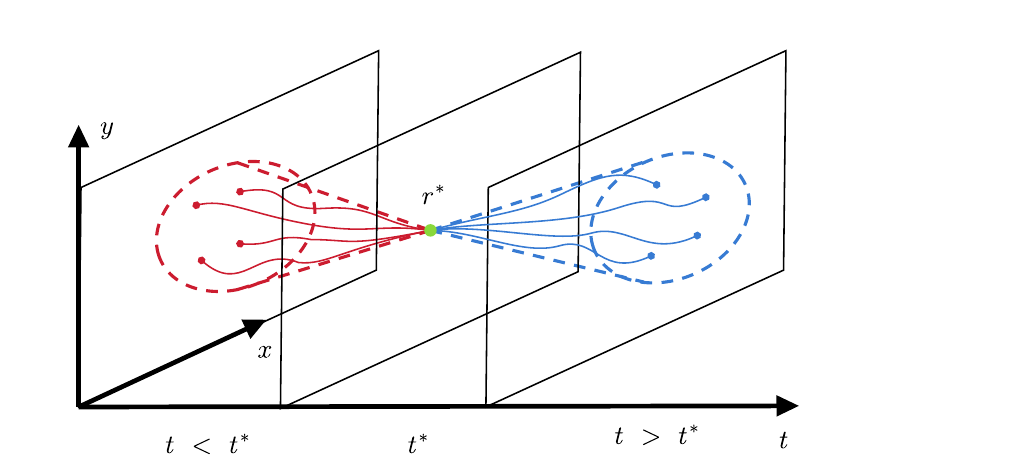}
\caption{Co-occurring past (red) and future (blue) Lagrangian lightcones at spacetime point $(r^*, t^*)$ shown with dashed lines. All possible Lagrangian trajectories (examples shown by solid lines) leading to and emanating from $(r^*, t^*)$ are contained within the lightcones.
	}
\label{fig:lightcones}
\end{figure}

Recall that the Lagrangian perspective is useful for identifying complex
organization in time-dependent fluid flows. In this case, \emph{Lagrangian
lightcones} may be used, which are collections of Lagrangian trajectories that
can possibly lead to or from the present point; see Figure~\ref{fig:lightcones}. That is, the past Lagrangian lightcone is the set of all points at prior times that could possibly reach the present point through Lagrangian advection. Similarly, the future Lagrangian lightcone is the set of all points at later times that can be reached from the present point by advection. 

Reference~\cite{Rupe19a} demonstrates that local causal states identify
coherent structures in complex flows that agree well with structures identified
by specialized geometric Lagrangian methods \cite{Hadj17a}. In particular,
structures identified as coherent sets using Perron-Frobenius operators produce
characteristic signatures in Lagrangian lightcones that then give rise to
unique local causal states corresponding to the coherent sets (coherent
trajectories create coherent lightcones). Similarly, the boundaries of coherent
sets are associated with repelling structures that act as transport barriers
\cite{Froy10b}, which Ref.~\cite{Rupe19a} also demonstrated for local causal states. Transport barriers create distinct signatures in Lagrangian lightcones on either side of the barrier, giving two separate local causal states on either side with their boundary lying on the barrier. Note though that the dynamics of localized structures moving through space over time, as well as finite lifespans of structures, are naturally encapsulated by local causal states without special modification. 

\begin{figure*}[t]
\centering
\includegraphics[width=0.66 \textwidth]{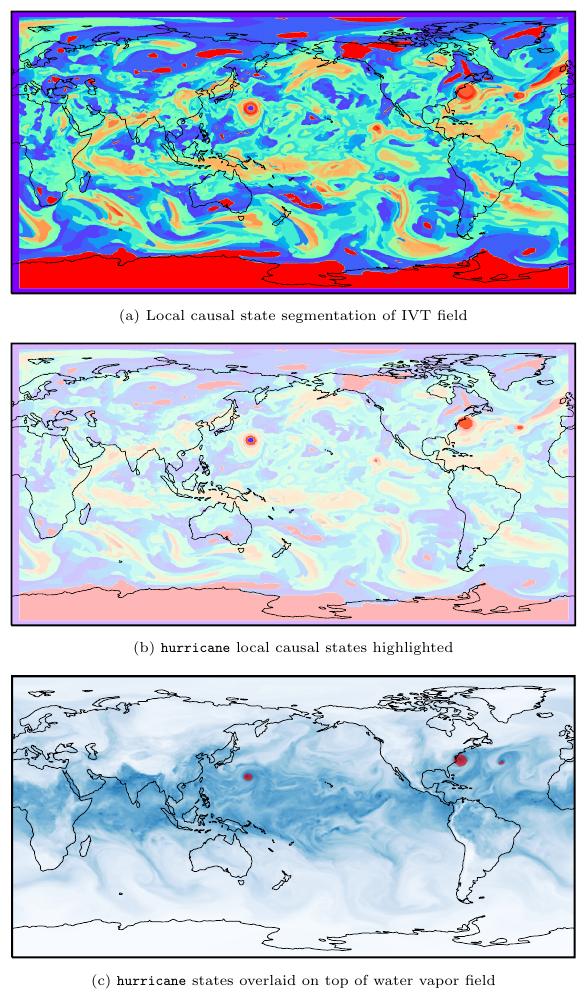}
\caption{Identification of hurricanes in high-resolution climate data using local causal states. 
Reproduced from Ref.~\cite{Rupe23a}. 
	}
\label{fig:hurricane-mask}
\end{figure*}

An example of identifying hurricanes in high-resolution climate data using local causal states is shown in Figure~\ref{fig:hurricane-mask}, reproduced from Ref.~\cite{Rupe23a}. The local causal states are approximated from the integrated vapor transport field (IVT), which measures water vapor flux and is often used by climate scientists for extreme weather analysis. The local causal states approximated from the IVT field are shown in (a) and, as highlighted in (b), there is a subset of local causal states that correspond to hurricanes. 
The water vapor field is shown in (c) with
spacetime points assigned to the subset of hurricane local causal states highlighted in red. Reference~\cite{Rupe23a} demonstrates the hurricanes identified in this way by local causal states agree with the specialized heuristics currently used by climate scientists to identify hurricanes.

\subsubsection{Recap}

Intrinsic computation decomposes dynamical processes into their minimal causal
components. These components---causal states---are defined as equivalence
classes of past trajectories that produce the same conditional distributions
over future trajectories. It has been shown that the causal states and their
internal transitions---the $\epsilon$-machines---are the unique minimal optimal
predictors of discrete stochastic processes \cite{Shal98a}. That is, they
uniquely represent the most compressed representation that is still fully
faithful to the original process. As a further argument for organization
through intrinsic computation, it has also been shown that $\epsilon$-machines
identify patterns that algebraically generalize exact symmetries \cite{rupe22a}. 

Recent progress extends intrinsic computation and predictive equivalence to continuous-time processes \cite{Marz17b,brod22a} and spatially-extended systems \cite{Rupe18a,Rupe23a}. The straightforward definition through predictive equivalence makes intrinsic computation actionable in practice through approximation or Bayesian inference \cite{Stre13a}.
Thus, organization can now be identified and analyzed in a wide array of systems through intrinsic computation.

\subsection{Outlook on Organization}

Having completed our review of modern mathematical approaches to organization,
let's ask again---what \emph{is} organization? Is it identified through
spectral decomposition of evolution operators or through equivalence classes
of predictive equivalence relations? In short, the existence of a universal
theory of organization remains an open question. Evolution operators and
intrinsic computation represent distinct but overlapping characterizations of
organization. For one, their domains of applicability do not fully overlap,
with evolution operators strictly defined for dynamical systems (deterministic
and stochastic), whereas predictive equivalence was originally defined for
discrete stochastic processes. 

That said, the extension of predictive equivalence to continuous field theories
brings connections with evolution operators. This is particularly the case with
coherent structures identified from Lagrangian flows, as just described.
Coherent structures are key forms of organization, and the agreement on their
identification between local causal states and evolution operators is
encouraging. Further connections remain to be worked out; for example, between
spectral measures of evolution operators \cite{colb21a} and spectra of
$\epsilon$-machines \cite{Crut13a,Jurg20c}. It may be that predictive
equivalence and evolution operators provide different views to the same
underlying notion of organization.

An important similarity between both approaches must be emphasized: They both
fall outside the constructionist paradigm. Evolution operators and their
spectral decompositions, as well as predictive equivalence classes and their
transitions, are not directly constructible apart from a handful of idealized
systems \cite{gasp95a,Laso13a}. To be clear on this, one cannot start with the
Navier-Stokes equations and write down the time-dependent Perron-Frobenius
operators generated by the Lagrangian flow map for some specified boundary
conditions, nor can one start with the predictive equivalence classes of
Lagrangian lightcones. However, as emphasized, evolution operators and
predictive equivalence provide a rigorous theoretical scaffolding on which to
build data-driven algorithms that can discover emergent organization from the
behaviors of complex systems.

With the concept of organization now addressed, what about principles? Since
the modern formulations of organization lie outside the constructionist
paradigm and the classical notions of ``principles'' are constructionist in
nature, we are largely in uncharted territory. General formulations of
scientific principles beyond constructionism is an enormous and ongoing
endeavor that is beyond our scope here. However, we will conclude by returning
to the concept of emergence---which, as we noted at the beginning, highlights
phenomena that thwart constructionist principles of organization.




\section{A Statistical Mechanics of Emergence}
\label{sec:StatMechOrganization}
%
Organization \emph{emerges} as a higher-Level description that cannot be
constructed from a system's lower Level transport equations. The seemingly
stochastic vortex gas of two-dimensional turbulence \cite{mcwil90a} and the
``particle dynamics'' of elementary cellular automata \cite{Hans95a} come
immediately to mind. It is natural to ask \emph{how} and \emph{why} new
organization appears. Constructionism is the standard paradigm for answering
``how'' and ``why'' questions in physics. While in line with colloquial uses of
the term, here we use \emph{emergent} to emphasize organization that cannot be
directly constructed. There is, at this time, no satisfactory paradigm to
answer ``how'' and ``why'' questions in the absence of constructionism.
Arguably, a central task of complexity science is to lay the foundations for
this new paradigm \cite{Crut93g}. As such, it is too big a task for us to
conquer here. We can say, however, that this new paradigm will be intrinsically
computational and data-driven, at least in part.

Rather than address theories of how and why organization may spontaneously
emerge, we instead close with a universal theory to formulate complete and
self-contained dynamics for higher-Level organization (and behaviors more
generally) that is consistent with the lower-Level from which it emerges. This
universal theory is, in fact, given by the causal states and their Markovian
dynamics. 

A key mandate of statistical mechanics is to show that higher-Level
thermodynamics is consistent with lower-Level particle kinetics. It does so by
directly constructing thermodynamic relations from the statistical behavior of
kinetic particles. Generally, our abstracted statistical mechanics cannot, of
course, analytically construct the dynamics of higher-Level organization from
the lower-Level dynamics. However, we will show how to define the causal states
and their dynamics in terms of Koopman and Perron-Frobenius evolution
operators. This provides the requisite physical consistency between Levels. 

In addition to theoretically and conceptually providing consistency, this
Section serves to introduce a physical foundation to data-driven dynamics of
emergent organization. As hopefully established by this point,
intrinsic-computational and data-driven approaches are and will continue to be
indispensable in the study of emergent organization. However, data-driven
methods are often seen as ``black box'' and potentially not
physically-grounded. The following shows that the causal states and their
dynamics are the physical underpinnings of a complete, consistent, and
self-contained dynamics of emergent organization. Here, the ``self-contained''
qualifier emphasizes that the emergent dynamics can be consistently
approximated directly from data.
For example, Refs.~\cite{costa19a} and \cite{costa23a} use the statistical mechanics of emergence for data-driven reconstruction of long-term behaviors that emerge from short-term movements of simple organisms. 

\subsection{Consistency of Emergent Behaviors}
Consider a system with a Level A description given as a deterministic dynamical system: 
\begin{align}
    a_{t+1} = \Phi(a_t)
    ~.
\label{eq:dyn_A}
\end{align}
For simplicity, we work in discrete time, but generalizing to continuous time
is straightforward. This Level A description includes, for example, particle
kinetics or nonequilibrium transport equations given by Eq.~(\ref{eq:gen_dyn}). 

We now want to define an emergent Level B description that is physically
\emph{consistent} with the Level A description. To do so, we simply define 
Level B through \emph{partial observations} of the Level A description. At each
time, the system state at Level B is given by a \emph{noninvertible}
function of the Level A system state, so that $b_t = B(a_t)$. The
noninvertibility of $B(\cdot)$ is crucial for identifying Level B as emergent,
since the system state at the Level A description cannot be directly recovered
from the Level B state description.

\begin{figure*}[t]
\centering
\includegraphics[width=0.9 \textwidth]{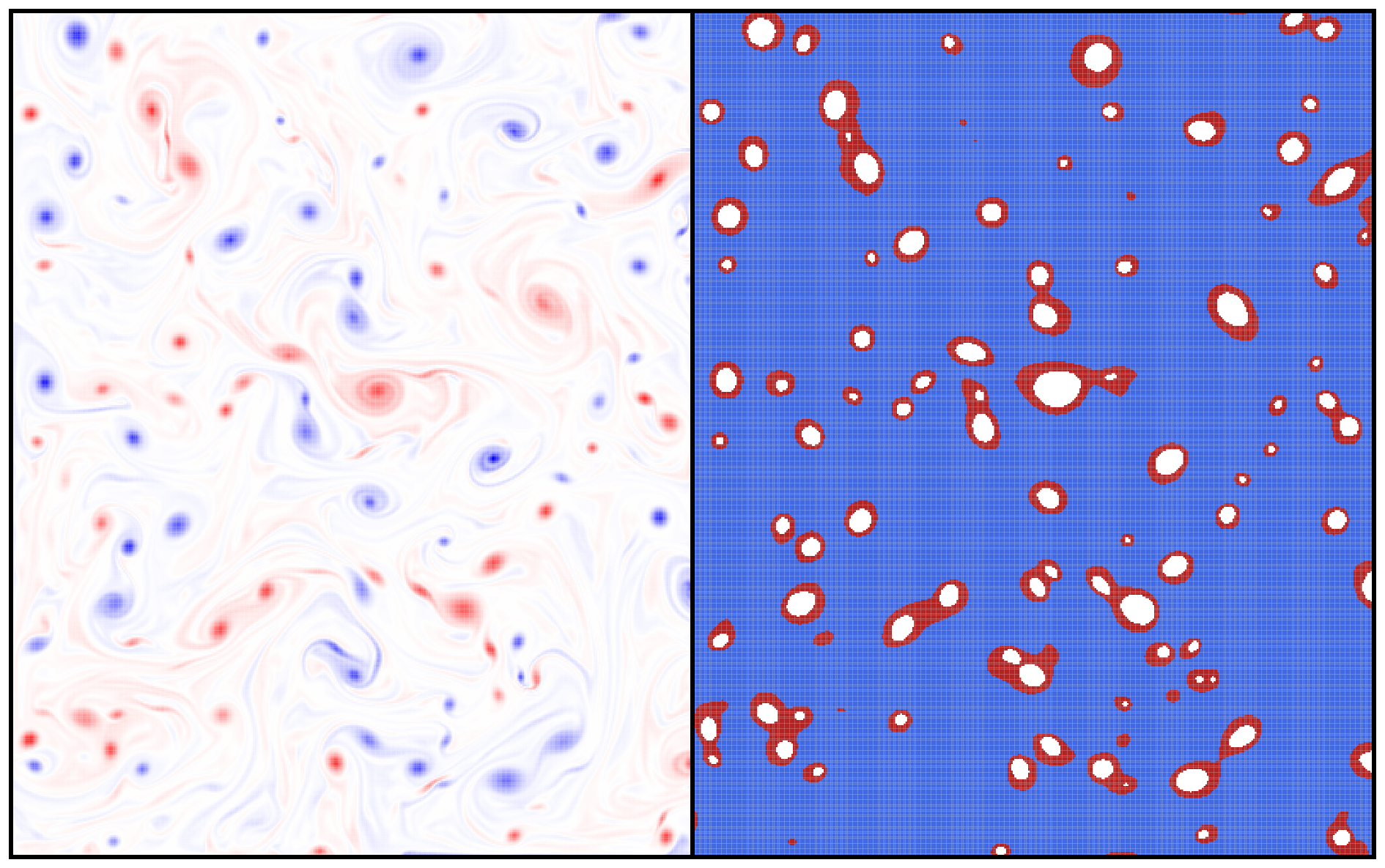}
\caption{A vorticity field $a_t$ (left) from two-dimensional turbulence and its local causal state field $S_t=\epsilon(a_t)$ (right). Taking only the white vortex-core local causal states gives an example of higher-Level emergent organization $b_t = B(a_t)$.
	}
\label{fig:vortices}
\end{figure*}

Furthermore, here we are interested in B Levels that describe the dynamics of
emergent organization. For the formalism that follows to apply to organized
structures, it is crucially important to have a function $B(\cdot)$ that
identifies such organization. The methods outlined above in
Section~\ref{sec:FormalizingOrganization} do exactly this. That is, we may
think of $B(\cdot)$ as being a set of local causal states defined from
$\epsilon(\cdot)$, or $B(\cdot)$ may be a collection of finite-time coherent
sets and the like.

As a guiding example in what follows, consider Level A as being (numerical
simulations of) the two-dimensional vorticity equation and Level B as the
dynamics of emergent vortices. Again, the function $B(\cdot)$ may be defined,
e.g., using $\epsilon(\cdot)$ and the local causal states it induces
\cite{Rupe23a}---as shown in Figure~\ref{fig:vortices}---or coherent structures
that arise in some Lagrangian method \cite{Hadj17a}. The collection of
emergent vortices $b_t$ at any given time is insufficient to determine the full
vorticity field $a_t$, so $B(\cdot)$ is not invertible. 

Our task is to define what it means to have a dynamical description that is
fully \emph{self-contained} at Level B, while also being physically
consistent with the dynamics given by Level A. Following our emergent
vortices example, we want a formulation of the evolution of the vortices $\{b_t,
b_{t+1}, b_{t+2}, \ldots \}$ that is self-contained, without direct reference
to the vorticity fields $\{a_t, a_{t+1}, a_{t+2}, \ldots\}$. However,
the Level B vortex evolution must still be physically consistent with the
Level A vorticity equation, such that there exists a sequence of vorticity
fields---satisfying the vorticity equation---that gives $\{b_t = B(a_t),
b_{t+1} = B(a_{t+1}), \ldots\}$. We will find that the Level B evolution is
in general stochastic and is defined in terms of predictive distributions.
Conditions for convergence to a deterministic Level B evolution, analogous to a
thermodynamic limit, can then be given.

The starting point is to observe that $B(\cdot)$ is a vector-valued system
observable, in the technical sense given above, and so it is evolved by Koopman
operators. A unit-step of the dynamics at Level B is given by:
\begin{align}
    b_{t+1} = [\Koop B](a_t)
    ~.
\end{align}
Note though, this requires the initial Level A state $a_t$. Knowing only $b_t$
on Level B at time $t$ means the system at the Level A description could be
in many states $a_t$ that are consistent with the Level B observation: $b_t =
B(a_t)$. Therefore, given $b_t$, many $b_{t+1}$ may follow that are consistent
with the Level A description: 
\begin{align}
    \{b_{t+1} = [\Koop B](a_t) \; \; \text{for all } a_t \in B^{-1}(b_t)\}
    ~,
    \label{eq:inst_dist_supp}
\end{align}
where $B^{-1}(b_t) = \{a_t: b_t = B(a_t)\}$ is the pre-image of $B(\cdot)$.

Therefore, the \emph{instantaneous dynamics} on Level B are stochastic:
\begin{align}
    \Pr(b_{t+1}) = M_0 \Pr(b_t)
    ~,
\end{align}
with some Markov operator $M_0$. The set in Eq.~(\ref{eq:inst_dist_supp}) can
be seen as the support of a conditional probability distribution $\Pr(b_{t+1} |
b_t)$. This is the \emph{instantaneous predictive distribution} that captures
the maximal information available about the next value $b_{t+1}$ given a
specific observed value $b_t$. 

Under the Level A dynamics, knowledge of state $a_t$ at any given time is sufficient
to fully determine the Level A state at later times. Clearly, this is not the
case for the Level B dynamics. There, state $b_t$ is associated with a
distribution over consistent future Level A states---the instantaneous
predictive distribution. A notable consequence is that prior observations
$b_{t-k}$ contain relevant additional information for predicting future values.

This follows from the \emph{Mori-Zwanzig equation} or generalized Langevin
equation: 
\begin{align}
    b_{t+1} = M_0(b_t) + \sum_{k=1}^t M_k(b_{t-k}) + \xi_{t+1}(a_0)
    ~.
    \label{eq:MZ}
\end{align}
This gives the \emph{exact} Level B evolution for a given evolution at Level A and is equivalent to:
\begin{align*}
b_{t+1} = [\Koop^{t+1} B](a_0)
\end{align*}
via the Dyson expansion of $\Koop^{t+1}$ using projection operators; see
Refs.~\cite{Wild98a,chor02a} for the full derivation.

The Mori-Zwanzig equation, originally developed as a first-principles theory of
nonequilibrium transport \cite{Zwan73a}, expresses as much of the Level B
dynamics as possible in terms of Level B states. The first term $M_0$ is
referred to as the \emph{Markov} term, as it gives the dependence of $b_{t+1}$
on its single prior value $b_t$. As discussed above, this is insufficient to
fully determine $b_{t+1}$. There is an additional temporal dependence,
given by the middle term. The $M_k$ are known as \emph{memory kernels}, and
they capture how information is propagated and dissipated between the A and B
Levels. The last term $\xi_{t+1}(a_t)$ is called the \emph{orthogonal
component} and it captures the remaining dependence on the initial unknown
Level A state $a_0$ that cannot be accounted for by the memory contributions.

In our framing of emergent organization, Level A states are not accessible, and
so we cannot include the orthogonal component $\xi_{t+1}(a_0)$. If this term is
nonzero, it is required to recover the exact deterministic Level B evolution.
And so, a self-contained and consistent Level B dynamics must be stochastic in
this case. 

At this point, changing perspective to consider trajectories or paths, as
central entities, rather than instantaneous states is helpful. This is a common
strategy in nonequilibrium statistical mechanics. Denoting the collection of
current and prior values of $b_t$ as its \emph{past} $\overleftarrow{b}_t =
\{b_{t'}\}$, $t' < t$, we can directly subsume the memory dependence of the
Mori-Zwanzig equation into a dependence on the past:
\begin{align}
    b_{t+1} = \overleftarrow{M}(\overleftarrow{b_t}) + \Xi(a_0)
    ~.
    \label{eq:WienerMZ}
\end{align}
In this, the Markov and memory effects are combined to give a single contribution from the past $\overleftarrow{b}_t$. This procedure is formally carried out using Wiener projections of the Koopman operator onto pasts of partial observations~\cite{Lin21a,Gila21a}. 

Since Eq.~(\ref{eq:WienerMZ}) is equivalent to the Mori-Zwanzig equation in Eq.
(\ref{eq:MZ}), the exact deterministic dynamics of $b_t$ requires a dependence
on its past and may exhibit a residual dependence $\Xi(a_0)$ on the initial A
state. As above, a self-contained Level B dynamics cannot include $\Xi(a_0)$
since information about Level A is taken to be unavailable. The
trajectory-oriented perspective compactly expresses the self-contained B
dynamics as $\overleftarrow{M}(\overleftarrow{b_t})$---a Wiener projection of
the action of the Koopman operator. The optimal projection takes the form of a
conditional expectation:
\begin{align}
    \overleftarrow{M}(\overleftarrow{b_t}) = \mathbb{E}[b_{t+1} | \overleftarrow{b}_t]
    ~.
\end{align}
This has the form of an expectation over a \emph{history-dependent predictive distribution}.

Appealing to the Maximum Caliber approach to nonequilibrium statistical
mechanics, we can fully specify the predictive distributions, rather than only
their expectations. In the absence of Level A and $\Xi(a_0)$, these
predictive distributions give \emph{all} of the possible future Level B
trajectories that are physically consistent with the Level A dynamics and their
relative likelihoods.

Note that the MZ equation, in either form, is exact only when the dependence on
the initial Level A state $a_0$ is included. When the orthogonal term
$\Xi(a_0)$ is not included, the conditional expectation
$\overleftarrow{M}(\overleftarrow{b_t}) = \mathbb{E}[b_{t+1} |
\overleftarrow{b}_t]$ taken over consistent trajectories may itself not be
consistent with the Level A dynamics. Thus, using the full predictive
distributions themselves, rather than their expectations, is crucial to
maintain physical consistency between Levels. We now detail these predictive
distributions and how they maintain consistency. This then leads us naturally
to the causal states and their stochastic dynamics.

Returning to the example of vortex dynamics in two-dimensional turbulence,
assume we observe a trajectory of evolving vortices without details of
the vorticity field or knowledge of the vorticity equation. Many
future trajectories will be physically consistent. In this case, this means
vortices undergo pairwise mergers such that the total number of vortices decay
over time as a power law \cite{mcwil90a}. Averaging over such trajectories may
produce the correct power-law decay in the total vortex number, but it may
not display the proper pairwise-merger mechanism generating the decay. 

\begin{figure*}[t]
\centering
\includegraphics[width=0.9 \textwidth, trim={1.2cm 1.5cm 0cm 1.4cm}, clip]{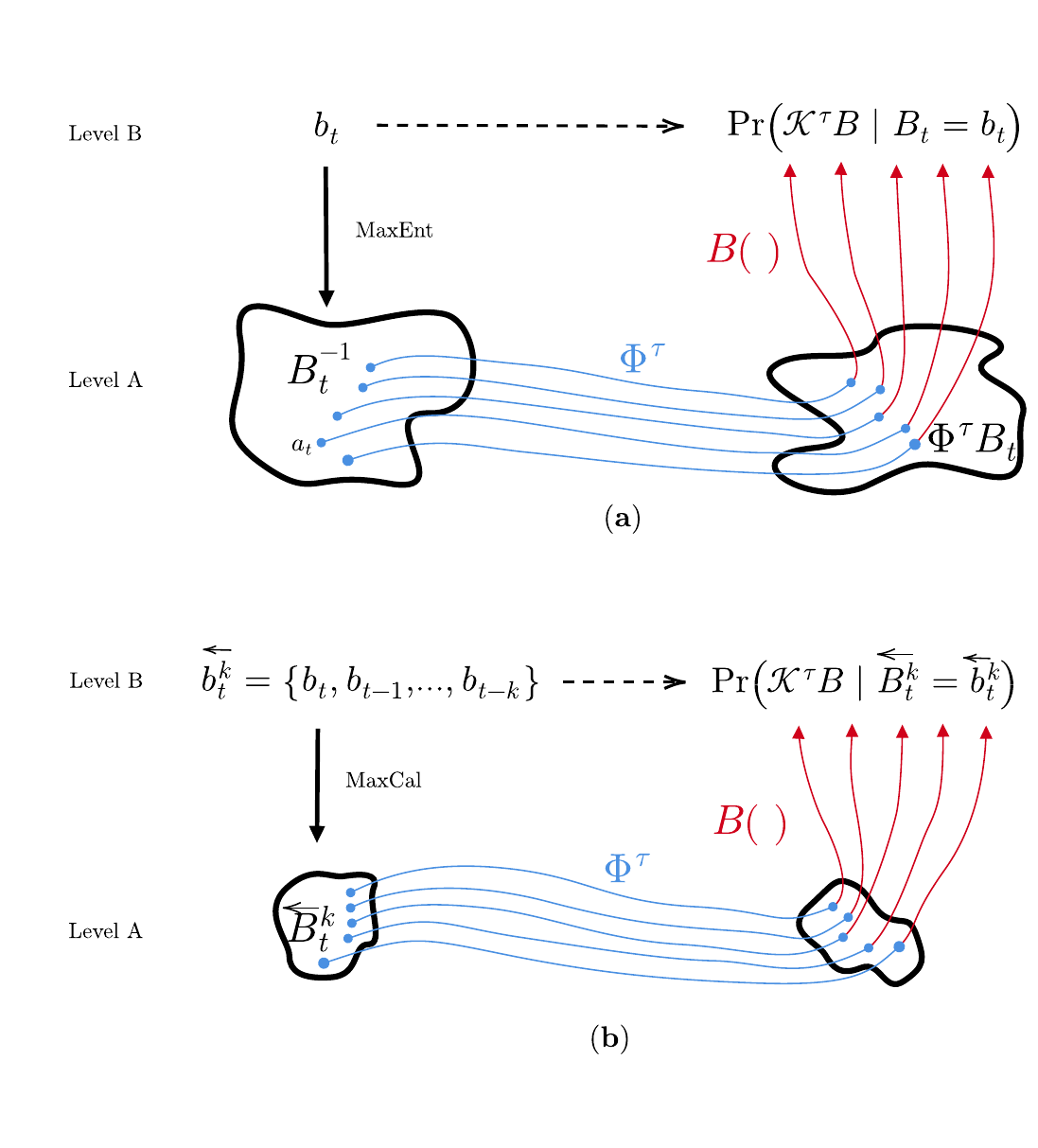}
\caption{Consistent predictive distributions: (a) Instantaneous case of a
	single Level B observation. (b) History-dependent case of a trajectory of
	Level B observations. MaxEnt (MaxCal) is used to infer the consistent set
	$B^{-1}_t$ ($\protect\overleftarrow{B}_t^k$) of Level A states from the
	Level B observation $b_t$ ($\protect\overleftarrow{b}_t^k$) in the
	instantaneous (history-dependent) case. The set of consistent Level A
	states will generally be smaller for more observations at Level B,
	leading to more accurate self-contained Level B predictions.
	}
\label{fig:pred-dists}
\end{figure*}

\subsection{Consistent Predictive Distributions and Causal States of Partially-Observed Processes}

Recall that the pre-image $B^{-1}(b_t)$ of a single instantaneous observation
$b_t$ gives the set of all $a_t$ consistent with $b_t = B(a_t)$. With a
subsequent observation $b_{t+1}$, it is not necessarily the case that the
dynamics of all $a_t$ in $B^{-1}$ are consistent with the new observation. That
is, there may be an $a^*_t \in B^{-1}$ such that $B\bigl(\Phi(a^*_t)\bigr)
\neq b_{t+1}$. Therefore, new observations of the Level B dynamics may
constrain the set of possible initial conditions on Level A that are
consistent with the Level B dynamics. For a past $\overleftarrow{b}_t^k$ at
time $t$ of length $k$, let $\overleftarrow{B}_t^k$ denote the set of initial A
states consistent with all B observations in $\overleftarrow{b}_t^k$.

Furthermore, recall that the action of the Koopman operator $\Koop^l$ on the
set $B^{-1}(b_t)$ of initial Level A states consistent with the instantaneous
Level B observation $b_t$ provides the set of all possible future Level B
observations $b_{t+l}$ that are consistent with the initial observation $b_t$.
From the above, a past history of observations $\overleftarrow{b}_t^k$
constrains the set of consistent initial Level A states. And, this also
constrains the set of consistent future Level B observations $b_{t+l}$.
Building on the original Mori-Zwanzig framework, this perspective again
underscores the importance of past histories in formulating self-contained
dynamics of emergent organization---dynamics that is consistent with the
underlying physics. If the full history is not taken into account, the
reconstructed self-contained dynamics may not be physically consistent.

To achieve an exact deterministic dynamics that is self-contained on
Level B it is sufficient that $\overleftarrow{B}_t^k$ contains one and only one
initial Level A state $a_{t-k}$ for all times $t$ (with $k$ potentially
limiting to $\infty$). Then there will always be one and only one:
\begin{align*}
b_{t+1} = [\Koop^{t+k+1} B](a_{t-k} = \overleftarrow{B}_t^k)
  ~.
\end{align*}
This is analogous to a thermodynamic limit where fluctuations about the mean
vanish in the limit. See Ref.~\cite{rupe22b} for further details.

It cannot be guaranteed that the residual dependence $\Xi(a_0) = 0$ in general,
even if we take the $k \rightarrow \infty$ limit of infinite memory. In this
case we can still give a complete and self-contained dynamic for Level B. It
will simply be stochastic rather than deterministic. This complete and
self-contained dynamic for Level B is given by the causal states and their
Markov transitions.

Thus far, we showed that the set of future Level B observations consistent with
a past history of Level B observations is given by the action of the Koopman
operator:
\begin{align}
    \{b_{t+l} = [\Koop^l B](a_t) \text{ for all } a_t \in \overleftarrow{B}_t\}
    ~,
    \label{eq:pred_dist_supp}
\end{align}
which holds for all distances $l$ into the future. Since history dependence may
persist arbitrarily far in the past and we can use the Koopman operators to
evolve arbitrarily far into the future, we can use the above to define the
support sets for \emph{predictive distributions} $\Pr(\overrightarrow{b}_t |
\overleftarrow{b}_t)$ in full generality, with semi-infinite pasts and futures. 

We can go further and add a natural probability distribution that directly
specifies these predictive distributions. Following Maximum Caliber, we assign
a uniform distribution to the Level A states $a_t \in \overleftarrow{B}_t$ that
are consistent with the observed Level B history $\overleftarrow{b}_t$ and
assign zero probability to Level A states not in $\overleftarrow{B}_t$---i.e.,
those not consistent with the observed B history. This distribution is then
evolved forward by Perron-Frobenius operators and pushed forward through the
observation map $B(\cdot)$ to give the predictive distribution
$\Pr(\overrightarrow{b}_t | \overleftarrow{b}_t)$; see
Figure~\ref{fig:pred-dists}.

With the predictive distributions in hand, it is straightforward to define
causal states in the same way as above in
Section~\ref{sec:FormalizingOrganization}---as sets of past histories with the
same predictive distributions. We can also define the Markov operator
$M_\epsilon$ for the stochastic dynamics over causal states using
Perron-Frobenius operators, given that the $\epsilon$-function is measurable.
The full measure-theoretic details are left for the future. 

Having established the continuous versions of causal states and their Markov
chain dynamics for the emergent dynamics of Level B, it follows from the
definition of predictive (causal) equivalence that this is an optimal
self-contained predictive model \cite[Theorem 1]{Shal98a} of the Level B
dynamics consistent with the physics of Level A. Here, we showed explicitly how
predictive distributions, and hence causal states, encapsulate the full
dynamical information self-contained on Level B that is consistent with
Level A. 

It is an important practical note that, while we can write down the general
form of predictive distributions, unsurprisingly we cannot directly construct
them in closed form in specific instances. However, as described above in
Section~\ref{sec:FormalizingOrganization}, the causal states and their dynamics
may be consistently approximated from data using predictive equivalence. This
makes the information contained in predictive distributions actionable in
practice.

We can now also import the tools and interpretations of intrinsic computation
to this continuous setting of emergent dynamics. The asymptotic degree of
temporal organization is given by the \emph{statistical complexity}
\cite{Crut12a}---the entropy of the steady-state distribution over causal
states. In other words, the statistical complexity is the amount of historical
information a system stores. And, if there are uncountably-many causal
states, the divergence rate of that stored information is given by the
\emph{statistical complexity dimension} \cite{Jurg20c}. 

In this way, the causal states provide a complete and self-contained
description for general emergent behaviors. From their definition in terms of
Koopman and Perron-Frobenius evolution operators, they are fully consistent
with lower-Level descriptions of the system dynamics. At the same time, they
decompose the dynamics into its (temporal) organizational structure. Recall
from above in Section~\ref{sec:FormalizingOrganization} D that the causal
states capture patterns as generalized symmetries through their semigroup
algebra. 

By combining evolution operators, nonequilibrium statistical mechanics
(MaxCal), and predictive equivalence, we arrived at a universal framework
for defining consistent and self-contained emergent dynamics in terms of causal
states and their Markov transitions. Thus, the approach simultaneously presents
emergent dynamics in terms of its structural organization.

\section{Conclusion and Looking Forward}

Historically, \emph{principles of organization} are given in the
\emph{constructionist} paradigm. Paraphrasing Feynman \cite{feyn11a}, what we
now call constructionism is described as: ``We look for a new law [principle]
by the following process. First, we guess it. Then, we compute the consequences
of the guess ... to see what it would imply. Then we compare the computation
results with experiment or observations.'' The nonlinear dynamics
bifurcation-theoretic derivation of the critical Rayleigh number that predicts
the first onset of \Benard convection cells is a prototypical example that
follows Feynman's dictum.

For equilibrium, we outlined how the \SecondLaw provides a variational
organization principle: equilibrium thermodynamic states, including organized
states, are those that maximize entropy subject to system constraints.
We described Classical Irreversible Thermodynamics as an attempt to generalize
the First and Second Laws of equilibrium thermodynamics to nonequilibrium field
theories. However, we showed that the entropy production density and its
associated balance equation---the nonequilibrium generalizations of entropy and
the \SecondLaw---are superfluous at best and, at worst, not physically justified. Indeed, extrema of entropy production do \emph{not} determine
nonequilibrium steady-states, analogous to extrema of entropy determining
equilibrium states. Ultimately, empirically-derived phenomenological laws and
energy balance---the nonequilibrium generalization of the \FirstLaw---are used
to derive the nonequilibrium transport equations used by nonlinear dynamics'
pattern formation theory. 

Prigogine et al.'s theory of \emph{dissipative} \emph{structures} claimed to be
a thermodynamic theory of pattern formation and self-organization. After
reviewing prior arguments and counterexamples that disproved dissipative
structures, we went further to show that certain foundational assumptions from
classical irreversible thermodynamics---on which dissipative structures theory
is built---are not valid for interacting systems with nonzero levels of
organization. Therefore, dissipative structures cannot be a theory of
self-organization, as it purported to be. The thermodynamics of pattern
formation and spontaneous self-organization thus \emph{remains an open problem}. 

What might a future theory, or principles, of organization look like? We argued
that the standard constructionist paradigm can fail for systems with
\emph{emergent} behavior. Since the intricate organization that forms in
far-from-equilibrium systems is emergent, it is no wonder that universal
principles of organization have remained elusive within the constructionist
paradigm. Analytic intractability and even uncomputability thwart our ability
to ``compute the consequences'' of a proposed mechanistic hypothesis. 

A key step forward is to mathematically \emph{identify} organization.
Implicitly, this implies moving beyond simple notions of exact symmetries (and
small deviations from them) and beyond subjectively selecting function bases
for representational ``dictionaries''.

We reviewed two modern approaches to this challenge---evolution operators and
intrinsic computation. Both lie outside constructionism, as neither can be
analytically computed in general. However, they provide a theoretical framework
for \emph{data-driven} discovery of pattern and organization. Koopman and
Perron-Frobenius evolution operators evolve system observables and probability
distributions, respectively. Their spectral decompositions can identify
statistical organization in time-independent systems, as well as coherent
structures that heavily dictate transport in time-dependent systems.

In a complementary way, intrinsic computation decomposes a system into its
minimal causal components using predictive equivalence. The semigroup algebra
of intrinsic computation identifies organization by generalizing exact
symmetries, and its extension to field theories using lightcones can also
identify coherent structures. 

Completing the relation between evolution operators and intrinsic computation
is a challenge currently. It is encouraging that they both converge on their
identification of coherent structures. At this time, though, it is not clear
whether these two represent two pieces of a larger, universal theory of
organization.  Whether or not such a universal theory of organization even
exists is an open question. 

A tantalizing path forward is offered by combining evolution operators with
predictive equivalence to formulate a statistical mechanics of emergence.
Emergence is what has stymied efforts to develop constructionist principles of
organization. The new mathematical tools that directly identify complex forms
of organization also come together in a universal theory that provides complete
and self-contained dynamics for a higher-Level emergent description of a system
that is physically consistent with its lower-Level description. While not
analytically constructible in practice, this theory provides a rigorous
foundation for data-driven approximation of emergent dynamics through
predictive equivalence.

We close with a brief remark on \emph{causal mechanism} discovery. While
nonequilibrium equations Eq. (\ref{eq:gen_dyn}) govern transport of energy and
matter, different equations---specifically those derived from different
phenomenological laws---represent distinct, but equally important mechanisms
governing system dynamics. To address emergent behaviors in such broader
classes of complex system, the added importance of \emph{information transport}
\cite{boss16a,Jame16a,smir13a} and related \emph{causal mechanisms}
\cite{pear18a} is increasingly appreciated. Identifying causal mechanisms
directly from system behaviors is an active and ongoing area of research
\cite{sugi12a,rung19a,sinh20a}. We once again see the data-driven paradigm to
scientific discovery filling the void left by the shortcomings of
constructionism for complex systems with emergent behaviors.

\section*{Acknowledgments}
The authors thank Alec Boyd, Nicolas Brodu, David Campbell, Antonio Carlos Costa, David Gier, Alex Jurgens, Chris Pratt, Kyle Ray, Mikhael Semaan, Kuen Wai Tang, Ariadna Venegas-Li, and Greg Wimsatt
for helpful comments and suggestions. 
The authors thank the Telluride Science Research Center for hospitality during
visits and the participants of the Information Engines Workshops there. 
Part of this research was performed while AR was visiting the Institute for
Pure and Applied Mathematics, which is supported by the National Science
Foundation grant DMS-1440415.
JPC
acknowledges the kind hospitality of the Santa Fe Institute, Institute for
Advanced Study at the University of Amsterdam, and California Institute of
Technology for their hospitality during visits. This material is based upon
work supported by, or in part by, FQXi Grant number FQXi-RFP-IPW-1902
and U.S. Army Research Laboratory and the U.S. Army Research Office under
grants W911NF-21-1-0048 and W911NF-18-1-0028.

%

\bibliography{chaos,spacetime}

\begin{thebibliography}{100}

\bibitem{Bena01a}
H.~B{\'e}nard.
\newblock {\em Les Tourbillons Cellulaires dans une nappe Liquide Propageant de
  la Chaleur par Convection: en R{\'e}gime Permanent}.
\newblock Gauthier-Villars, 1901.

\bibitem{Rayl16a}
Lord (J. W.~Strutt) Rayleigh.
\newblock On convection currents in a horizontal layer of fluid, when the
  higher temperature is on the under side.
\newblock {\em Phil. Mag. (Series 6)}, 32(192):529--546, 1916.

\bibitem{Chan68a}
S.~Chandrasekhar.
\newblock {\em Hydrodynamic and Hydromagnetic Stability}.
\newblock Oxford, Clarendon Press, 1968.

\bibitem{Buss78a}
F.~H. Busse.
\newblock Non-linear properties of thermal convection.
\newblock {\em Reports on Progress in Physics}, 41(12):1929, 1978.

\bibitem{Fens79a}
P.~R. Fenstermacher, H.~L. Swinney, and J.~P. Gollub.
\newblock Dynamical instabilities and the transition to chaotic {T}aylor vortex
  flow.
\newblock {\em J. Fluid Mech.}, 94(1):103--128, 1979.

\bibitem{Stei85}
V.~Steinberg, G.~Ahlers, and D.~S. Cannell.
\newblock Pattern formation and wave-number selection by rayleigh-benard
  convection in a cylindrical container.
\newblock {\em Physica Scripta}, T9:97, 1985.

\bibitem{Zure01a}
W.~H. Zurek.
\newblock Sub-planck structure in phase space and its relevance for quantum
  decoherence.
\newblock {\em Nature}, 412:712--717, 2001.

\bibitem{Turi52}
A.~M. Turing.
\newblock The chemical basis of morphogenesis.
\newblock {\em Trans. Roy. Soc., Series B}, 237:5, 1952.

\bibitem{Weyg09a}
R.~v.~de Weygaert, B.~J.~T. Jones, E.~Platen, and M.~A. Aragon-Calvo.
\newblock Geometry and morphology of the cosmic web: Analyzing spatial patterns
  in the universe.
\newblock In {\em ISVD09 (Intl. Symp. Voronoi Diagrams Engin.)}. 2009.

\bibitem{Schi14a}
H.~Y. Schive, T.~Chiueh, and T.~Broadhurst.
\newblock Cosmic structure as the quantum interference of a coherent dark wave.
\newblock {\em Nature Physics}, 10:496--499, 2014.

\bibitem{Newe19a}
A.~C. Newell and S.~C. Venkataramani.
\newblock Pattern universes.
\newblock {\em Comptes Rendus Mechanique}, 347:318--331, 2019.

\bibitem{Cros09a}
M.~Cross and H.~Greenside.
\newblock {\em Pattern Formation and Dynamics in Nonequilibrium Systems}.
\newblock Cambridge University Press, Cambridge, United Kingdom, 2009.

\bibitem{Hoyl06a}
R.~Hoyle.
\newblock {\em Pattern Formation: {An} Introduction to Methods}.
\newblock Cambridge University Press, New York, 2006.

\bibitem{Heis67a}
W.~Heisenberg.
\newblock Nonlinear problems in physics.
\newblock {\em Physics Today}, 20:23--33, 1967.

\bibitem{Ruel71a}
D.~Ruelle and F.~Takens.
\newblock On the nature of turbulence.
\newblock {\em Comm. Math. Phys.}, 20:167--192, 1971.

\bibitem{Bran83}
A.~Brandstater, J.~Swift, Harry~L. Swinney, A.~Wolf, J.~D. Farmer, E.~Jen, and
  J.~P. Crutchfield.
\newblock Low-dimensional chaos in a hydrodynamic system.
\newblock {\em Phys. Rev. Lett.}, 51:1442, 1983.

\bibitem{Cros93a}
M.~C. Cross and P.~C. Hohenberg.
\newblock Pattern formation outside of equilibrium.
\newblock {\em Rev. Mod. Phys.}, 65(3):851--1112, 1993.

\bibitem{boltz1896a}
L.~Boltzmann.
\newblock Entgegnung auf die w{\"a}rmetheoretischen betrachtungen des hrn. e.
  zermelo.
\newblock {\em Annalen der physik}, 293(4):773--784, 1896.

\bibitem{brush16a}
St.~G. Brush.
\newblock {\em Kinetic theory: Irreversible processes}.
\newblock Elsevier, 2016.

\bibitem{Mack92a}
M.~C. Mackey, editor.
\newblock {\em Time's Arrow: The Origins of Thermodynamic Behavior}.
\newblock Springer-Verlag, New York, 1992.

\bibitem{lebo93a}
J.~L. Lebowitz.
\newblock Boltzmann's entropy and time's arrow.
\newblock {\em Physics today}, 46(9):32--38, 1993.

\bibitem{Newe74a}
A.~C. Newell.
\newblock Envelope equations.
\newblock {\em Lectures in Applied Mathematics}, 15(157):4, 1974.

\bibitem{Eman91a}
K.~A. Emanuel.
\newblock The theory of hurricanes.
\newblock {\em Ann. Rev. Fluid Mech.}, 23(1):179--196, 1991.

\bibitem{Swin00a}
H.~L. Swinney.
\newblock Emergence and evolution of patterns.
\newblock In {\em AIP Conference Proceedings}, volume 501, pages 3--22.
  American Institute of Physics, 2000.

\bibitem{Ball99a}
P.~Ball.
\newblock {\em The Self-Made Tapestry: Pattern Formation in Nature}.
\newblock Oxford University Press, New York, 1999.

\bibitem{vics95a}
T.~Vicsek, A.~Czir{\'o}k, E.~Ben-Jacob, I.~Cohen, and O.~Shochet.
\newblock Novel type of phase transition in a system of self-driven particles.
\newblock {\em Phys. Rev. Let.}, 75(6):1226, 1995.

\bibitem{tone5a}
J.~Toner and Y.~Tu.
\newblock Long-range order in a two-dimensional dynamical {XY} model: How birds
  fly together.
\newblock {\em Phys. Rev. Let.}, 75(23):4326, 1995.

\bibitem{taji01a}
Y.~Tajima and T.~Nagatani.
\newblock Scaling behavior of crowd flow outside a hall.
\newblock {\em Physica A: Statistical Mechanics and its Applications},
  292(1-4):545--554, 2001.

\bibitem{aren08a}
A.~Arenas, A.~D{\'\i}az-Guilera, J.~Kurths, Y.~Moreno, and C.~Zhou.
\newblock Synchronization in complex networks.
\newblock {\em Physics Reports}, 469(3):93--153, 2008.

\bibitem{Gras83a}
P.~Grassberger.
\newblock New mechanism for deterministic diffusion.
\newblock {\em Phys. Rev. A}, 28:3666, 1983.

\bibitem{Grass86b}
P.~Grassberger.
\newblock Long-range effects in an elementary cellular automaton.
\newblock {\em J. Stat. Physics}, 45(1-2):27--39, 1986.

\bibitem{Crut93a}
J.~P. Crutchfield and J.~E. Hanson.
\newblock Turbulent pattern bases for cellular automata.
\newblock {\em Physica D}, 69:279 -- 301, 1993.

\bibitem{Hans95a}
J.~E. Hanson and J.~P. Crutchfield.
\newblock Computational mechanics of cellular automata: An example.
\newblock {\em Physica D}, 103:169--189, 1997.

\bibitem{cart01a}
A.~H. Carter.
\newblock {\em Classical and Statistical Thermodynamics}.
\newblock Prentice Hall, New Jersey, 2001.

\bibitem{joule1850a}
J.~P. Joule.
\newblock On the mechanical equivalent of heat.
\newblock {\em Phil. Trans. Roy. Soc. London}, pages 61--82, 1850.

\bibitem{carn1824a}
S.~Carnot.
\newblock {\em Reflections on the motive power of fire, and on machines fitted
  to develop that power}.
\newblock Paris: Bachelier, 1824.

\bibitem{crop86a}
W.~H. Cropper.
\newblock Rudolf clausius and the road to entropy.
\newblock {\em American journal of physics}, 54(12):1068--1074, 1986.

\bibitem{Winf84a}
A.~T. Winfree.
\newblock The prehistory of the {B}elousov-{Z}habotinsky oscillator.
\newblock {\em Journal of Chemical Education}, 61(8):661, 1984.

\bibitem{Zhab91a}
A.~M. Zhabotinsky.
\newblock A history of chemical oscillations and waves.
\newblock {\em Chaos}, 1(4):379--386, 1991.

\bibitem{kive98a}
S.~A. Kivelson, E.~Fradkin, and V.~J. Emery.
\newblock Electronic liquid-crystal phases of a doped {M}ott insulator.
\newblock {\em Nature}, 393(6685):550--553, 1998.

\bibitem{zhen17a}
B.-X. Zheng, C.-M. Chung, P.~Corboz, G.~Ehlers, M.-P. Qin, R.~M. Noack, H.~Shi,
  S.~R. White, S.~Zhang, and G.~K.-L. Chan.
\newblock Stripe order in the underdoped region of the two-dimensional
  {H}ubbard model.
\newblock {\em Science}, 358(6367):1155--1160, 2017.

\bibitem{Qian17a}
Q.~Qian, J.~Nakamura, S.~Fallahi, G.~C. Gardner, and M.~J. Manfra.
\newblock Possible nematic to smectic phase transition in a two-dimensional
  electron gas at half-filling.
\newblock {\em Nature Comm.}, 8(1):1--6, 2017.

\bibitem{fern14a}
R.~M. Fernandes, A.~V. Chubukov, and J.~Schmalian.
\newblock What drives nematic order in iron-based superconductors?
\newblock {\em Nature physics}, 10(2):97--104, 2014.

\bibitem{Land69a}
L.~D. Landau.
\newblock On the theory of phase transitions.
\newblock In {\em L. D. Landau Collected Papers}, volume~1, pages 234--252.
  Nauka, Moscow, 1969.
\newblock Originally published in Zh. Eksp. Teor. Fiz. 7, pp. 19--32 (1937).

\bibitem{Prig68a}
I.~Prigogine.
\newblock {\em Introduction to Thermodynamics of Irreversible Processes}.
\newblock John Wiley and Sons, New York, third edition, 1968.

\bibitem{DeGr62a}
S.~R.~de Groot and P.~Mazur.
\newblock {\em Non-Equilibrium Thermodynamics}.
\newblock North Holland, Amsterdam, 1962.

\bibitem{Jayn57a}
E.~T. Jaynes.
\newblock Information theory and statistical mechanics.
\newblock {\em Phys. Rev.}, 106(4):620--630, 1957.

\bibitem{Gran08a}
W.~T. Grandy.
\newblock {\em Entropy and The Time Evolution of Macroscopic Systems},
  volume~10.
\newblock Oxford University Press, 2008.

\bibitem{Onsa49}
L.~Onsager.
\newblock The effects of shape on the interaction of colloidal particles.
\newblock {\em Ann. New York Acad. Sci.}, 51(4):627--659, 1949.

\bibitem{Moor03a}
T.~A. Moore.
\newblock {\em Six Ideas that Shaped Physics; Unit T: Some Processes are
  Irreversible}.
\newblock McGraw-Hill, New York, second edition, 2003.

\bibitem{Naim08a}
A.~Ben-Naim.
\newblock {\em A Farewell to Entropy: Statistical Thermodynamics Based on
  Information}.
\newblock World Scientific, 2008.

\bibitem{brush76a}
S.~G. Brush.
\newblock {\em The kind of motion we call heat}, volume~1.
\newblock North-Holland Amsterdam, Oxford, 1976.

\bibitem{boltz1909}
L.~Boltzmann.
\newblock {\em Wissenschaftliche Abhandlungen: Bd. 1865-1874}.
\newblock Verlag von Johann Ambrosius Barth, Leipzig, 1909.

\bibitem{gibbs1875a}
J.~W. Gibbs.
\newblock On the equilibrium of heterogeneous substances.
\newblock {\em Trans. Conn. Acad. Sci.}, III:108, 343, 1875-1878.

\bibitem{path11a}
R.~K. Pathria and P.~D. Beale.
\newblock {\em Statistical mechanics}.
\newblock Amsterdam. Elsevier, 2011.

\bibitem{schr21a}
D.~V. Schroeder.
\newblock {\em {An Introduction to Thermal Physics}}.
\newblock Oxford University Press, 2021.

\bibitem{wehr78a}
A.~Wehrl.
\newblock General properties of entropy.
\newblock {\em Reviews of Modern Physics}, 50(2):221, 1978.

\bibitem{Shan48a}
C.~E. Shannon.
\newblock A mathematical theory of communication.
\newblock {\em Bell Sys. Tech. J.}, 27:379--423, 623--656, 1948.

\bibitem{Cove06a}
T.~M. Cover and J.~A. Thomas.
\newblock {\em Elements of Information Theory}.
\newblock Wiley-Interscience, New York, second edition, 2006.

\bibitem{Call85a}
H.~B. Callen.
\newblock {\em Thermodynamics and an Introduction to Thermostatistics}.
\newblock Wiley, New York, second edition, 1985a.

\bibitem{Jame11a}
R.~G. James, C.~J. Ellison, and J.~P. Crutchfield.
\newblock Anatomy of a bit: {Information} in a time series observation.
\newblock {\em CHAOS}, 21(3):037109, 2011.

\bibitem{jayn65a}
E.~T. Jaynes.
\newblock Gibbs vs boltzmann entropies.
\newblock {\em American Journal of Physics}, 33(5):391--398, 1965.

\bibitem{ehren90a}
P.~Ehrenfest and T.~Ehrenfest.
\newblock {\em The conceptual foundations of the statistical approach in
  mechanics}.
\newblock Dover, 1990.

\bibitem{gao19a}
X.~Gao, E.~Gallicchio, and A.~E. Roitberg.
\newblock The generalized {B}oltzmann distribution is the only distribution in
  which the {G}ibbs-{S}hannon entropy equals the thermodynamic entropy.
\newblock {\em The Journal of Chemical Physics}, 151(3):034113, 2019.

\bibitem{Onsa31a}
L.~Onsager.
\newblock Reciprocal relations in irreversible processes. {I.}
\newblock {\em Phys. Rev.}, 37(4):405, 1931.

\bibitem{Onsa31b}
L.~Onsager.
\newblock Reciprocal relations in irreversible processes. {II.}
\newblock {\em Phys. Rev.}, 38(12):2265, 1931.

\bibitem{Kond14a}
D.~Kondepudi and I.~Prigogine.
\newblock {\em Modern Thermodynamics: From Heat Engines to Dissipative
  Structures}.
\newblock John Wiley \& Sons, 2014.

\bibitem{thom74b}
W.~Thomson.
\newblock A mechanical theory of thermo-electric currents.
\newblock {\em Proc. Royal Society of Edinburgh}, 3:91--98, 1857.

\bibitem{Sing22a}
M.~S. Singh and M.~E. O'Neill.
\newblock The climate system and the second law of thermodynamics.
\newblock {\em Rev. Mod. Phys.}, 94:015001, Jan 2022.

\bibitem{Keiz87a}
J.~Keizer.
\newblock {\em Statistical Thermodynamics of Nonequilibrium Processes}.
\newblock Springer, 1987.

\bibitem{Seif12a}
U.~Seifert.
\newblock Stochastic thermodynamics, fluctuation theorems and molecular
  machines.
\newblock {\em Rep. Prog. Phys.}, 75:126001, 2012.

\bibitem{Kirc48a}
G.~D. Kirchhoff.
\newblock {\em Ann. Phys.}, 75:1891, 1848.

\bibitem{Gibb06a}
J.~W. Gibbs.
\newblock {\em The Scientific Papers of J. Willard Gibbs}.
\newblock Longmans, Green, New York, New York, 1906.

\bibitem{Maxw54a}
J.~C. Maxwell.
\newblock {\em A Treatise on Electricity and Magnetism, vol. {I} and {II}}.
\newblock Dover Publications, Inc., New York, New York, third edition, 1954.

\bibitem{helm68a}
H.~von Helmholtz.
\newblock Zur theorie der station{\"a}ren str{\"o}me in reibenden
  fl{\"u}ssigkeiten.
\newblock {\em Wissenschaftliche Abhandlugen}, 1:223--230, 1868.

\bibitem{Rayl77a}
Lord (J. W.~Strutt) Rayleigh.
\newblock {\em The theory of sound}.
\newblock Macmillan \& Company, 1877.

\bibitem{lore96a}
H.~A. Lorentz.
\newblock The theorem of {P}oynting concerning the energy in the
  electromagnetic field and two general propositions concerning the propagation
  of light.
\newblock {\em Amsterdammer Akademie der Wetenschappen}, 4:176, 1896.

\bibitem{Land75a}
R.~Landauer.
\newblock Inadequacy of entropy and entropy derivatives in characterizing the
  steady state.
\newblock {\em Physical Review A}, 12(2):636, 1975.

\bibitem{Jayn80a}
E.~T. Jaynes.
\newblock The minimum entropy production principle.
\newblock {\em Annu. Rev. Phys. Chem.}, 31:579--601, 1980.

\bibitem{barb99a}
E.~Barbera.
\newblock On the principle of minimal entropy production for
  {Navier-Stokes-Fourier} fluids.
\newblock {\em Continuum Mechanics and Thermodynamics}, 11(5):327--330, 1999.

\bibitem{Palf01a}
P.~Palffy-Muhoray.
\newblock Comment on “{A} check of {P}rigogine’s theorem of minimum entropy
  production in a rod in a nonequilibrium stationary state” by {I}rena
  {D}anielewicz-{F}erchmin and {A}. {R}yszard {F}erchmin [{A}m. {J}. {P}hys. 68
  (10), 962--965 (2000)].
\newblock {\em Am. J. Physics}, 69(7):825--826, 2001.

\bibitem{Ross05a}
J.~Ross and M.~O. Vlad.
\newblock Exact solutions for the entropy production rate of several
  irreversible processes.
\newblock {\em J. Physical Chemistry A}, 109(46):10607--10612, 2005.

\bibitem{Glan71a}
P.~Glansdorff and I.~Prigogine.
\newblock Thermodynamic theory of structure, stability and fluctuations.
\newblock 1971.

\bibitem{Nico77a}
G.~Nicolis and I.~Prigogine.
\newblock {\em Self-Organization in Nonequilibrium Systems}.
\newblock Wiley, New York, 1977.

\bibitem{Keiz74a}
J.~Keizer and R.~F. Fox.
\newblock Qualms regarding the range of validity of the {Glansdorff-Prigogine}
  criterion for stability of non-equilibrium states.
\newblock {\em Proc. Natl. Acad. Sci. USA}, 71(1):192--196, 1974.

\bibitem{Glan74a}
P~Glansdorff, G~Nicolis, and I~Prigogine.
\newblock The thermodynamic stability theory of non-equilibrium states.
\newblock {\em Proceedings of the National Academy of Sciences},
  71(1):197--199, 1974.

\bibitem{Fox79a}
R.~F. Fox.
\newblock Irreversible processes at nonequilibrium steady states.
\newblock {\em Proc. Natl. Acad. Sci. USA}, 76(5):2114--2117, 1979.

\bibitem{Nico79a}
G.~Nicolis and I.~Prigogine.
\newblock Irreversible processes at nonequilibrium steady states and lyapounov
  functions.
\newblock {\em Proceedings of the National Academy of Sciences},
  76(12):6060--6061, 1979.

\bibitem{Fox80a}
R.~F. Fox.
\newblock The ``excess entropy'' around nonequilibrium steady
  states,$(\delta^2s)_{ss}$, is not a {Liapunov} function.
\newblock {\em Proc. Natl. Acad. Sci. USA}, 77(7):3763--3766, 1980.

\bibitem{Ande87a}
P.~W. Anderson and D.~L. Stein.
\newblock Broken symmetry, emergent properties, dissipative structures, life.
\newblock In {\em Self-organizing systems}, pages 445--457. Springer, 1987.

\bibitem{Reic80a}
L.~E. Reichl.
\newblock {\em A Modern Course in Statistical Mechanics}.
\newblock University of Texas Press, Austin, Texas, 1980.

\bibitem{Keiz76a}
J.~Keizer.
\newblock Fluctuations, stability, and generalized state functions at
  nonequilibrium steady states.
\newblock {\em The Journal of Chemical Physics}, 65(11):4431--4444, 1976.

\bibitem{Gasp04a}
P.~Gaspard.
\newblock Time-reversed dynamical entropy and irreversibility in markovian
  random processes.
\newblock {\em J. Stat. Phys.}, 117(3/4):599--615, 2004.

\bibitem{Jame16a}
R.~G. James, N.~Barnett, and J.~P. Crutchfield.
\newblock Information flows? {A} critique of transfer entropies.
\newblock {\em Physical review letters}, 116(23):238701, 2016.

\bibitem{Parr15a}
J.~M.~R. Parrondo, J.~M. Horowitz, and T.~Sagawa.
\newblock Thermodynamics of information.
\newblock 11:131--139, 2015.

\bibitem{Will10a}
P.~L. Williams and R.~D. Beer.
\newblock Nonnegative decomposition of multivariate information.
\newblock 2010.
\newblock arXiv:1004.2515 [cs.IT].

\bibitem{Jame17a}
R.~G. James and J.~P. Crutchfield.
\newblock Multivariate dependence beyond {Shannon} information.
\newblock {\em Entropy}, 19(10):531, 2017.

\bibitem{Fala18a}
G.~Falasco, R.~Rao, and M.~Esposito.
\newblock Information thermodynamics of {Turing} patterns.
\newblock {\em Phys. Rev. Let.}, 121:108301, 2018.

\bibitem{Jayn85a}
E.~T. Jaynes.
\newblock Macroscopic prediction.
\newblock In {\em Complex Systems---Operational Approaches in Neurobiology,
  Physics, and Computers}, pages 254--269. Springer, 1985.

\bibitem{Press13a}
S.~Press{\'e}, K.~Ghosh, J.~Lee, and K.~A Dill.
\newblock Principles of maximum entropy and maximum caliber in statistical
  physics.
\newblock {\em Reviews of Modern Physics}, 85(3):1115, 2013.

\bibitem{Ghos20a}
K.~Ghosh, P.~D. Dixit, L.~Agozzino, and K.~A. Dill.
\newblock The maximum caliber variational principle for nonequilibria.
\newblock {\em Annual Review of Physical Chemistry}, 71:213--238, 2020.

\bibitem{Atta12a}
P.~Attard.
\newblock Optimising principle for non-equilibrium phase transitions and
  pattern formation with results for heat convection.
\newblock {\em arXiv:1208.5105}.

\bibitem{Hake12a}
H.~Haken.
\newblock {\em Advanced Synergetics: Instability Hierarchies of Self-Organizing
  Systems and Devices}, volume~20.
\newblock Springer, 2012.

\bibitem{Hake16a}
H.~Haken.
\newblock {\em Information and Self-Organization}.
\newblock Springer, New York, 2016.

\bibitem{Ande72a}
P.~W. Anderson.
\newblock More is different.
\newblock {\em Science}, 177(4047):393--396, 1972.

\bibitem{lebo99a}
J.~L. Lebowitz and H.~Spohn.
\newblock A {Gallavotti-Cohen-type} symmetry in the large deviation functional
  for stochastic dynamics.
\newblock {\em J. Stat. Phys.}, 95:333, 1999.

\bibitem{Crut92c}
J.~P. Crutchfield.
\newblock The calculi of emergence: Computation, dynamics, and induction.
\newblock {\em Physica D}, 75:11--54, 1994.

\bibitem{carr17a}
S.~Carroll.
\newblock {\em The big picture: on the origins of life, meaning, and the
  universe itself}.
\newblock Penguin, 2017.

\bibitem{bone00a}
F.~Bonetto, J.~L. Lebowitz, and L.~Rey-Bellet.
\newblock Fourier's law: a challenge to theorists.
\newblock In {\em Mathematical physics 2000}, pages 128--150. World Scientific,
  2000.

\bibitem{grad63a}
H.~Grad.
\newblock Asymptotic theory of the {B}oltzmann equation.
\newblock {\em The Physics of Fluids}, 6(2):147--181, 1963.

\bibitem{slem18a}
M.~Slemrod.
\newblock Hilbert’s sixth problem and the failure of the {B}oltzmann to
  {E}uler limit.
\newblock {\em Philosophical Transactions of the Royal Society A: Mathematical,
  Physical and Engineering Sciences}, 376(2118):20170222, 2018.

\bibitem{Bloc56a}
M.~J. Block.
\newblock Surface tension as the cause of {B}{\'e}nard cells and surface
  deformation in a liquid film.
\newblock {\em Nature}, 178(4534):650--651, 1956.

\bibitem{Pear58a}
J.~R.~A. Pearson.
\newblock On convection cells induced by surface tension.
\newblock {\em J. Fluid Mechanics}, 4(5):489--500, 1958.

\bibitem{Scha95a}
M.~F. Schatz, S.~J. VanHook, W.~D. McCormick, J.~B. Swift, and H.~L. Swinney.
\newblock Onset of surface-tension-driven {B}{\'e}nard convection.
\newblock {\em Physical review letters}, 75(10):1938, 1995.

\bibitem{merl14a}
T.~M. Merlis.
\newblock Interacting components of the top-of-atmosphere energy balance affect
  changes in regional surface temperature.
\newblock {\em Geophysical Research Letters}, 41(20):7291--7297, 2014.

\bibitem{Eman87a}
K.~A. Emanuel.
\newblock The dependence of hurricane intensity on climate.
\newblock {\em Nature}, 326(6112):483, 1987.

\bibitem{Wals97a}
K.~Walsh and I.G. Watterson.
\newblock Tropical cyclone-like vortices in a limited area model: Comparison
  with observed climatology.
\newblock {\em J. Climate}, 10(9):2240--2259, 1997.

\bibitem{Wehn10a}
M.~F. Wehner, G.~Bala, P.~Duffy, A.~A. Mirin, and R.~Romano.
\newblock Towards direct simulation of future tropical cyclone statistics in a
  high-resolution global atmospheric model.
\newblock {\em Advances in Meteorology}, 2010, 2010.

\bibitem{Moor11a}
C.~Moore and S.~Mertens.
\newblock {\em The nature of computation}.
\newblock Oxford, 2011.

\bibitem{Gu09a}
M.~Gu, C.~Weedbrook, A.~Perales, and M.~A. Nielsen.
\newblock More really is different.
\newblock {\em Physica D: Nonlinear Phenomena}, 238(9-10):835--839, 2009.

\bibitem{Turi36}
A.~M. Turing.
\newblock On computable numbers, with an application to the
  entsheidungsproblem.
\newblock {\em Proc. Lond. Math. Soc. Ser. 2}, 42:230, 1936.

\bibitem{Lewi98a}
H.~R. Lewis and C.~H. Papadimitriou.
\newblock {\em Elements of the Theory of Computation}.
\newblock Prentice-Hall, Englewood Cliffs, N.J., second edition, 1998.

\bibitem{Hopc06a}
J.~E. Hopcroft, R.~Motwani, and J.~D. Ullman.
\newblock {\em Introduction to Automata Theory, Languages, and Computation}.
\newblock Prentice-Hall, New York, third edition, 2006.

\bibitem{Sips14a}
M.~Sipser.
\newblock {\em Introduction to the Theory of Computation}.
\newblock Cengage Learning, New York, third edition, 2014.

\bibitem{Cook04a}
M.~Matthew.
\newblock Universality in elementary cellular automata.
\newblock {\em Complex Systems}, 15(1):1--40, 2004.

\bibitem{Hans90a}
J.~E. Hanson and J.~P. Crutchfield.
\newblock The attractor-basin portrait of a cellular automaton.
\newblock {\em J. Stat. Phys.}, 66:1415 -- 1462, 1992.

\bibitem{Conw70a}
J.~Conway.
\newblock The game of life.
\newblock {\em Scientific American}, 223(4):4, 1970.

\bibitem{Kari94a}
J.~Kari.
\newblock Rice's theorem for the limit sets of cellular automata.
\newblock {\em Theoretical computer science}, 127(2):229--254, 1994.

\bibitem{Moor90a}
C.~Moore.
\newblock Unpredictability and undecidability in dynamical systems.
\newblock {\em Phys. Rev. Lett.}, 64:2354, 1990.

\bibitem{Moor97c}
C.~Moore.
\newblock Majority-vote cellular automata, {I}sing dynamics, and
  {P}-completeness.
\newblock {\em Journal of Statistical Physics}, 88(3-4):795--805, 1997.

\bibitem{Near06a}
T.~Neary and D.~Woods.
\newblock P-completeness of cellular automaton {R}ule 110.
\newblock In {\em International Colloquium on Automata, Languages, and
  Programming}, pages 132--143. Springer, 2006.

\bibitem{Moor97b}
C.~Moore.
\newblock Quasilinear cellular automata.
\newblock {\em Physica D: Nonlinear Phenomena}, 103(1-4):100--132, 1997.

\bibitem{carn91a}
G.~F. Carnevale, J.~C. McWilliams, Y~Pomeau, J.~B. Weiss, and W.~R. Young.
\newblock Evolution of vortex statistics in two-dimensional turbulence.
\newblock {\em Phys. Rev. Let.}, 66(21):2735, 1991.

\bibitem{mcwil90a}
J.~C. McWilliams.
\newblock The vortices of two-dimensional turbulence.
\newblock {\em J. of Fluid Mech.}, 219:361--385, 1990.

\bibitem{Binn92a}
J.~J. Binney, N.~J. Dowrick, A.~J. Fisher, and M.~E.~J. Newman.
\newblock {\em The Theory of Critical Phenomena}.
\newblock Oxford University Press, Oxford, 1992.

\bibitem{Feld97a}
D.~P. Feldman and J.~P. Crutchfield.
\newblock Measures of statistical complexity: {Why}?
\newblock {\em Phys. Lett. A}, 238:244--252, 1998.

\bibitem{Vita89a}
M.~Li and P.~M.~B. Vitanyi.
\newblock Kolmogorov complexity and its applications.
\newblock Technical Report CS-R8901, Centruum voor Wiskunde en Informatica,
  Universiteit van Amsterdam, 1989.

\bibitem{Kolm65}
A.~N. Kolmogorov.
\newblock Three approaches to the concept of the amount of information.
\newblock {\em Prob. Info. Trans.}, 1:1, 1965.

\bibitem{Chai66}
G.~Chaitin.
\newblock On the length of programs for computing finite binary sequences.
\newblock {\em J. ACM}, 13:145, 1966.

\bibitem{Brud83}
A.~A. Brudno.
\newblock Entropy and the complexity of the trajectories of a dynamical system.
\newblock {\em Trans. Moscow Math. Soc.}, 44:127, 1983.

\bibitem{Benn86}
C.~H. Bennett.
\newblock On the nature and origin of complexity in discrete, homogeneous
  locally-interacting systems.
\newblock {\em Found. Phys.}, 16:585--592, 1986.

\bibitem{Kopp87a}
M.~Koppel.
\newblock Complexity, depth, and sophistication.
\newblock {\em Complexity}, 1:1087--1091, 1987.

\bibitem{Benn88}
C.~H. Bennett.
\newblock Dissipation, information, computational complexity, and the
  definition of organization.
\newblock In D.~Pines, editor, {\em Emerging Syntheses in the Sciences}.
  Addison-Wesley, Redwood City, 1988.

\bibitem{Kopp91a}
M.~Koppel and H.~Atlan.
\newblock An almost machine-independent theory of program-length complexity,
  sophistication, and induction.
\newblock {\em Information Sciences}, 56(1-3):23--33, 1991.

\bibitem{Crut98d}
J.~P. Crutchfield and C.~R. Shalizi.
\newblock Thermodynamic depth of causal states: {O}bjective complexity via
  minimal representations.
\newblock {\em Physical Review E}, 59(1):275--283, 1999.

\bibitem{Lemp76a}
A.~Lempel and J.~Ziv.
\newblock On the complexity of individual sequences.
\newblock {\em IEEE Trans. Info. Th.}, IT-22:75, 1976.

\bibitem{Ziv78a}
J.~Ziv and A.~Lempel.
\newblock Compression of individual sequences via variable-rate encoding.
\newblock {\em IEEE Trans. Info. Th.}, IT-24:530, 1978.

\bibitem{Riss78a}
J.~Rissanen.
\newblock Modeling by shortest data description.
\newblock {\em Automatica}, 14:462, 1978.

\bibitem{hint93a}
G.~E. Hinton and R.~Zemel.
\newblock Autoencoders, minimum description length and {H}elmholtz free energy.
\newblock {\em Advances in neural information processing systems}, 6, 1993.

\bibitem{Tair17a}
K.~Taira, S.~L. Brunton, S.~T.~M. Dawson, C.~W. Rowley, T.~Colonius, B.~J.
  McKeon, O.~T. Schmidt, S.~Gordeyev, V.~Theofilis, and L.~S. Ukeiley.
\newblock Modal analysis of fluid flows: An overview.
\newblock {\em Aiaa Journal}, 55(12):4013--4041, 2017.

\bibitem{Alga69a}
V.~Algazi and D.~Sakrison.
\newblock On the optimality of the {Karhunen-Lo{\`e}ve} expansion (corresp.).
\newblock {\em IEEE Trans. Info. Th.}, 15(2):319--321, 1969.

\bibitem{Rowl04a}
C.~W. Rowley, T.~Colonius, and R.~M. Murray.
\newblock Model reduction for compressible flows using {POD} and {Galerkin}
  projection.
\newblock {\em Physica D: Nonlinear Phenomena}, 189(1-2):115--129, 2004.

\bibitem{Holm12a}
P.~Holmes, J.L. Lumley, G.~Berkooz, and C.W. Rowley.
\newblock {\em Turbulence, Coherent Structures, Dynamical Systems and
  Symmetry}.
\newblock Cambridge University Press, Cambridge, United Kingdom, 2012.

\bibitem{maul20a}
R.~Maulik, A.~Mohan, B.~Lusch, S.~Madireddy, P.~Balaprakash, and D.~Livescu.
\newblock Time-series learning of latent-space dynamics for reduced-order model
  closure.
\newblock {\em Physica D: Nonlinear Phenomena}, 405:132368, 2020.

\bibitem{wang20a}
Q.~Wang, N.~Ripamonti, and J.~S. Hesthaven.
\newblock Recurrent neural network closure of parametric {POD}-{G}alerkin
  reduced-order models based on the {M}ori-{Z}wanzig formalism.
\newblock {\em J. Computational Physics}, 410:109402, 2020.

\bibitem{Laso13a}
A.~Lasota and M.~C. Mackey.
\newblock {\em Chaos, fractals, and noise: stochastic aspects of dynamics},
  volume~97.
\newblock Springer Science \& Business Media, 2013.

\bibitem{reyn72a}
W.~C. Reynolds and A.~K. M.~F. Hussain.
\newblock The mechanics of an organized wave in turbulent shear flow. {P}art 3.
  theoretical models and comparisons with experiments.
\newblock {\em J. Fluid Mechanics}, 54(2):263--288, 1972.

\bibitem{Mezi13a}
I.~Mezi{\'c}.
\newblock Analysis of fluid flows via spectral properties of the {Koopman}
  operator.
\newblock {\em Ann. Rev. Fluid Mechanics}, 45:357--378, 2013.

\bibitem{huer90a}
P.~Huerre and P.~A. Monkewitz.
\newblock Local and global instabilities in spatially developing flows.
\newblock {\em Ann. Rev. Fluid Mechanics}, 22(1):473--537, 1990.

\bibitem{Mezi05a}
I.~Mezi{\'c}.
\newblock Spectral properties of dynamical systems, model reduction and
  decompositions.
\newblock {\em Nonlinear Dynamics}, 41(1):309--325, 2005.

\bibitem{Koop31a}
B.~O. Koopman.
\newblock Hamiltonian systems and transformation in {H}ilbert space.
\newblock {\em Proc. Natl. Acad. Sci. USA}, 17(5):315, 1931.

\bibitem{brun16b}
S.~L. Brunton, B.~W. Brunton, J.~L. Proctor, and J.~N. Kutz.
\newblock Koopman invariant subspaces and finite linear representations of
  nonlinear dynamical systems for control.
\newblock {\em PloS one}, 11(2):e0150171, 2016.

\bibitem{brun22a}
S.~L. Brunton, M.~Budi{\v{s}}i{\'c}, E.~Kaiser, and J.~N. Kutz.
\newblock Modern {K}oopman theory for dynamical systems.
\newblock {\em SIAM Review}, 64(2), 2022.

\bibitem{Tu14a}
J.~H. Tu, C.~W. Rowley, D.~M. Luchtenburg, S.~L. Brunton, and J.~N. Kutz.
\newblock On dynamic mode decomposition: Theory and applications.
\newblock {\em Journal of Computational Dynamics}, 1(2):391--421, 2014.

\bibitem{Klus16a}
S.~Klus, P.~Koltai, and C.~Schutte.
\newblock On the numerical approximation of the {Perron-Frobenius} and
  {Koopman} operator.
\newblock {\em J. Computational Dynamics}, 3(1):51--79, 2016.

\bibitem{colb21a}
M.~J. Colbrook and A.~Townsend.
\newblock Rigorous data-driven computation of spectral properties of {K}oopman
  operators for dynamical systems.
\newblock {\em arXiv preprint arXiv:2111.14889}, 2021.

\bibitem{luca23a}
V.~Lucarini and M.~Chekroun.
\newblock Hasselmann's program and beyond: New theoretical tools for
  understanding the climate crisis.
\newblock {\em arXiv preprint arXiv:2303.12009}, 2023.

\bibitem{Hall15a}
G.~Haller.
\newblock Lagrangian coherent structures.
\newblock {\em Ann. Rev. Fluid Mech.}, 47:137--162, 2015.

\bibitem{Hadj17a}
A.~Hadjighasem, M.~Farazmand, D.~Blazevski, G.~Froyland, and G.~Haller.
\newblock A critical comparison of {L}agrangian methods for coherent structure
  detection.
\newblock {\em Chaos}, 27(5):053104, 2017.

\bibitem{gasp05b}
P.~Gaspard.
\newblock {\em Chaos, scattering and statistical mechanics}.
\newblock Cambridge Univ. Press, 2005.

\bibitem{Klus19a}
S.~Klus, B.~E. Husic, M.~Mollenhauer, and F.~No\'{e}.
\newblock Kernel methods for detecting coherent structures in dynamical data.
\newblock {\em Chaos: An Interdisciplinary Journal of Nonlinear Science},
  29(12):123112, 2019.

\bibitem{Froy03a}
G.~Froyland and M.~Dellnitz.
\newblock Detecting and locating near-optimal almost-invariant sets and cycles.
\newblock {\em SIAM Journal on Scientific Computing}, 24(6):1839--1863, 2003.

\bibitem{Froy09a}
G.~Froyland and K.~Padberg.
\newblock Almost-invariant sets and invariant manifolds — connecting
  probabilistic and geometric descriptions of coherent structures in flows.
\newblock {\em Physica D: Nonlinear Phenomena}, 238(16):1507 -- 1523, 2009.

\bibitem{Noe13a}
F.~No{\'e} and F.~Nuske.
\newblock A variational approach to modeling slow processes in stochastic
  dynamical systems.
\newblock {\em Multiscale Modeling \& Simulation}, 11(2):635--655, 2013.

\bibitem{Nuske16a}
F.~N{\"u}ske, R.~Schneider, F.~Vitalini, and F.~No{\'e}.
\newblock Variational tensor approach for approximating the rare-event kinetics
  of macromolecular systems.
\newblock {\em J. Chemical Physics}, 144(5):054105, 2016.

\bibitem{Froy10a}
G.~Froyland, S.~Lloyd, and N.~Santitissadeekorn.
\newblock Coherent sets for nonautonomous dynamical systems.
\newblock {\em Physica D: Nonlinear Phenomena}, 239(16):1527--1541, 2010.

\bibitem{Froy10b}
G.~Froyland, N.~Santitissadeekorn, and A.~Monahan.
\newblock Transport in time-dependent dynamical systems: Finite-time coherent
  sets.
\newblock {\em Chaos: An Interdisciplinary Journal of Nonlinear Science},
  20(4):043116, 2010.

\bibitem{froy21a}
G.~Froyland and P.~Koltai.
\newblock Detecting the birth and death of finite-time coherent sets.
\newblock {\em arXiv preprint arXiv:2103.16286}, 2021.

\bibitem{Schmi10a}
P.~J. Schmid.
\newblock Dynamic mode decomposition of numerical and experimental data.
\newblock {\em J. Fluid Mechanics}, 656:5--28, 2010.

\bibitem{bagh13a}
S.~Bagheri.
\newblock Koopman-mode decomposition of the cylinder wake.
\newblock {\em Journal of Fluid Mechanics}, 726:596--623, 2013.

\bibitem{hann07a}
A.~Hannachi, I.~T. Jolliffe, and D.~B. Stephenson.
\newblock Empirical orthogonal functions and related techniques in atmospheric
  science: A review.
\newblock {\em Inl. J. Climatology: J. Roy. Meteorological Society},
  27(9):1119--1152, 2007.

\bibitem{Tant15a}
A.~Tantet, F.~R. van~der Burgt, and H.~A. Dijkstra.
\newblock An early warning indicator for atmospheric blocking events using
  transfer operators.
\newblock {\em Chaos: An Interdisciplinary Journal of Nonlinear Science},
  25(3):036406, 2015.

\bibitem{froy21b}
G.~Froyland, D.~Giannakis, B.~R. Lintner, M.~Pike, and J.~Slawinska.
\newblock Spectral analysis of climate dynamics with operator-theoretic
  approaches.
\newblock {\em Nature Communications}, 12(1):6570, 2021.

\bibitem{Will15a}
M.~O. Williams, I.~G. Kevrekidis, and C.~W. Rowley.
\newblock A data-driven approximation of the {K}oopman operator: Extending
  dynamic mode decomposition.
\newblock {\em Journal of Nonlinear Science}, 25(6):1307--1346, 2015.

\bibitem{Arba17a}
H.~Arbabi and I.~Mezic.
\newblock Ergodic theory, dynamic mode decomposition, and computation of
  spectral properties of the {Koopman} operator.
\newblock {\em SIAM Journal on Applied Dynamical Systems}, 16(4):2096--2126,
  2017.

\bibitem{Crut12a}
J.~P. Crutchfield.
\newblock Between order and chaos.
\newblock {\em Nature Physics}, 8(January):17--24, 2012.

\bibitem{Lind95a}
D.~Lind and B.~Marcus.
\newblock {\em An Introduction to Symbolic Dynamics and Coding}.
\newblock Cambridge University Press, New York, 1995.

\bibitem{Mors38a}
M.~Morse and G.~A. Hedlund.
\newblock Symbolic dynamics.
\newblock {\em Am. J. Math.}, 60(4):815--866, 1938.

\bibitem{Smal67a}
S.~Smale.
\newblock Differentiable dynamical systems.
\newblock {\em Bull. Amer. Math. Soc.}, 73:797--817, 1967.

\bibitem{Turi37a}
A.~Turing.
\newblock On computable numbers, with an application to the
  {Entschiedungsproblem}.
\newblock {\em Proc. Lond. Math. Soc.}, 42, 43:230--265, 544--546, 1937.

\bibitem{Chur36a}
A.~Church.
\newblock A note on the entscheidungsproblem.
\newblock {\em J. Symbolic Logic}, 1:40--41, 1936.

\bibitem{Post21a}
E.~Post.
\newblock Introduction to the general theory of elementary propositions.
\newblock {\em Am. J. Math.}, 43:163--185, 1921.

\bibitem{Nage68a}
E.~Nagel and J.~R. Newman.
\newblock {\em {G\"{o}del's} Proof}.
\newblock New York University Press, New York, 1968.

\bibitem{ante96a}
C.~Anteneodo and A.R. Plastino.
\newblock Some features of the {L}{\'o}pez-{R}uiz-{M}ancini-{C}albet ({LMC})
  statistical measure of complexity.
\newblock {\em Physics Letters A}, 223(5):348--354, 1996.

\bibitem{Weis73}
B.~Weiss.
\newblock Subshifts of finite type and sofic systems.
\newblock {\em Monastsh. Math.}, 77:462, 1973.

\bibitem{Kitc86a}
B.~Kitchens and S.~Tuncel.
\newblock Semi-groups and graphs.
\newblock {\em Israel. J. Math.}, 53:231, 1986.

\bibitem{ginz68a}
A.~Ginzburg.
\newblock {\em Algebraic theory of automata}.
\newblock Academic Press, 1968.

\bibitem{rupe22a}
A.~Rupe and J.~P. Crutchfield.
\newblock Algebraic theory of patterns as generalized symmetries.
\newblock {\em Symmetry}, 14(8):1636, 2022.

\bibitem{Mins67}
M.~Minsky.
\newblock {\em Computation: Finite and Infinite Machines}.
\newblock Prentice-Hall, Englewood Cliffs, New Jersey, 1967.

\bibitem{Huff59a}
D.~Huffman.
\newblock Canonical forms for information-lossless finite-state logical
  machines.
\newblock {\em IRE Trans. Circ. Th.}, 6:41--59, 1959.

\bibitem{Shal98a}
C.~R. Shalizi and J.~P. Crutchfield.
\newblock Computational mechanics: Pattern and prediction, structure and
  simplicity.
\newblock {\em J. Stat. Phys.}, 104:817--879, 2001.

\bibitem{Crut08a}
J.~P. Crutchfield, C.~J. Ellison, and J.~R. Mahoney.
\newblock Time's barbed arrow: {Irreversibility}, crypticity, and stored
  information.
\newblock {\em Phys. Rev. Lett.}, 103(9):094101, 2009.

\bibitem{Rupe23a}
A.~Rupe, K.~Kashinath, N.~Kumar, and J.~P. Crutchfield.
\newblock Unsupervised discovery of extreme weather events using universal
  representations of emergent organization.
\newblock 2023.
\newblock arXiv:2206.15050.

\bibitem{Rupe18a}
A.~Rupe and J.~P. Crutchfield.
\newblock Local causal states and discrete coherent structures.
\newblock {\em Chaos}, 28(7):1--22, 2018.

\bibitem{Rupe18b}
A.~Rupe and J.~P. Crutchfield.
\newblock Spacetime symmetries, invariant sets, and additive subdynamics of
  cellular automata.
\newblock {\em arXiv preprint arXiv:1812.11597}, 2018.

\bibitem{Rupe19a}
A.~Rupe, N.~Kumar, V.~Epifanov, K.~Kashinath, O.~Pavlyk, F.~Schlimbach,
  M.~Patwary, S.~Maidanov, V.~Lee, Prabhat, and J.~P. Crutchfield.
\newblock Disco: Physics-based unsupervised discovery of coherent structures in
  spatiotemporal systems.
\newblock In {\em 2019 IEEE/ACM Workshop on Machine Learning in High
  Performance Computing Environments (MLHPC)}, pages 75--87. IEEE, 2019.
\newblock arXiv:1909.11822 [physics.comp-ph].

\bibitem{Marz17b}
S.~Marzen and J.~P. Crutchfield.
\newblock Structure and randomness of continuous-time discrete-event processes.
\newblock {\em J. Stat. Physics}, 169(2):303--315, 2017.

\bibitem{brod22a}
N.~Brodu and J.~P. Crutchfield.
\newblock Discovering causal structure with reproducing-kernel {H}ilbert space
  $\epsilon$-machines.
\newblock {\em Chaos: An Interdisciplinary Journal of Nonlinear Science},
  32(2):023103, 2022.

\bibitem{Stre13a}
C.~C. Strelioff and J.~P. Crutchfield.
\newblock Bayesian structural inference for hidden processes.
\newblock {\em Phys. Rev. E}, 89:042119, 2014.

\bibitem{Crut13a}
J.~P. Crutchfield, C.~J. Ellison, and P.~Riechers.
\newblock Exact complexity: {The Spectral} decomposition of intrinsic
  computation.
\newblock {\em Physics Letters A}, 380(9-10):998--1002, 2016.

\bibitem{Jurg20c}
A.~Jurgens and J.~P. Crutchfield.
\newblock Divergent predictive states: The statistical complexity dimension of
  stationary, ergodic hidden {Markov} processes.
\newblock {\em Chaos}, 31(8):0050460, 2021.

\bibitem{gasp95a}
P.~Gaspard, G.~Nicolis, A.~Provata, and S.~Tasaki.
\newblock Spectral signature of the pitchfork bifurcation: {L}iouville equation
  approach.
\newblock {\em Physical Review E}, 51(1):74, 1995.

\bibitem{Crut93g}
J.~P. Crutchfield.
\newblock Is anything ever new? {C}onsidering emergence.
\newblock In G.~Cowan, D.~Pines, and D.~Melzner, editors, {\em Complexity:
  Metaphors, Models, and Reality}, volume XIX of {\em Santa Fe Institute
  Studies in the Sciences of Complexity}, pages 479--497, Reading, MA, 1994.
  Addison-Wesley.
\newblock Santa Fe Institute Technical Report 94-03-011; reprinted in
  Emergence: Contemporary Readings in Philosophy and Science, M. A. Bedau and
  P. Humphreys, editors, Bradford Book, MIT Press, Cambridge, MA (2008)
  269-286.

\bibitem{costa19a}
A.~C. Costa, T.~Ahamed, and G.~J. Stephens.
\newblock Adaptive, locally linear models of complex dynamics.
\newblock {\em Proceedings of the National Academy of Sciences},
  116(5):1501--1510, 2019.

\bibitem{costa23a}
A.~C. Costa, T.~Ahamed, D.~Jordan, and G.~J. Stephens.
\newblock {Maximally predictive states: From partial observations to long
  timescales}.
\newblock {\em Chaos: An Interdisciplinary Journal of Nonlinear Science},
  33(2):023136, 02 2023.

\bibitem{Wild98a}
R.~Wilde and S.~Singh.
\newblock {\em Statistical Mechanics: Fundamentals and Modern Applications}.
\newblock Wiley \& Sons, New York, first edition, 1998.

\bibitem{chor02a}
A.~J. Chorin, O.~H. Hald, and R.~Kupferman.
\newblock Optimal prediction with memory.
\newblock {\em Physica D: Nonlinear Phenomena}, 166(3-4):239--257, 2002.

\bibitem{Zwan73a}
R.~Zwanzig.
\newblock Nonlinear generalized {Langevin} equations.
\newblock {\em J. Statistical Physics}, 9(3):215--220, 1973.

\bibitem{Lin21a}
K.~K. Lin and F.~Lu.
\newblock Data-driven model reduction, {Wiener} projections, and the
  {Koopman-Mori-Zwanzig} formalism.
\newblock {\em J. Computational Physics}, 424:109864, 2021.

\bibitem{Gila21a}
F.~Gilani, D.~Giannakis, and J.~Harlim.
\newblock Kernel-based prediction of {non-Markovian} time series.
\newblock {\em Physica D: Nonlinear Phenomena}, 418:132829, 2021.

\bibitem{rupe22b}
A.~Rupe, V.~V. Vesselinov, and J.P. Crutchfield.
\newblock Nonequilibrium statistical mechanics and optimal prediction of
  partially-observed complex systems.
\newblock {\em New Journal of Physics}, 24(10):103033, 2022.

\bibitem{feyn11a}
R.P. Feynman.
\newblock Feynman on scientific method.
\newblock \url{https://www.youtube.com/watch?v=EYPapE-3FRw}, Accessed:
  2023-06-25.

\bibitem{boss16a}
T.~Bossomaier, L.~Barnett, M.~Harr{\'e}, and J.~T. Lizier.
\newblock {\em Transfer Entropy}.
\newblock Springer, 2016.

\bibitem{smir13a}
D.~A. Smirnov.
\newblock Spurious causalities with transfer entropy.
\newblock {\em Phys. Rev. E}, 87(4):042917, 2013.

\bibitem{pear18a}
J.~Pearl and D.~Mackenzie.
\newblock {\em The book of why: the new science of cause and effect}.
\newblock Basic books, 2018.

\bibitem{sugi12a}
G.~Sugihara, R.~May, Hao Ye, Chih-hao H., E.~Deyle, M.~Fogarty, and S.~Munch.
\newblock Detecting causality in complex ecosystems.
\newblock {\em Science}, 338(6106):496--500, 2012.

\bibitem{rung19a}
J.~Runge, P.~Nowack, M.~Kretschmer, S.~Flaxman, and D.~Sejdinovic.
\newblock Detecting and quantifying causal associations in large nonlinear time
  series datasets.
\newblock {\em Science Advances}, 5(11), 2019.

\bibitem{sinh20a}
S.~Sinha and U.~Vaidya.
\newblock On data-driven computation of information transfer for causal
  inference in discrete-time dynamical systems.
\newblock {\em J. Nonlinear Science}, 30(4):1651--1676, 2020.

\end{thebibliography}

\end{document}